\documentclass[12pt]{iopart}
\usepackage{epsfig}
\usepackage{graphicx}
\usepackage{iopams}
\def\aA{$\alpha$-nucleus }
\def\AA{nucleus-nucleus }
\def\nA{nucleon-nucleus }
\def\oo{$^{16}$O+$^{16}$O }
\def\cc{$^{12}$C+$^{12}$C }
\def\oc{$^{16}$O+$^{12}$C }
\def\aC{$\alpha+^{12}$C }
\def\Li6C{$^6$Li+$^{12}$C }
\def\LiC11{$^{11}$Li+$^{12}$C }
\def\He6C{$^{6}$He+$^{12}$C }
\def\aNi{$\alpha+^{58}$Ni }
\def\aCa{$\alpha$+$^{40}$Ca\ }
\def\aZr{$\alpha$+$^{90}$Zr\ }
\def\7o5o{$^{17}$O+$^{15}$O\ }
\def\o17o15{$^{16}$O($^{16}$O,$^{17}$O)$^{15}$O\ }
\def\x7o5o{$^{17}$O+$^{15}$O$^*$\ }
\def\xo17o15{$^{16}$O($^{16}$O,$^{17}$O)$^{15}$O$^*$\ }
\def\InOO{$^{16}{\rm O}(^{16}{\rm O},^{16}$O$^{'})^{16}{\rm O}^*$\ }

\def\2x{$^{16}$O+$^{16}$O$_{2^+}$\ }
\def\3x{$^{16}$O+$^{16}$O$_{3^-}$\ }
\begin{document}

\topical[ ]{Nuclear rainbow scattering and nucleus-nucleus potential}

\author{Dao T Khoa\dag\S\ W von Oertzen\dag\ddag\ H G Bohlen\dag\ and
 S Ohkubo\P\ }

\address{\dag\ Hahn-Meitner-Institut Berlin, Glienicker Str.100, D-14109 Berlin,
 Germany}
\address{\S\ Institute for Nuclear Science and
 Technique, VAEC, P.O. Box 5T-160,\\ Nghia Do, Hanoi, Vietnam}
\address{\ddag\ Freie Universit\"at Berlin, Fachbereich Physik, Berlin, Germany}
\address{\P\ Department of Applied Science and Environment,
 Kochi Women's University, Kochi 780-8515, Japan}
\ead{khoa@vaec.gov.vn, oertzen@hmi.de, bohlen@hmi.de,
 shigeo@cc.kochi-wu.ac.jp}

\begin{abstract} Elastic scattering of $\alpha$-particle and some
tightly-bound light nuclei has shown the pattern of \emph{rainbow scattering}
at medium energies, which is due to the \emph{refraction} of the incident wave
by a strongly attractive nucleus-nucleus potential. This review gives an
introduction to the physics of the nuclear rainbow based essentially on the
optical model description of the elastic scattering. Since the realistic \AA
optical potential (OP) is the key to explore this interesting process, an
overview of the main methods used to determine the \AA OP is presented. Given
the fact that the absorption in a rainbow system is much weaker than that
usually observed in elastic heavy-ion scattering, the observed rainbow patterns
were shown to be linked directly to the density overlap of the two nuclei
penetrating each other in the elastic channel, with a total density reaching up
to twice the nuclear matter saturation density $\rho_0$. For the calculation of
the \AA OP in the double-folding model, a realistic density dependence has been
introduced into the effective M3Y interaction which is based originally on the
G-matrix elements of the Reid- and Paris nucleon-nucleon (NN) potentials. Most
of the elastic rainbow scattering data were found to be best described by a
\emph{deep} real OP like the folded potential given by this density dependent
M3Y interaction. Within the Hartree-Fock formalism, the same NN interaction
gives consistently a \emph{soft} equation of state of cold nuclear matter which
has an incompressibility constant $K \approx 230-260$ MeV. Our folding analysis
of numerous rainbow systems has shown that the elastic \aA and \AA refractive
rainbow scattering is indeed a very helpful experiment for the determination of
the realistic $K$ value. The refractive rainbow-like structures observed in
other quasi-elastic scattering reactions have also been discussed. Some
evidences for the refractive effect in the elastic scattering of unstable
nuclei are presented and perspectives for the future studies are discussed.
\end{abstract}

\submitto{\JPG} \maketitle

\section{Introduction}
\subsection{Atmospheric Rainbow}
The commonly known atmospheric rainbow is observed whenever there are water
droplets illuminated by the sun light. It can be seen during the rain with the
sunshine not completely covered by the clouds (see Fig.~1) or from a fountain,
when the sunlight enters from behind the point of observation. Besides the
fascinating effect of colour splitting due to the dependence of the refraction
on the wavelength, the more interesting physics effect is the \emph{increased}
light intensity around the rainbow angle $\Theta_{\rm R}$ and the
\emph{shadow}-region lying beyond $\Theta_{\rm R}$.
\begin{figure}[ht]
\begin{center}
\hspace{-0.5cm}
\includegraphics[angle=270,scale=0.50]{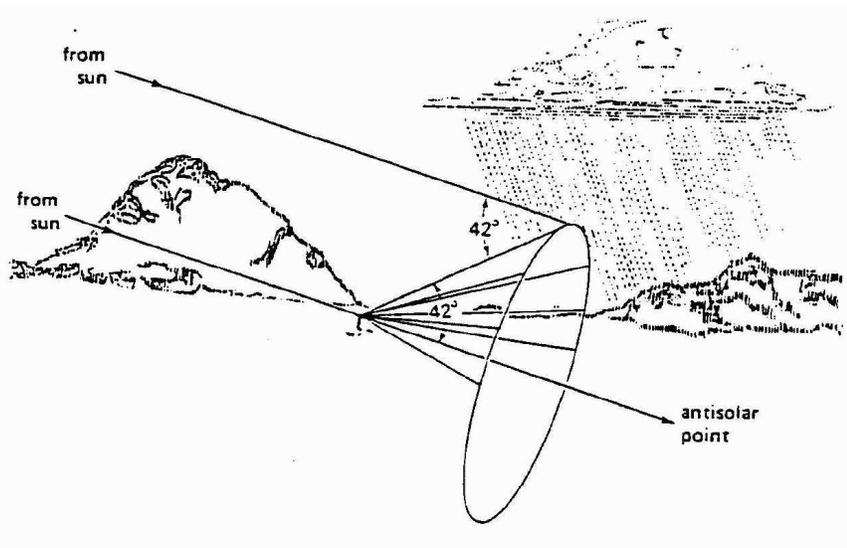}
\caption{Descartes traced the light ray reflected from a uniform rain drop and
found a ``critical'' ray which has a minimal deflection angle of about
$138^\circ$, whose supplementary $\Theta_{\rm R} \simeq 42^\circ$ is the
largest and known nowadays as the ``rainbow'' angle. The atmospheric rainbow is
produced by the piling up of the light rays near $\Theta_{\rm R}$ which is
slightly larger than $42^\circ$ for red light and smaller for blue, and hence
the colour splitting of the ``white'' sunlight. Illustration taken from
Ref.~\cite{Greenler}.}
 \label{fig1}
\end{center}
\end{figure}

\begin{figure}[ht]
\vspace{-1cm}
 \begin{center}
\hspace{6cm}
\includegraphics[angle=270,scale=0.5]{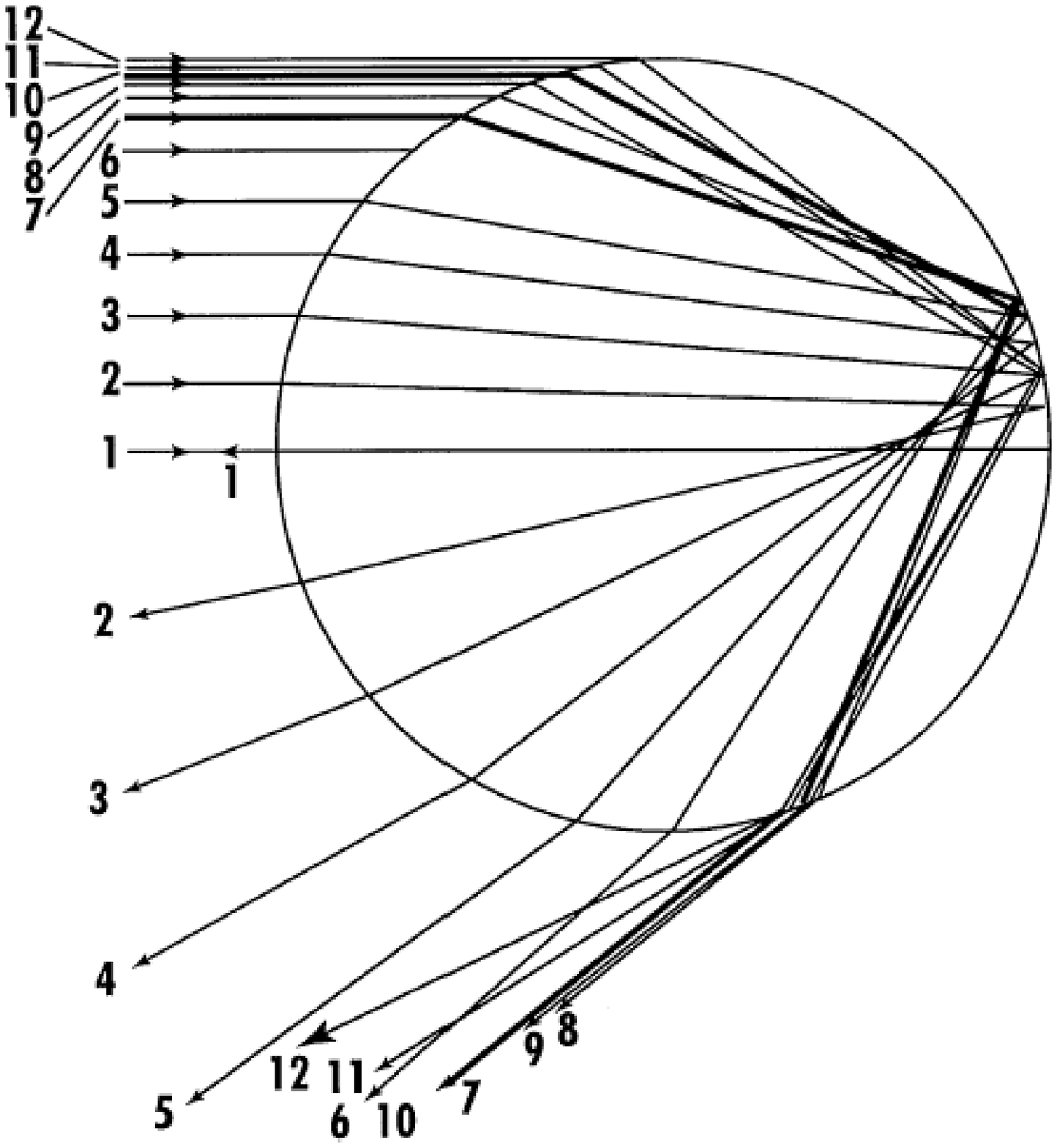}
\caption{The paths of light rays entering a spherical water drop at different
        impact parameters and leaving it after a
        refraction - reflection - refraction
        sequence. The ray numbered 7 is the rainbow ray, at which
        light is deflected to a maximal (negative) scattering angle
        of around $42^\circ$. Illustration taken from Ref.~\cite{Adam}.}
   \end{center}
  \label{fig2}
\end{figure}

The first modern explanation of the atmospheric rainbow was given by Descartes
in 1637 in his book ``Les Meteores''. An illustration of the atmospheric rainbow
according to Descartes' interpretation is shown in Fig.~1. Historically, Noah
(Genesis 9:13 ``\emph{I have set my rainbow in the clouds, and it will be the
sign of convening between me and earth}''), and his family can be identified as
the first ``recorded'' observers of the rainbow, since the person speaking in
singular ``I'' should have known the rainbow well before. Since Descartes' time
up to the present, the physics of the atmospheric rainbow has been described by
different models, ranging from a simple classical geometrical ray optics to the
quantum mechanical complex angular momentum theory for the scattering of
electromagnetic waves \cite{Nussenzweig,Adam} with higher and higher
mathematical sophistication. For interested readers there exists in the
literature a number of excellent monographs and reviews on the physics of the
atmospheric rainbow, like the book by Greenler on rainbows, halos and glories
\cite{Greenler} or the review articles by Nussenzweig \cite{Nussenzweig} and
Adam \cite{Adam}. Especially, one can learn in the latter about all the main
physical and mathematical approaches used sofar to study the physics of the
atmospheric rainbow. Educational information about the atmospheric rainbow
is also available in the internet (see, e.g.,
\emph{http://www.meteoros.de/rainbow/rainbow.htm}). In our topical review we do
not attempt to cover all this development, but will concentrate on the
interference pattern of the refractive nucleus-nucleus scattering which gives
rise to the nuclear rainbow. Note that the rainbow-like interference pattern has
been observed also in molecular and atomic scattering \cite{Bra96}.
\begin{figure}[ht]
 \begin{center}
  \includegraphics[angle=0,scale=0.5]{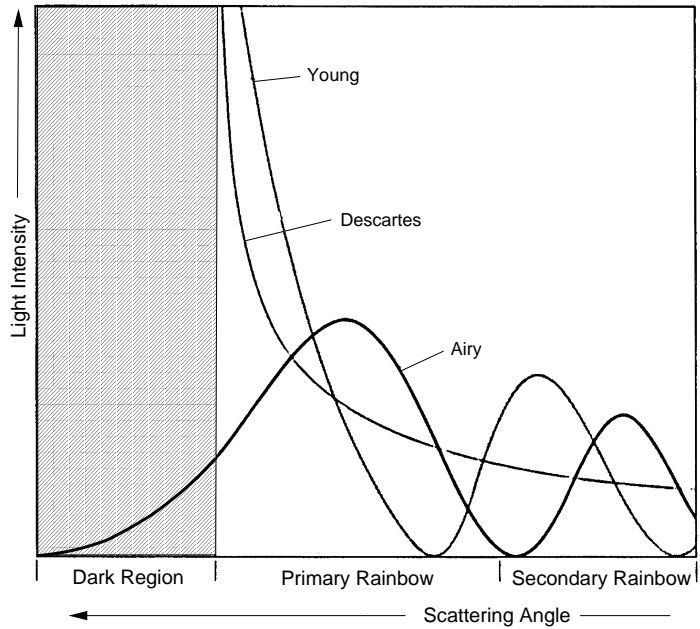}
\caption{Different descriptions of the light intensity of the atmospheric
rainbow. While the ``classical'' solutions by Descartes, Newton and Young give a
divergence at the rainbow angle, Airy's description shows clearly the wave
nature of light which penetrates also into the shadow region beyond the rainbow
angle. The rainbow structure inside the main rainbow, see Fig.~5, is due to
the second order Airy structure.}
  \end{center}
 \label{fig3}
\end{figure}

The light focusing near the rainbow angle, as a result of a refraction -
reflection - refraction process by the light rays in a spherical water drop, is
illustrated in Fig.~2. The first refraction occurs when the light enters the
water drop, the refracted light undergoes then a reflection and is refracted
again before leaving the drop. One can see in Fig.~2 an interesting variation of
the ``deflection angle'' as function of the impact parameter $b$, from the
``head-on'' ray with a maximal deflection at 180$^\circ$ to the ``rainbow'' ray
with a minimal deflection at about 138$^\circ$. For further discussion, it is
convenient that we discuss this refraction - reflection process in terms of
negative deflection $\Theta$ of the light ray, i.e., the supplementary of the
deflection angle so that the rainbow ray is deflected at $\Theta_{\rm R}\simeq
42^\circ$. The interesting physics effect here is the enhanced light intensity
(concentration of many light rays, e.g., those numbered 6 to 12 in Fig.~2) near
the rainbow angle $\Theta_{\rm R}$, which is followed by a ``shadow'' region.
Classically  \cite{Bra97}, this shadow is produced by the maximum of the
deflection function $\Theta(b)$ because the intensity (or in our cases: cross
section $\displaystyle\frac{d\sigma}{d\Omega}$) of the scattered light is
proportional to the inverse of the first derivative of $\Theta(b)$, as can be
seen in the following expression
\begin{equation}
 \frac{d\sigma}{d\Omega}=\sum_b\frac{b}{\sin\Theta(b)|d\Theta(b)/db|}.
\end{equation}
Here the sum is taken over the light rays entering the water drop at different
impact parameters $b$, some of them scatter to the same angle $\Theta$, like
those numbered 7, 8 and 9 in Fig.~2, with the maximal (negative) deflection
occurring near the rainbow angle $\Theta_{\rm R}$. This simple expression of
light intensity has a \emph{divergence} at $\Theta_{\rm R}$ (with
$d\Theta(b)/db$=0), which is also known as a caustic in optics. The
existence of the divergence of the light intensity  at the rainbow angle could
not be remedied in both the Descartes' theory and the more advanced version of
ray optics by Newton and Young (see Fig.~3). Moreover, the observation of the
supernumerary bows inside the primary rainbow (see Fig.~5 below) persistently
pointed to the inadequacy of Descartes' and Newton's models which could not
explain the origin of the supernumeraries. All this remained a mathematical
challenge until Airy, in the 19-th century, provided the first mathematical
model of the rainbow based on the light wave diffraction and interference
\cite{Airy}.
\begin{figure}[ht]
 \begin{center}\vspace{-1.5cm}
  \includegraphics[angle=270,width=1.1\textwidth]{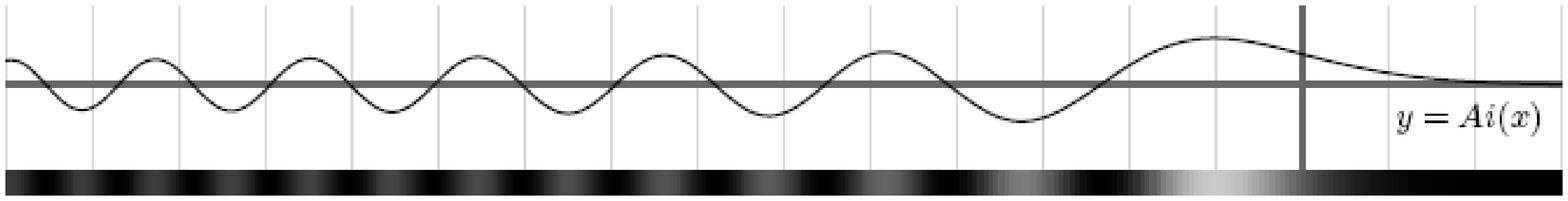}\vspace{-8cm}
   \caption{Graph of the Airy rainbow integral $Ai(x)$. The argument $x$ is
   proportional to $\Theta-\Theta_{\rm R}$ so that $x=0$ corresponds to
   $\Theta=\Theta_{\rm R}$ and positive $x$ is on the dark side of the rainbow.
   Below, according to Airy's theory, illumination is proportional to $Ai(x)^2$
   which gives rise to the \emph{primary} bow along with several
 \emph{supernumerary} bows. Illustration taken from Ref.~\cite{Adam}.}
 \label{fig4}
 \end{center}
\end{figure}

By using the standard Huygens' concept of the light wavefront rather than the
optical light rays, Airy has shown the self-interference of such a wavefront as
it becomes folded onto itself during the refraction and reflection within the
rain drop. As a result, the primary rainbow is the first interference maximum,
the second and third maxima being the first and second supernumerary bows,
respectively. Airy described the local intensity of scattered light wave with
the help of a ``rainbow integral'', which is now known as the Airy function
$Ai(x)$ (see Fig.~4). Airy's model not only removed the divergence of the light
intensity at the rainbow angle as shown in Fig.~3, but also explained
successfully the existence of the supernumerary bows which appear naturally as
the maxima of $Ai(x)$. Although more precise and sophisticated mathematical
physics approaches  to describe the atmospheric rainbow have been developed
later in the 20-th century \cite{Nussenzweig,Adam}, Airy's model remains a
simple but very realistic approximate description of the rainbow pattern, which
has been used widely to identify the rainbow features observed in molecular,
atomic and nuclear scattering \cite{Bra96,Bra97}.

\begin{figure}[ht]
 \begin{center}
  \includegraphics[angle=270,scale=0.58]{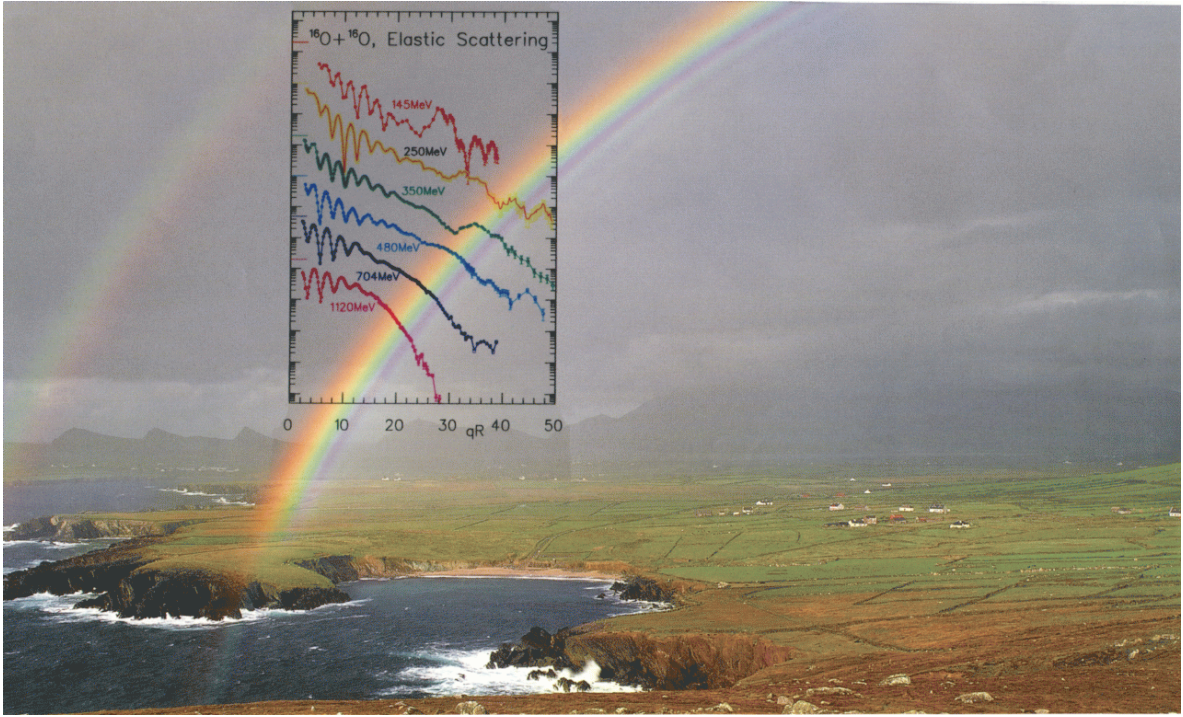}
  \vspace{1cm}
\caption{Photographic image of the atmospheric rainbow where both the primary
and secondary rainbows can be seen. The faint bows located inside the primary
rainbow are the supernumeraries which were first explained in 1838 by Airy
\cite{Airy}. The secondary rainbow is formed by light rays undergoing a second
reflection in the rain drops and hence is fainter and has a reversed sequence
of colours. The inset along the primary bow shows the elastic \oo scattering
data (plotted as function of momentum transfer) measured at different
laboratory energies, where the most pronounced rainbow pattern associated with
the first Airy maximum has been observed at 350 MeV. At lower energies the
secondary Airy maximum was also observed. The rainbow angle $\Theta_{\rm R}$
found for the elastic \oo scattering, at energies of 250 to 1120 MeV, is
located near the maximum of the light intensity in the atmospheric rainbow.
Illustration taken from Ref.~\cite{vOe00}.}
  \end{center}
 \label{fig5}
\end{figure}

\subsection{Nuclear Rainbow}
Nuclei are known to have wave properties and they can be diffracted, refracted
and suffer interference just like the sun light. Consequently, the \AA
scattering may also display rainbow features depending on the scattering
conditions and binding structure of the projectile and target. In terms of the
wave scattering theory, if the scattering proceeds only elastically the total
flux would remain unchanged. Therefore, the nuclear rainbow should be strongest
in the elastic scattering channel if a system with small absorption is chosen.
Indeed, the nuclear rainbow pattern has been first observed during the 70's of
the last century in the elastic \aA scattering \cite{Go72,Go73,Go74}, and later on
in the elastic scattering measured for some strongly bound light heavy-ion
systems like \cc \cite{Stokstad,Boh82,Boh85} or \oo \cite{Sti89,Boh93,Bar96}.
For these systems the absorption due to nuclear reactions was sufficiently low
for the rainbow effect to appear. The concept of the wave \emph{refraction}
implies that the wavelength changes whenever the wave penetrates from one
medium into another. In a realistic case of the elastic \AA scattering, the de
Broglie wavelength $\lambda_{\rm B}$ of the scattered wave is changed as the
projectile penetrates the target nucleus at the internuclear distance $R$.
There, the strong projectile-target interaction occurs, due to the local
strength $V(R)$ of the (real) \AA optical potential (OP), and the wavelength
$\lambda_{\rm B}(R)$ is determined as
\begin{equation}
 \lambda_{\rm B}(R)=h/\sqrt{2\mu[E-V(R)]},
 \label{eq1}
\end{equation}
where $E$ is the centre-of-mass energy of the scattering system, $\mu$ is the
reduced mass. It is obvious that the nuclear OP, used in the optical model (OM)
calculation to describe the \AA elastic scattering, is the most important
physics input in the study of nuclear rainbow scattering. We further determine
the quantal analog of the index of refraction $n(R)$ for nuclear scattering in
the same way as in the classical optics (ratio of the velocity of the scattered
wave to that of free wave) as
\begin{equation}
 n(R)=\sqrt{1-V(R)/E}.
 \label{eq2}
\end{equation}
The refractive index is larger unity ($n>1$), because the potentials are
attractive ($V(R)<0$), and can reach values well above 2 at low energies. The
big difference between the index of refraction used in the classical optics and
its quantal counterpart (\ref{eq2}) is that the latter is not uniformly
constant but depends on the strength of the OP at given interaction distance
$R$. This is the main reason, why the nuclear rainbow is much more difficult to
observe and to identify. It also becomes clear from Eqs.~(\ref{eq1}) and
(\ref{eq2}) that the nuclear rainbow can only be properly analyzed and
identified based on a correct choice of the nuclear OP. The \AA OP is,
therefore, the key physics quantity being discussed throughout this topical
review.
\begin{figure}[ht]
 \vspace{-2cm}
  \begin{center}
  \includegraphics[angle=270,scale=0.65]{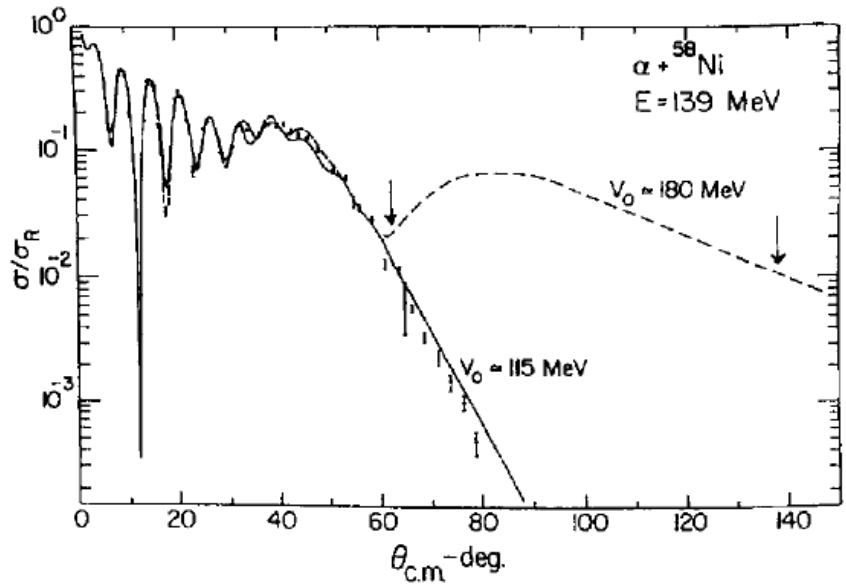}
  \vspace{-1.5cm}
\caption{Elastic \aNi scattering data at $E_{\rm lab}=139$ MeV \cite{Go73} and
the OM descriptions given by two different choices of the real OP of
Woods-Saxon shape (with the depths $V_0$ indicated) which give the rainbow
angle of $\Theta_{\rm R}\approx 61^\circ$ and 139$^\circ$, respectively.
Illustration
 taken from Ref.~\cite{Bra97}.}
  \end{center}
 \label{fig6}
\end{figure}

As already mentioned, the first observation of nuclear rainbow was made some 30
years ago by Goldberg {\sl et al.} \cite{Go72,Go73,Go74} in their experiments
on elastic \aA scattering at $E_{\rm lab}\approx 140$ MeV (see Fig.~6). They
have found that a strong and broad maximum of the elastic cross section at
large scattering angles is of refractive nature, and thus can be identified as
a nuclear rainbow \cite{De77}. Moreover, the most significant physics effect
established by these first experiments on the nuclear rainbow is that the
extension of the elastic scattering data well beyond the rainbow angle
$\Theta_{\rm R}$ (marked in Fig.~6 by the first arrow) allowed the elimination
of discrete ambiguities in the depth of the \aA OP. In the \aNi case, the data
up to $\Theta\approx 60^\circ$ could be described by a number of Woods-Saxon
(WS) shaped potentials with discretely different central depth parameters $V_0$
of the real OP. The observation of the exponential fall-off of the rainbow
maximum well beyond $60^\circ$ allowed the unique selection of $V_0=115$ MeV
for the depth of the potential as the most appropriate one.

Consequently, extracting important information about the \AA OP at short
distances and, hence, testing the validity of different theoretical models of
the \AA OP has always been one of the main goals of numerous experimental and
theoretical studies of the nuclear rainbow during the last three decades.

\section{Rainbows in the elastic \AA scattering }
\label{sec2}
\subsection{Hindrance by a strong absorption}

In general, the elastic \AA scattering is  associated with a strong absorption,
i.e., the partial loss of the incident wave from the elastic scattering channel
into various non-elastic reaction channels during the \AA collision. As a
result, the \AA or heavy-ion (HI) optical potential always has an imaginary
part, $W$, which describes the absorption from the elastic channel. This
absorption can suppress significantly the refractive structure of the elastic
scattering and no rainbow-like effect has been observed in strongly absorbing
HI systems. Since the total optical potential (denoted hereafter as $U$) is
complex, $U=V+iW$, with the real part $V$ describing elastic scattering and
the imaginary part $W$ describing absorption, we can draw an analogy between the
nuclear absorption and the absorption of light by a cloudy crystal ball using a
complex index of refraction. It is natural that the stronger the absorption,
the smaller the chance to observe the rainbow pattern. Due to the strong
absorption, most of the elastic HI scattering is dominated by the surface
scattering and the information about the \AA interaction is obtained for
peripheral trajectories only. Here, the term ``surface" means the region where
the nuclear forces begin to act strongly \cite{Bra97}. The location of this
surface region can be represented by a strong absorption radius $R_{\rm sa}$
which can be identified with the apsidal distance on a Rutherford orbit with
the same angular momentum as that for which the transmission coefficient is
one-half. In terms of the distance between the centres of the two colliding
nuclei with mass numbers $A_{1}$ and $A_{2}$, $R_{\rm sa}$ can be parameterized
\cite{Bra97} as
\begin{equation}
 R_{\rm sa} = r_0\big(A^{1/3}_{1} + A^{1/3}_{2}\big) + \Delta.
 \label{eq3}
\end{equation}
Since $r_0\approx 1.1$ fm, the first term represents the sum of the radii of
the density distributions of the two nuclei and the second term $\Delta$ is the
separation of their surfaces. Values of $\Delta$ between 2 and 3 fm are typical
separations at energies of 10 to 20 MeV/nucleon. The value of the radius
$R_{\rm sa}$, and hence the separation $\Delta$, decreases slowly as the energy
increases \cite{RO87,RO88} ($\Delta$ has reduced to values of  1-2 fm at
energies around 100 MeV/nucleon). Thus, a strong absorption usually takes place
well before there is any substantial overlap of the two nuclear density
distributions. The strong absorption makes it difficult or impossible to gain
any information on the \AA OP at short distances, where the two nuclei begin to
overlap appreciably. This situation is typical for most of the HI elastic
scattering systems \cite{Sat79}, especially those involving medium to heavy
nuclei.
\begin{figure}[ht]
 \vspace{-1.5cm}
  \begin{center}
  \includegraphics[angle=270,scale=0.5]{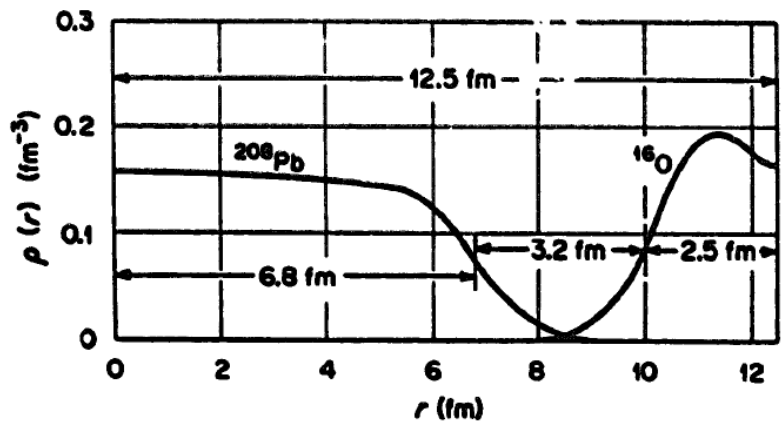}\vspace{-2cm}
\caption{The densities of the nuclei $^{16}$O and $^{208}$Pb when their centres
are separated by the strong absorption radius $R_{\rm sa}$ deduced for the
elastic scattering at incident energies of 100 - 200 MeV. Illustration taken
from Ref.~\cite{Sat79}.}
  \end{center}
 \label{fig7}
\end{figure}

In a semiclassical representation, as soon as a \AA collision occurs, the
composite projectile will induce various (inelastic) reactions and ``dissolve
like a piece of salt (or sugar) in water''. The reaction rates depend on the
``chemical potentials" of the interacting systems. These non-elastic reaction
channels give rise to the imaginary part $W$ of the \AA OP. The mean free path
$\lambda_{\rm mfp}$ for the penetration of the projectile into the target can
be determined in the WKB approximation \cite{BrogliaW} as
\begin{equation}
\lambda_{\rm mfp}(R)
 =\frac{h}{\sqrt{\mu}}\frac{\sqrt{2[E-V(R)]}}{|W(R)|}=v(R)\tau(R),
\label{lambda}
\end{equation}
where $v(R)$ is the projectile velocity and $\tau(R)$ is the average life time
of the projectile in the nuclear medium (the average time until a nuclear
reaction occurs which changes the ground-state structure of the projectile
and/or target). As an example, $\lambda_{\rm mfp}(R)$ values determined for the
elastic \oo scattering at 350 MeV are shown in lower part of Fig.~8.

Let us consider now the quantum mechanical expansion of the elastic \AA
scattering amplitude into a partial-wave series \cite{Sat83}
\begin{equation}
 f(\Theta)=f_{\rm C}(\Theta)+\frac{i}{2k}\sum_l (2l+1)\exp{(2i\sigma_l)}
 (1-S_l)P_l(\cos\Theta),
 \label{Smat}
\end{equation}
where $f_{\rm C}(\Theta)$ is the amplitude of the Coulomb scattering,
$\sigma_l$ - the Coulomb phase shift, $k$ - the wave number, $S_l$ - the
scattering matrix element for the $l$-th partial wave, and $P_l(\cos\Theta)$ -
the Legendre polynomial. In such a representation, the magnitude of $|S_l|$ gives
us the measure of the absorption strength at a given impact parameter or
internuclear distance $R\approx (l+1/2)\hbar/k$. For a strong absorbing HI
system one usually has $|S_l|\leq 10^{-4}$ for $l<l_{\rm g}\approx kR_{\rm g}$,
where $R_{\rm g}$ is the critical or grazing distance at which the
colliding pair begin to experience the strong nuclear interaction acting between
them. Then, the transmission coefficient $T_{l} = 1 - |S_l|^2$ is close to  zero and
represents complete absorption to within 10$^{-8}$ \cite{Bra97}.

\subsection{Incomplete absorption and the rise of the nuclear rainbow}

The situation is, however, different for \aA and light HI systems consisting of
strongly bound nuclei, where the refractive or rainbow pattern has been
observed. In such a case, due to a weaker absorption in the \AA system, one can
observe elastic scattering events occurring at sub-surface distances (with
$l<l_{\rm g}$). The elastic cross section is larger at large scattering angles
and carries information on the \AA OP at smaller partial waves or shorter
distances. To illustrate this effect, we have plotted in Fig.~8 the mean free
path $\lambda_{\rm mfp}$ for the \oo system, using a best-fit complex OP which
reproduces the rainbow pattern observed at $E_{\rm lab}=350$ MeV \cite{Sti89}.
One can see that $\lambda_{\rm mfp}$ is reduced to its (asymptotic) minimal
value at $R\leq 5$ fm. This distance is significantly smaller than the strong
absorption radius $R_{\rm sa}\approx 7-8$ fm as given by the global systematics
(\ref{eq3}) from Ref.~\cite{Bra97}, and the measured elastic \oo data
\cite{Sti89} at this energy also have one of the most pronounced rainbow
pattern ever observed in HI elastic scattering. The elastic \oo data at 350
MeV and those at lower energies show consistently a well defined Airy pattern
which is very helpful in reducing the ambiguity of the shape and depth of the
OP \cite{Kon90,Kho00}.
\begin{figure}[ht]
 \begin{center}
  \includegraphics[width=0.5\textwidth]{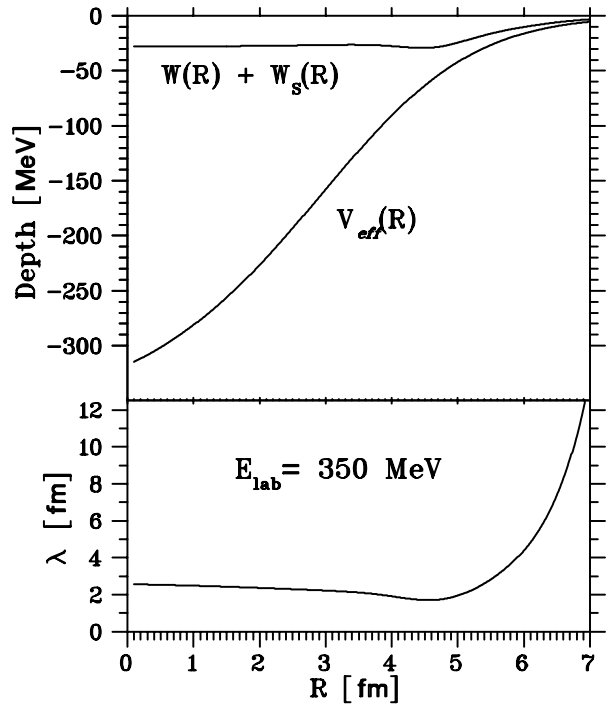}
   \caption{Upper part: Real (V) and imaginary (W) parts of the \oo optical
potential obtained from the OM fit to the elastic \oo scattering at
 $E_{\rm lab}=350$~MeV. Lower part: The mean free path $\lambda_{\rm mfp}$,
obtained from Eq.~(\ref{lambda}) with this complex OP, versus the distance $R$
between the centres of the two $^{16}$O nuclei.}
   \end{center}
  \label{fig8}
\end{figure}

In the semiclassical representation, a weak absorption allows us to keep the
underlying trajectory picture for the scattering system. Fig.~9 illustrates
some typical trajectories for the elastic wave scattered by an attractive
nuclear potential plus a repulsive Coulomb potential.  The scattering angle as
a function of impact parameter $R$, or angular momentum $(l+1/2)\hbar=kR$, is
called the \emph{deflection function} $\Theta(l)$ and is shown on the right
part of Fig.~9. In the semiclassical or WKB approximation, the deflection
function is expressed through the real scattering phase shifts $\delta(l)$ as
\begin{equation}
\Theta(l)=2\frac{d\delta(l)}{dl}. \label{deflect}
\end{equation}
The most important contribution by Airy was to show, that the atmospheric
rainbow originates from the extremum of the deflection function of the
scattered light wave. In a complete (optical) analogy for the \AA scattering,
the two extrema of $\Theta(l)$ shown in Fig.~9 can be identified as the Coulomb
and nuclear rainbow angles, respectively. While the Coulomb rainbow is well
described by the known nuclear potential at the surface and the Coulomb
interaction between the two ions, the nuclear rainbow can only be properly
identified and described based on a realistic choice for the nuclear part of
the \AA OP. Our further consideration is, therefore, concentrated on the
nuclear rainbow which can provide us with some information about the hadronic
interaction between the two colliding nuclei. In Fig.~9, the peripheral
trajectories with positive $\Theta$, which are dominated by the Coulomb
repulsion, contribute mainly to the \emph{nearside} scattering, while those
drawn to negative angles $\Theta$ (dominated by the attractive nuclear
potential) represent the \emph{farside} scattering. Thus, the more pronounced
the nuclear rainbow, the stronger the farside scattering and the more
information about the \AA OP can be deduced from the OM analysis of the elastic
scattering. If the scattering conditions (incident energy, absorption...) are
appropriate, the Airy pattern of the supernumeraries shown in Fig.~4 for the
atmospheric rainbow can also be present in the pattern of the nuclear rainbow.
Namely, the Airy oscillation pattern is observed in the farside scattering
cross section as a result of the interference between the two branches
(specified below as $l_<$ and $l_>$) of the deflection function on either side
of its minimum at $\Theta_{\rm R}$.
\begin{figure}[ht]
 \begin{center}\vspace{-1cm}
  \includegraphics[angle=270,width=0.8\textwidth]{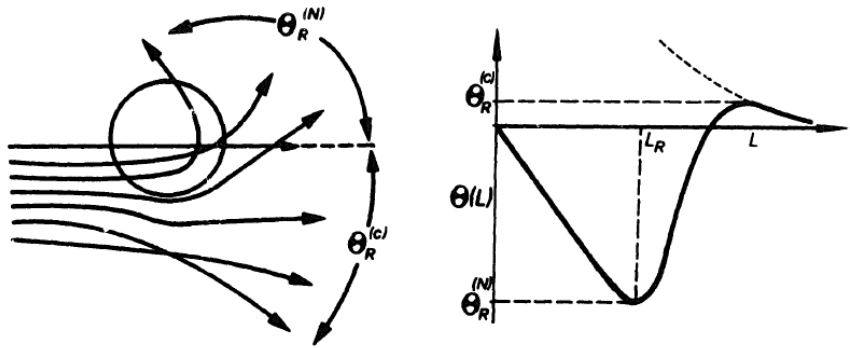}\vspace{-2cm}
   \caption{Left: Classical trajectories of the nuclear wave scattered
   elastically by a short-range attractive nuclear potential and a long-range
   repulsive Coulomb potential which lead to the nuclear (N) and Coulomb (C)
   rainbows, respectively. Right: The corresponding deflection function.
   Illustration taken from Ref.~\cite{Sat83c}.}
   \end{center}
  \label{fig9}
\end{figure}

\subsection{The farside scattering and the Airy oscillation pattern}
To understand the structure of the Airy supernumeraries (or Airy oscillations)
of the nuclear rainbow pattern, it is necessary to consider explicitly the
contributions from the nearside and farside scattering trajectories shown in
Fig.~9. For a strongly refractive (weakly absorptive) \AA system, the
amplitudes of the nearside and farside scattering can be determined from a
nearside/farside decomposition of the elastic scattering amplitude, a method
developed by Fuller \cite{Ful75}. Namely, by decomposing the Legendre function
$P_l(\cos\Theta)$ into waves travelling in $\Theta$ which are running in
opposite directions around the scattering centre, the nuclear part of the
scattering amplitude (\ref{Smat}) can be decomposed into the nearside ($f_{\rm
N}$) and farside ($f_{\rm F}$) components as
\begin{equation}
 f_{\rm N}(\Theta)+f_{\rm F}(\Theta)=\frac{i}{2k}\sum_l (2l+1)A_l
 \left[\tilde Q_l^{(-)}(\cos\Theta)+\tilde Q_l^{(+)}(\cos\Theta)\right],\
 \label{NFdec}
 \end{equation}
  \begin{equation}
 {\rm where}\ \ \tilde Q_l^{(\mp)}(\cos\Theta)={1\over 2}
  \left[P_l(\cos\Theta)\pm {2i\over\pi}Q_l(\cos\Theta)\right],
  \end{equation}
and $Q_l(\cos\Theta)$ are the Legendre functions of the second kind. The
nearside amplitude $f_{\rm N}(\Theta)$ represents contributions from waves
deflected to the direction of $\Theta$ on the near side of the scattering
centre and the farside amplitude $f_{\rm F}(\Theta)$ represents contributions
from waves travelling from the opposite (far) side of the scattering centre to
the same angle $\Theta$. The nearside/farside decomposition of the Rutherford
amplitude can be obtained analytically \cite{Ful75} using the explicit form of
$f_{\rm C}(\Theta)$ (available from, e.g., Ref.~\cite{Sat83}).
\begin{figure}[ht]
 \begin{center}\vspace{-1cm}
  \includegraphics[height=0.7\textheight]{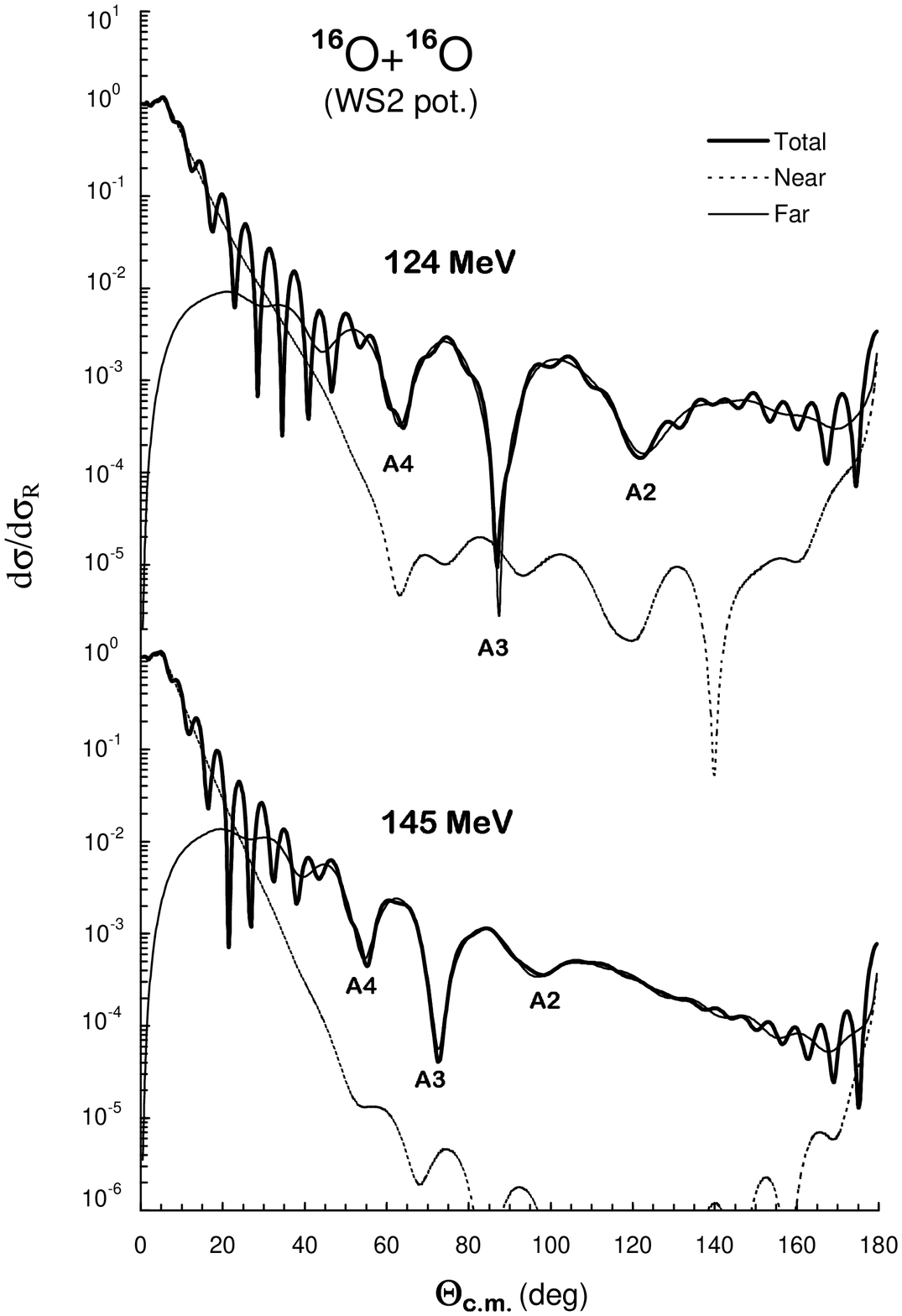}\vspace{-1cm}
   \caption{Decomposition of the unsymmetrized \oo elastic scattering cross
   section (thick solid curves) at $E_{\rm lab}=124$ and 145 MeV into the nearside
(dotted curves) and farside (solid curves) components using Fuller's method
\cite{Ful75}. The Woods-Saxon squared (WS2) potentials, which give the best fit
to the data measured at these energies by Sugiyama {\sl et al.}
\cite{Sug93,Kon96}, have been used in the OM calculation. A$k$ indicates the
$k$-th order of the Airy minimum in the farside cross section. Illustration
taken from Ref.~\cite{Kho00}.}
   \end{center}
  \label{fig10}
\end{figure}
Note that Fuller's method of nearside/farside decomposition has been
recently improved by Anni {\sl et al.} \cite{Anni} to underline the refractive
nature of the elastic scattering cross section in the nuclear rainbow cases. In
terms of scattering trajectories shown in Fig.~9, the real \AA OP has the
refractive effect of a converging lens \cite{McVoy84}, so that for a `detector'
located at angle $\Theta$ it pulls the nearside trajectories towards the
forward direction and the farside trajectories away from it. The nuclear
rainbow pattern is produced exclusively by the farside trajectories which are
governed by the strong nuclear interaction (through the attractive \AA OP).
Therefore, nuclear rainbows are absent in any scattering that does not involve
a strong nuclear interaction, like, e.g., the electron-nucleus scattering. Even
for the nucleon-nucleus scattering the real OP turns out to be too weak to
produce the Airy pattern \cite{Bra96}. The rainbow pattern can appear only in
the refractive \aA or \AA scattering with a deep, strongly attractive real OP
\cite{Bra97,vOe00}.
\begin{figure}[ht]
 \begin{center}
  \includegraphics[angle=270,width=0.85\textwidth]{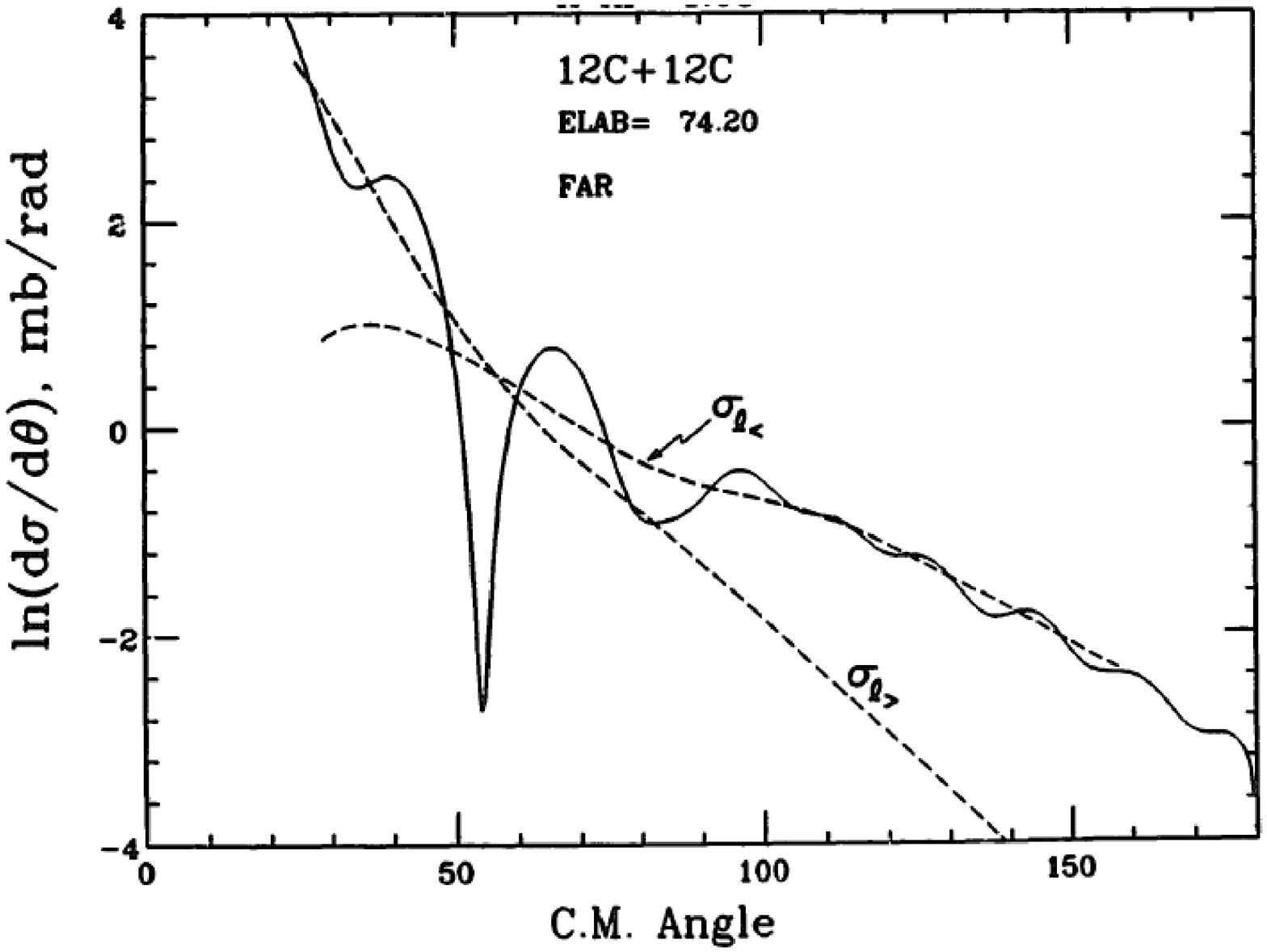}
  \vspace{-0.5cm}
\caption{The decomposition of the farside cross section of the elastic \cc
scattering at $E_{\rm lab}=74.2$ MeV into its $l_<$ and $l_>$ components, using
an OP of the Woods-Saxon shape which gives the best fit to the data
\cite{Stokstad}. The two minima at $\Theta_{\rm c.m.}\approx 53^\circ$ and
$80^\circ$ were identified by McVoy and Brandan \cite{McVoy92} as the fourth
(A4) and third (A3) Airy minima, respectively. Illustration taken from
Ref.~\cite{McVoy92}.}
   \end{center}
  \label{fig11}
\end{figure}

As an example, the nearside/farside decomposition of the elastic \oo scattering
amplitude (using Fuller's method \cite{Ful75}) is shown in Fig.~10. It can be
seen that the Fraunhofer diffraction pattern observed at small angles is due to
an interference between the nearside and farside amplitudes in forward
direction. At large angles, the elastic scattering pattern is determined
dominantly by the farside amplitude, and the Airy oscillation pattern can be
well observed when the absorption is weak (like that found for the \oo system).
In such a case, if one can measure accurately the scattering cross section down
to $d\sigma/d\sigma_{\rm R}\approx 10^{-5}$, very valuable information on the
real \AA OP is obtained. The (broad) Airy oscillation pattern seen at large
angles in Fig.~10 originates from an interference between the $l_<$ and $l_>$
components of the farside amplitude which correspond to trajectories scattered
at the same angle $\Theta$ with angular momenta $l<l_{\rm R}$ and $l>l_{\rm
R}$, where $l_{\rm R}$ is the angular momentum associated with the rainbow
angle $\Theta_{\rm R}$. From the simple relation $(l+1/2)\hbar=kR$ one can find
that the $l_<$ and $l_>$ trajectories are related to smaller and larger impact
parameters, respectively (compared with that given by $l_{\rm R}$).

The absorption due to all non-elastic processes is always present in the elastic
\AA scattering and reduces strongly the $l_<$ amplitudes relative to the $l_>$
ones. Therefore, in a strongly absorbing HI system, the $l_<$ contributions are
totally suppressed and the elastic scattering cross section decreases
exponentially with increasing scattering angles, without showing any
interference structure. In terms of the complex \AA OP, the imaginary
(absorptive) potential must be \emph{weak} enough for the $l_<$ component of
the farside amplitude to survive in the elastic channel and the real OP must be
strong enough to deflect trajectories to large 'negative' scattering angles,
thus giving rise to the Airy interference between the $l_<$ and $l_>$
amplitudes. Based on a realistic choice of the OP for a refractive rainbow
system, it is possible to extract from the farside scattering amplitude the
explicit contributions from the $l_<$ and $l_>$ amplitudes and provide a
complete description of the observed Airy interference structure (see, e.g.,
the decomposition done for the \cc system by Brandan and McVoy \cite{McVoy92}
in Fig.~11). The Airy oscillation arising from the $l_<$ and $l_>$ interference
is shown schematically in Fig.~12 where one can explain the oscillation pattern
in the scattering cross section (shown in Figs.~10 and 11) using a three-slit
interference mechanism.
\begin{figure}[ht]
 \begin{center}
 \vspace{-1cm}
  \includegraphics[angle=270,scale=0.55]{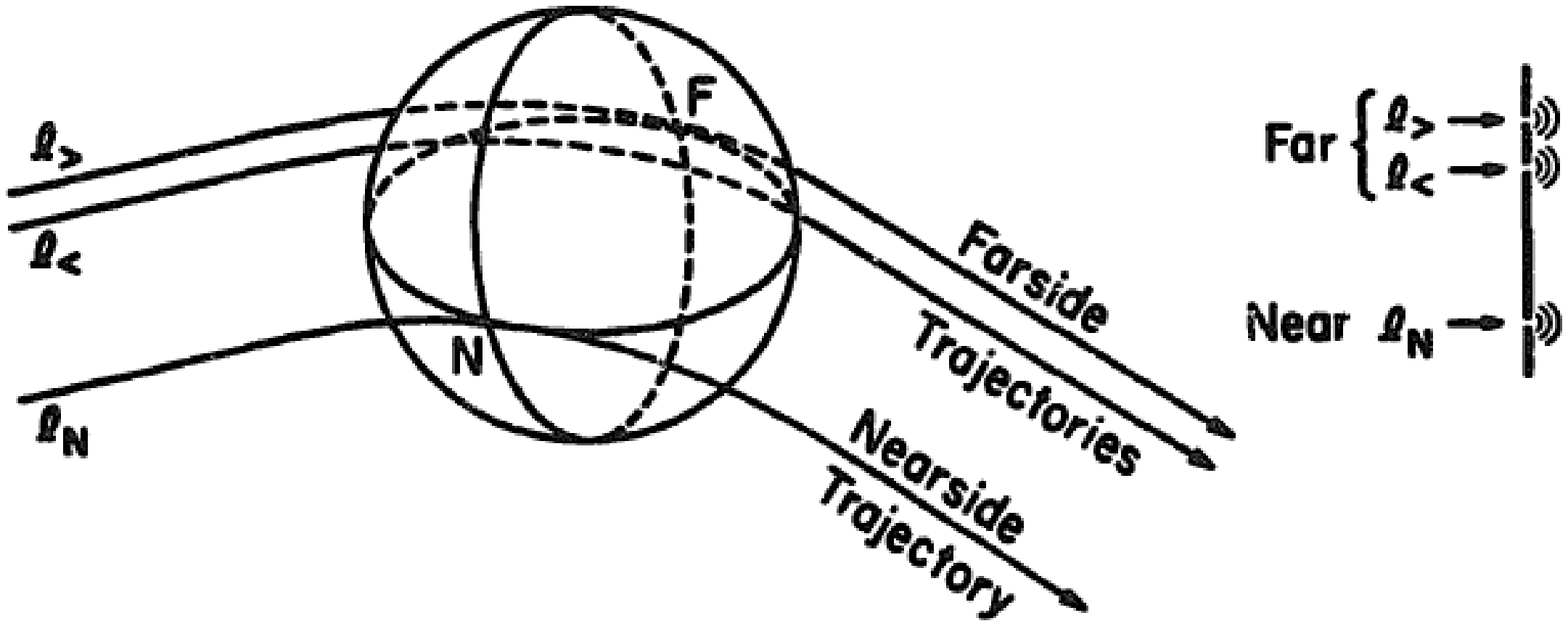}\vspace{-2cm}
   \caption{Schematic representation of three trajectories being deflected
  to the same scattering angle. The right-hand part shows the analogy to a
  three-slit interference pattern. Illustration taken from Ref.~\cite{McVoy92}.}
   \end{center}
  \label{fig12}
\end{figure}

With the proper choice for the OP (and its energy dependence) not only a
complete description of the observed Airy interference structure is obtained
but also the evolution of the Airy pattern with the incident energies is well
explained. For example, the nearside/farside decomposition of the elastic \oo
scattering amplitude at different energies \cite{Kho00} has revealed a
consistent evolution of the Airy interference pattern in the \oo system with
increasing incident energies. While higher-order Airy minima were identified at
low energies (see Fig.~10), the first Airy minimum A1 could be clearly seen
only in the elastic \oo scattering data at higher energies, with the pronounced
primary rainbow maximum following A1 established at $E_{\rm lab}=350$ MeV (see
Fig.~13). The observed broad bump of the primary rainbow at $\Theta_{\rm
c.m.}\approx 50^\circ$ is quite sensitive to the \oo optical potential at small
distances.
\begin{figure}[ht]
 \begin{center}
 \vspace{-2cm}
  \includegraphics[width=0.7\textwidth]{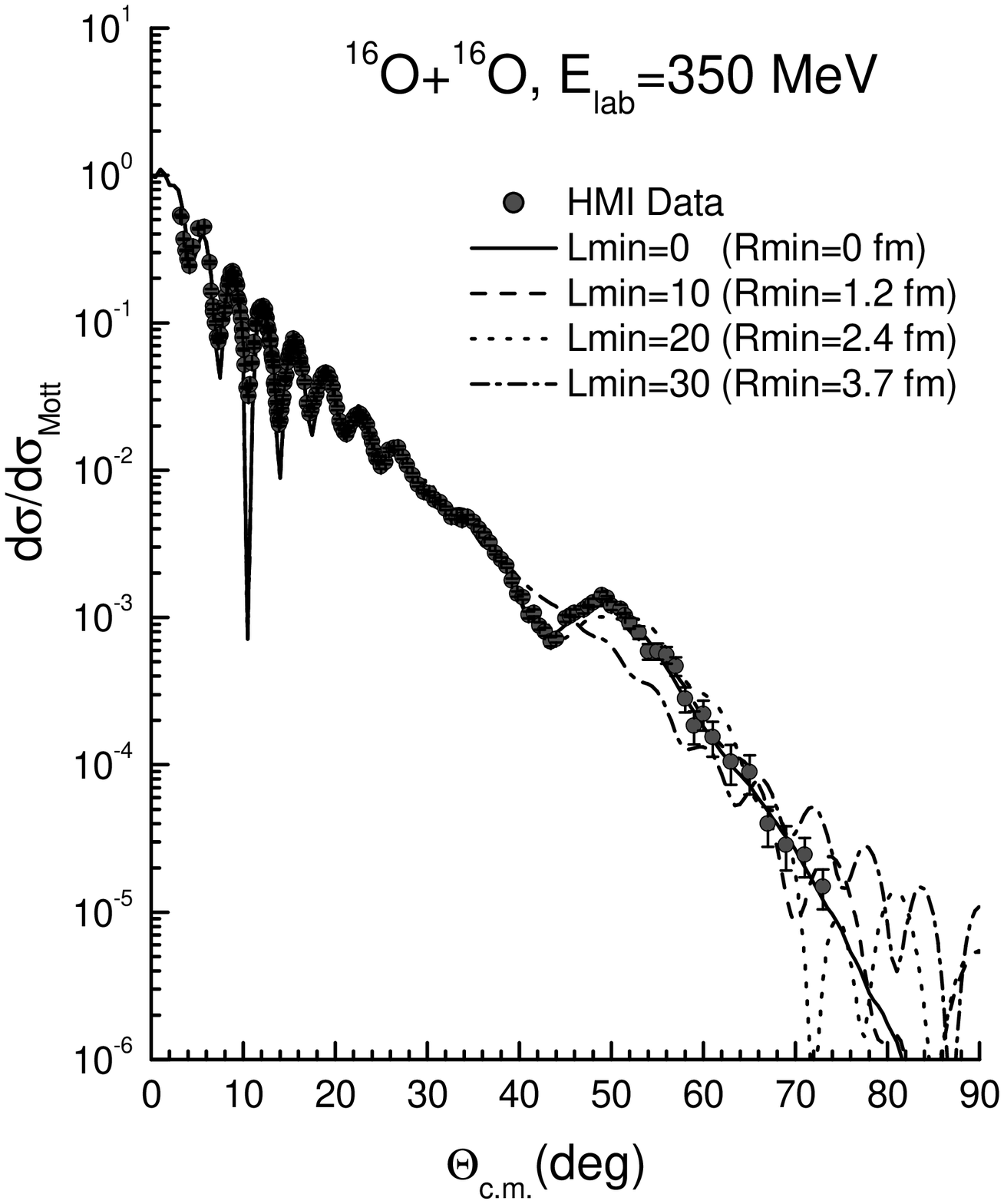}
  \vspace{-1.5cm}
   \caption{Elastic \oo scattering data \cite{Sti89,Boh93} at
   $E_{\rm lab}=350$ MeV in comparison with OM description using the real
  folded potential and WS imaginary potential \cite{Kho00}, and different
  cutoffs for the lowest partial wave $L_{\rm min}$; the corresponding impact
  parameters are $R_{\rm min}$. The observed minimum at $\Theta\approx 44^\circ$
  has been established \cite{Kho00} as the first Airy minimum (A1) which is
  followed by a broad bump of the primary rainbow.}
   \end{center}
  \label{fig13}
\end{figure}

In Fig.~13, we show the OM calculations of the elastic \oo scattering at
$E_{\rm lab}=350$ MeV, using the real OP given by the folding model, with
different cutoff values of the lowest partial wave in the expansion
(\ref{Smat}) of the elastic scattering amplitude. One can see that the data
points at large angles are indeed sensitive to very low partial waves which
correspond to the distances as small as $R\approx 2-4$ fm. This result shows
again that the considered elastic \oo data provide us with a valuable test of
the \AA interaction at small distances.
\begin{figure}[ht]
 \begin{center}
 \vspace{-0.5cm}
  \includegraphics[width=0.75\textwidth]{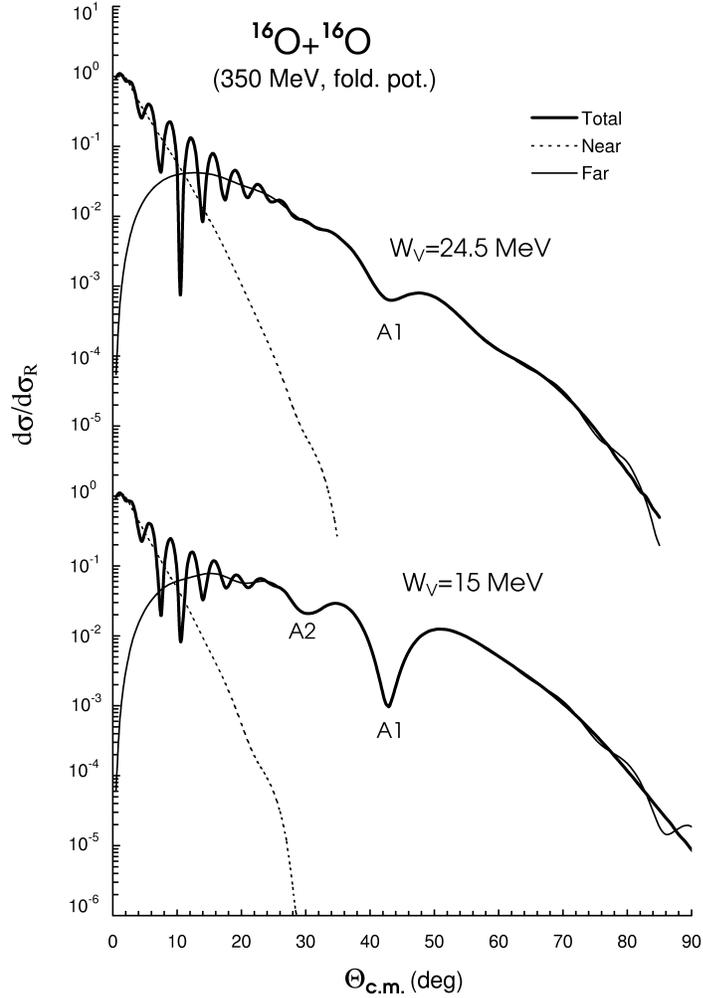}
  \vspace{-1cm}
   \caption{Decomposition of the unsymmetrized \oo elastic scattering
   cross section (thick solid curves) at $E_{\rm lab}=350$ MeV into the nearside
   (dotted curves) and farside (solid curves) components using Fuller's method
   \cite{Ful75}. Upper part: given by the same real folded and WS imaginary
   potentials as those used in Fig.~13; lower part: given by the same real folded
   potential but with a reduced strength of the WS imaginary potential,
   A1 and A2 are the first and second Airy minima. Illustration taken
   from Ref.~\cite{Kho00}.}
   \end{center}
  \label{fig14}
\end{figure}
The role of the weak absorption in observing the rainbow structure is
illustrated in Fig.~14, where the secondary Airy minimum (A2) preceding the
first bump (lower part of Fig.~14) can be revealed only if the strength of the
imaginary part of the OP is reduced in the OM calculation. The energy dependence
of the position of the first Airy minimum in the farside cross section of
elastic \oo scattering \cite{Kho00} shows that the incident energy should be
between 300 and 450 MeV for the first Airy minimum A1 to appear in the most
favorable angular range ($\Theta_{\rm c.m.}\approx 30^\circ - 60^\circ$). In
this angular region, the distorting effects by the Mott interference in the
symmetric \oo system are minimal and diffractive structures are absent. Thus,
the experiment of elastic \oo scattering at 350 MeV \cite{Sti89,Boh93} turned
out to be a perfect choice for the observation of the primary nuclear rainbow.

Only after the evolution of the Airy interference pattern in the \cc and \oo
systems with energy is well understood, one could go further and explain the
shape of the $90^\circ$ excitation functions of the elastic scattering at
energies from the Coulomb barrier up to about 20-30 MeV/nucleon. Such studies
of the evolution of the Airy interference pattern in the excitation function of
elastic scattering have been done for the \oo system by Kondo {\sl et al.}
\cite{Kon96} and for the \cc system by McVoy and Brandan \cite{McVoy92}. In
particular, the various deep and shallow minima in the excitation function
measured for the \cc system, which gave rise to the famous ``elephant
interpretation" of the excitation function (see Fig.~15 or Fig.~1 in
Ref.~\cite{McVoy92}) have remained an unsolved mystery for some 20 years. This
was solved with the most realistic families of the \cc optical potential
\cite{McVoy92} which give the correct and consistent description of the rainbow
structure in the elastic scattering cross section at different energies. In
terms of the nuclear rainbow scattering, the gross structures of the \cc
excitation function shown in Fig.~15 are caused by the refractive structures
passing through $90^\circ$, and the sharp minima between the elephants
correspond to the Airy minima of different orders drifting through $90^\circ$
\cite{Bra97}. Actually, the minima at $E_{\rm lab}\approx 102$ and 124 MeV are
the second (A2) and first (A1) Airy minima, respectively, so that there will
not be a fourth elephant at higher energies.
\begin{figure}[ht]
 \begin{center}\vspace{-3cm}
  \includegraphics[angle=270,scale=0.6]{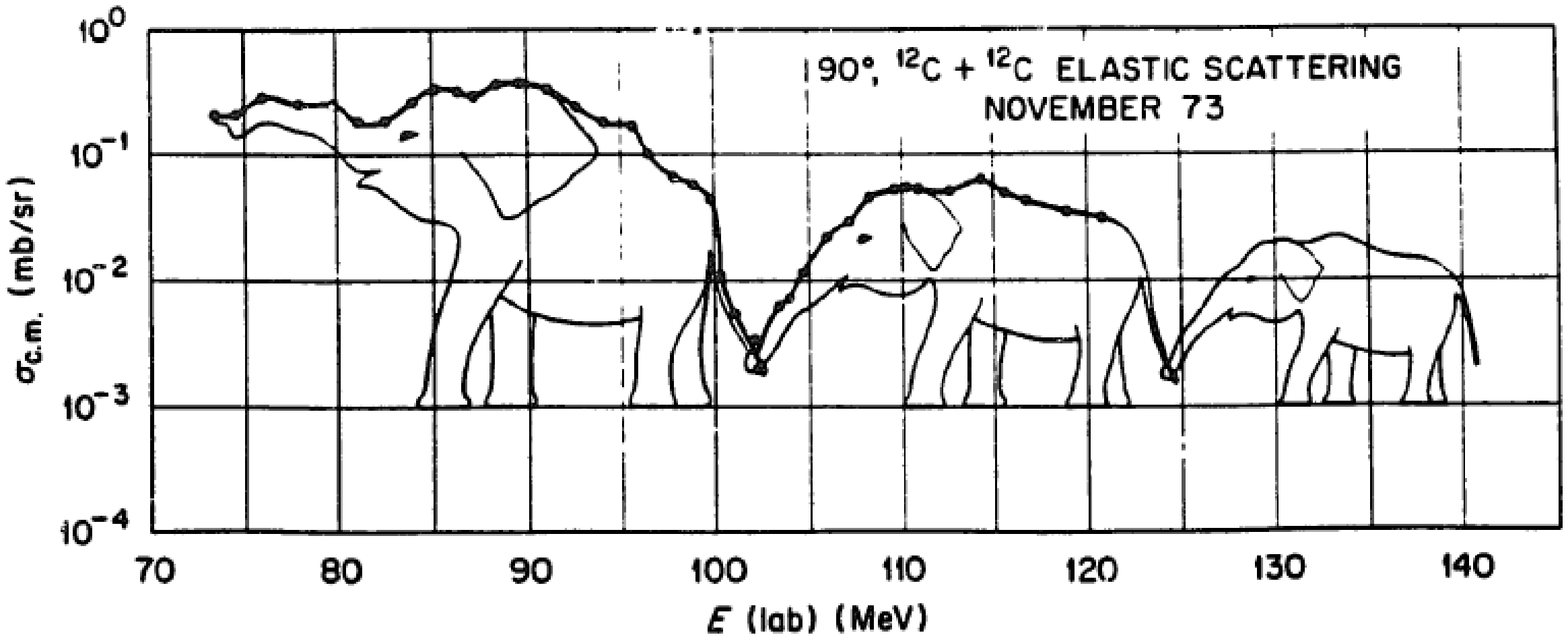}
  \vspace{-2.5cm}
   \caption{The ``elephant interpretation" of the shape of the $90^\circ$
   excitation function measured for the \cc system. Illustration taken
   from Ref.~\cite{McVoy92}.}
   \end{center}
  \label{fig15}
\end{figure}

Although the nearside/farside decomposition of the elastic scattering cross
section, like those shown in Figs.~10 and 14, can be done \emph{exactly} with
the quantal scattering amplitude using Fuller's method \cite{Ful75}, the
oscillating interference pattern at small angles can also be given by a
semiclassical, but quite illustrative, approach developed by Hussein and McVoy
\cite{Hus84} based on the strong absorption model. Namely, the observed
oscillating cross sections seen at forward angles, e.g., in Figs.~10 and 14, is
caused by the diffractive scattering of the incident wave into the classically
forbidden region where both $f_{\rm N}(\Theta)$ and $f_{\rm F}(\Theta)$
decrease exponentially with $\Theta$ and can be approximately expressed
\cite{Sat83,Hus84} as
\begin{eqnarray}
 \sqrt{2\pi\sin\Theta}\ f_{\rm N}(\Theta) & \sim & \exp(-i\lambda_{\rm g}\Theta)
 \exp(-\gamma_{\rm N}(\Theta-\Theta_{\rm g}), \nonumber \\
 \sqrt{2\pi\sin\Theta}\ f_{\rm F}(\Theta) & \sim & \exp(i\lambda_{\rm g}\Theta)
 \exp(-\gamma_{\rm F}(\Theta+\Theta_{\rm g}).
\label{ek7}
\end{eqnarray}
Here $\Theta_{\rm g}$ is the grazing angle associated with the grazing angular
momentum $\hbar\lambda_{\rm g}=\hbar(l_{\rm g}+1/2)$. The oscillating pattern
of the elastic angular distribution at small angles is due to the interference
between $f_{\rm N}(\Theta)$ and $f_{\rm F}(\Theta)$ amplitudes, with a
characteristic spacing of $\Delta\Theta\simeq\pi/\lambda_{\rm g}$. It is
obvious that the interference pattern depends strongly on the slope parameters
of the nearside ($\gamma_{\rm N}$) and farside ($\gamma_{\rm F}$) amplitudes. A
real \emph{attractive} OP refractively enhances $f_{\rm F}(\Theta)$ over
 $f_{\rm N}(\Theta)$ (with $\gamma_{\rm N}>\gamma_{\rm F}$) and allows there
to be an angle
\begin{equation}
 \bar\Theta=\Theta_{\rm g}(\gamma_{\rm N}+\gamma_{\rm F})/(\gamma_{\rm N}
  - \gamma_{\rm F}),
\label{ek8}
\end{equation}
known as `Fraunhofer crossover', where $|f_{\rm N}(\bar\Theta)|=|f_{\rm
F}(\bar\Theta)|$ and the nearside/farside interference oscillation reach its
maximum amplitude. At small scattering angles ($\Theta<\bar\Theta$) which
correspond to peripheral impact parameters, the nearside component is dominant,
with positive-angle scattering caused by the \emph{repulsion} from scattering
centre.  At large scattering angles ($\Theta>\bar\Theta$) which correspond to
small impact parameters, $f_{\rm F}(\Theta)$ becomes stronger than
 $f_{\rm N}(\Theta)$, with negative-angle scattering caused by the
\emph{attraction} toward scattering centre. Thus, an accurate experimental
observation of Fraunhofer crossover $\bar\Theta$ would give us a good measure
of the refraction and attractive strength of the real OP causing it. The
elastic angular distribution is usually characterized by a deep interference
minimum in the vicinity of $\bar\Theta$, while the minima become progressively
less marked as $\Theta$ moves from $\bar\Theta$ on either side as shown in
Fig.~14. In the \oo case, the measured elastic angular distribution at 350 Mev
shows a deep minimum at $\Theta\approx 10^\circ$ which is in fact associated
with the Fraunhofer crossover $\bar\Theta$ given by the realistic OP for this
system (see Figs.~13 and 14).

As a complimentary method to the nearside/farside decomposition, Michel {\sl et
al.} \cite{Michel} have pointed out that the observed Airy interference pattern
in the elastic \oo scattering can also be described by a
barrier-wave/internal-wave (B/I) decomposition of the scattering amplitude
using the same OP as that used in the nearside/farside decomposition of
Ref.~\cite{Kho00}. Such a method was first proposed by Brink and Takigawa
\cite{Brink} and has been shown by Michel {\sl et al.} to be a very useful tool
to disentangle the scattering trajectories into the surface and internal
components. For the farside trajectories which probe the \AA OP at small
distances, the B/I decomposition method was shown to give about the same Airy
interference pattern as that given by the decomposition of the farside
amplitude into the $l_<$ and $l_>$ components (compare Fig.~10 and Fig.~16).
The B/I decomposition method was also used to study the \cc system
\cite{Michel04} and the sequence of the Airy minima which made up the
``elephants" shown in Fig.~15 has been confirmed. Therefore, the two methods
\cite{Kho00,Michel} are very complementary to each other and give us a
complete physical understanding of the nuclear rainbow scattering phenomenon.
\begin{figure}[ht]
 \begin{center}\vspace{0cm}
  \includegraphics[width=0.7\textwidth]{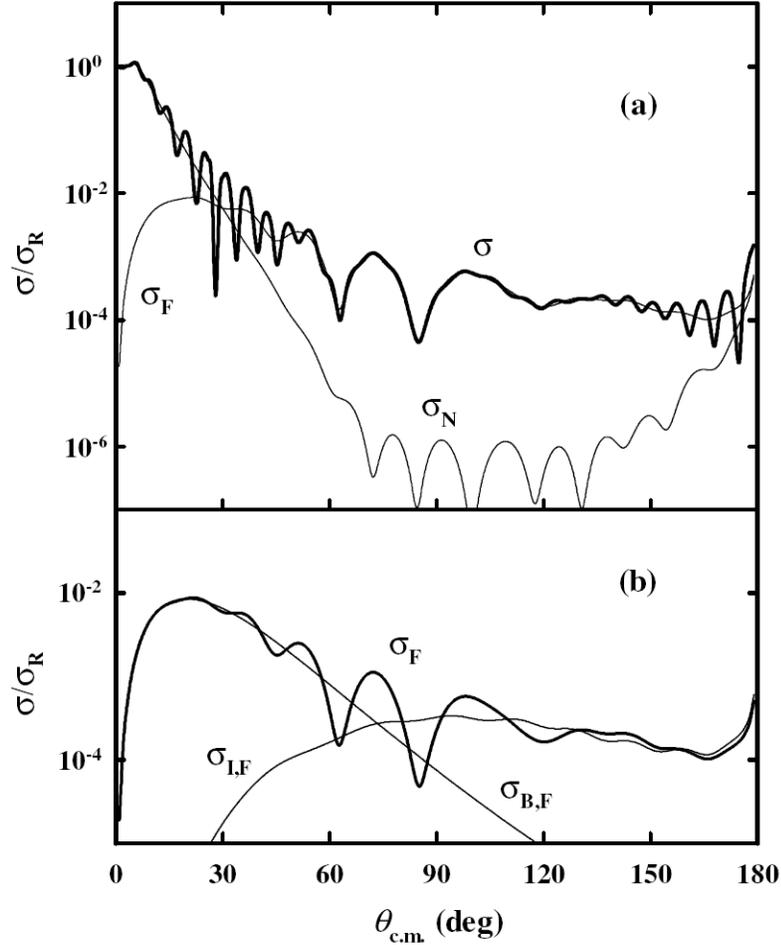}\vspace{-1cm}
   \caption{Decomposition of the unsymmetrized \oo elastic scattering
cross section (thick solid curves) at $E_{\rm lab}=124$ MeV \cite{Michel} into
the nearside and farside components (a), and a further decomposition (b) of the
farside cross section into the barrier-wave (B,F) and the internal-wave (I,F)
components based on a semiclassical method by Brink and Takigawa \cite{Brink}.
Illustration taken from Ref.~\cite{Michel}.}
   \end{center}
  \label{fig16}
\end{figure}

\section{Brief systematics of the nuclear rainbow scattering data}
\label{sec3}
\subsection{$^{16}$O+$^{16}$O system}
This is the ``heaviest" HI system sofar that has shown a prominent rainbow
pattern in the elastic scattering cross section. We give here a brief survey of
the experimental elastic \oo scattering data which show consistently the
refractive (rainbow) structure over a wide range of energies. At low energies,
the accurate data have been measured (up to sufficiently large angles) at IreS
in Strasbourg \cite{Nicoli} at $E_{\rm lab}=75\to 124$ MeV (see Fig.~17).
\begin{figure}[ht]
 \begin{center}\vspace{-0.5cm}
  \includegraphics[width=0.7\textwidth]{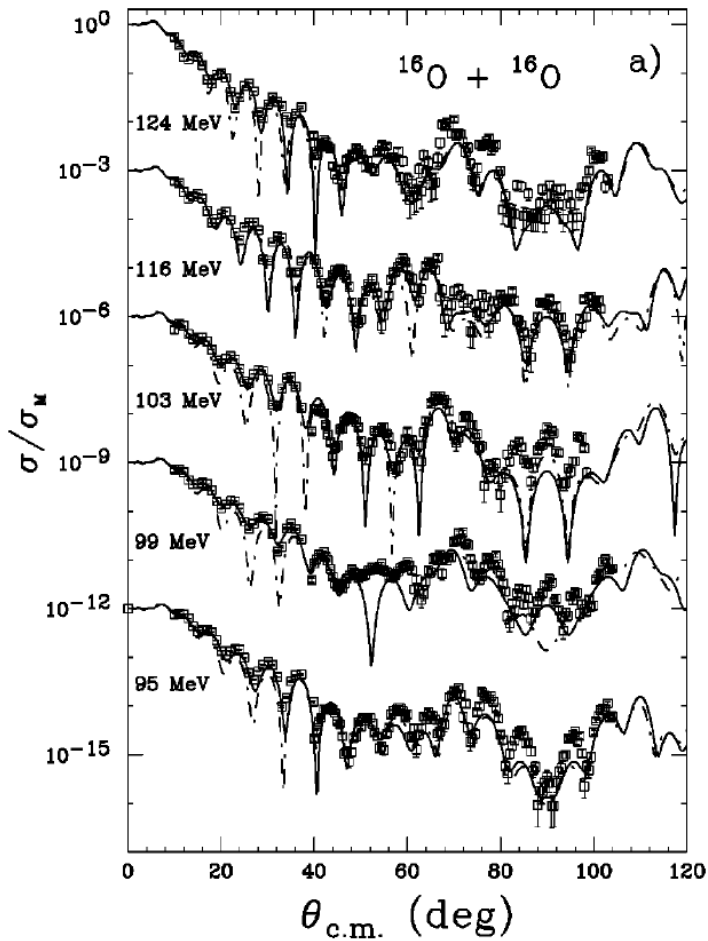}
   \vspace{-1cm}
  \caption{Elastic $^{16}$O+$^{16}$O scattering data measured at $E_{\rm lab}=
    95\to 124$ MeV in comparison with OM fits using the optical potentials
    of (quadratic) Woods-Saxon shape (solid curves) and that obtained with the
    folding model (dotted curves). Illustration taken from Ref.~\cite{Nicoli}.}
\end{center}
\label{f17}
\end{figure}
Although these data show a prominent oscillating Airy structure of the elastic
cross section, this Airy pattern is strongly distorted on either side of
$\Theta_{\rm c.m.}=90^\circ$ by the Mott interference. The elastic scattering
data at $E_{\rm lab}=250, 350$ and 480 MeV were measured using the Q3D magnetic
spectrometer at the cyclotron of the Hahn-Meitner Institute (HMI) in Berlin
\cite{Sti89,Boh93,Bar96}, with the most pronounced primary rainbow maximum
observed at 350 MeV. The data at higher energies of $E_{\rm lab}=704$ and 1120
MeV were measured at GANIL using the SPEG magnetic spectrograph
\cite{Bar96,Nuo98} (see also revision of the 704 MeV data in
Ref.~\cite{Kho00}). In addition, the elastic \oo data at $E_{\rm lab}=124$ and
145 MeV have been measured by Sugiyama {\sl et al.} at JAERI (Tokai)
\cite{Sug93,Kon96}. The JAERI data were used to investigate the evolution of
the Airy interference pattern in the excitation function at lower energies
\cite{Kon96}. The data sets obtained at JAERI, HMI and GANIL are summarized in
Fig.~18.

We emphasize that in order to reveal the rainbow structure, a tremendous
experimental effort is needed to measure data points at large angles where the
elastic cross sections become extremely small ($d\sigma/d\sigma_{\rm R}\leq
10^{-5}$). A remarkable feature in the \oo case is that we can follow the
evolution of the primary and secondary Airy structures from high energies of
$E_{\rm lab}=350$ and 480 MeV, where the first Airy minimum is clearly seen,
down to 124 MeV and lower, where the primary rainbow maximum moves beyond the
observable angular range of $\Theta_{\rm c.m.}=0^\circ\to 90^\circ$. At lower
energies, the higher-order Airy oscillation appears in the angular range of
$\Theta_{\rm c.m.}< 90^\circ$. It should be made clear that the
\emph{secondary} Airy maximum is \emph{not} the analog of the secondary
atmospheric rainbow seen in the nature with the reversed colour sequence due to
a second reflection inside of the water droplets. The second and higher-order
Airy minima observed in the low-energy $^{16}$O+$^{16}$O elastic data are, in
fact, the analogs of the first and higher-order supernumeraries observed in the
atmospheric rainbow (see the faint bows located below the primary bow in
Fig.~5). We note that the Mott interference caused by the boson symmetry
between the two identical $^{16}$O nuclei leads in addition to a rapidly
oscillating elastic cross section at angles around $\Theta_{\rm
c.m.}=90^\circ$, which obscures the Airy structure in this angular region.
Therefore, the Airy pattern can best be seen in an OM calculation, where the
boson symmetrization is artificially removed, as shown in Figs.~10 and 16 for
the case of $E_{\rm lab}=124$~MeV.
\begin{figure}[ht]
 \begin{center}
\includegraphics[width=0.6\textwidth]{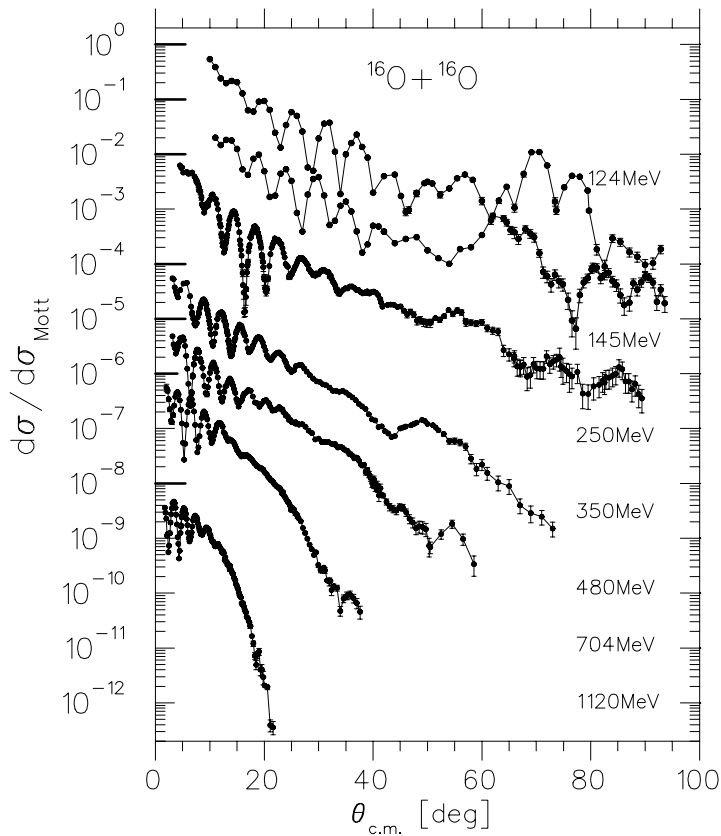}
  \caption{Elastic $^{16}$O+$^{16}$O scattering data measured at $E_{\rm lab}=
  124, 145$ MeV \cite{Sug93,Kon96}, 250, 350, 480 MeV \cite{Sti89,Boh93,Bar96},
  704 and 1120 MeV \cite{Bar96,Kho00,Nuo98}. The primary rainbow maximum at 350
  MeV is located at $\Theta_{\rm c.m.}\approx 50^\circ$. The lines are to guide
  the eye. Illustration taken from Ref.~\cite{vOe03}.}
\end{center}
\label{f18}
\end{figure}

\subsection{Other systems}
It should be recalled that the pioneering nuclear scattering experiment that
lead to the observation of the nuclear rainbow was the study of the elastic \aA
scattering at $E_{\rm lab}\approx 140$ MeV by Goldberg {\sl et al.}
\cite{Go73,Go74}. Our today's understanding of the nuclear rainbow as a
refractive phenomenon has been established based on detailed OM studies of the
elastic \aA scattering. While $^4$He can be considered as the lightest HI, it
is a very ``robust" projectile (with the nucleon separation energy of 21 MeV)
and it can penetrate rather deep into the interior of the target nucleus
without being absorbed. If the elastic \aA scattering data by Goldberg {\sl et
al.} \cite{Go73,Go74} remain the best evidence for the nuclear rainbow
scattering observed for different targets at a given incident energy, the
high-precision $\alpha+^{90}$Zr data measured by Put and Paans \cite{Pu77}
several years later (see Fig.~19) present a unique picture of how the primary
rainbow pattern associated with the first Airy minimum evolves with the energy.
While the farside scattering begins to dominate the large-angle scattering
already at the $\alpha$-particle energy of 59 MeV, the most pronounced primary
rainbow shoulder is observed at the energies of 80 MeV and higher. These data,
together with the data of $\alpha+^{90}$Zr measured at 141.7 MeV \cite{Go74},
provide us with a very accurate test ground for theoretical models of the \aA
OP.
\begin{figure}[ht]
 \begin{center}\vspace{-1cm}
 \includegraphics[width=0.95\textwidth]{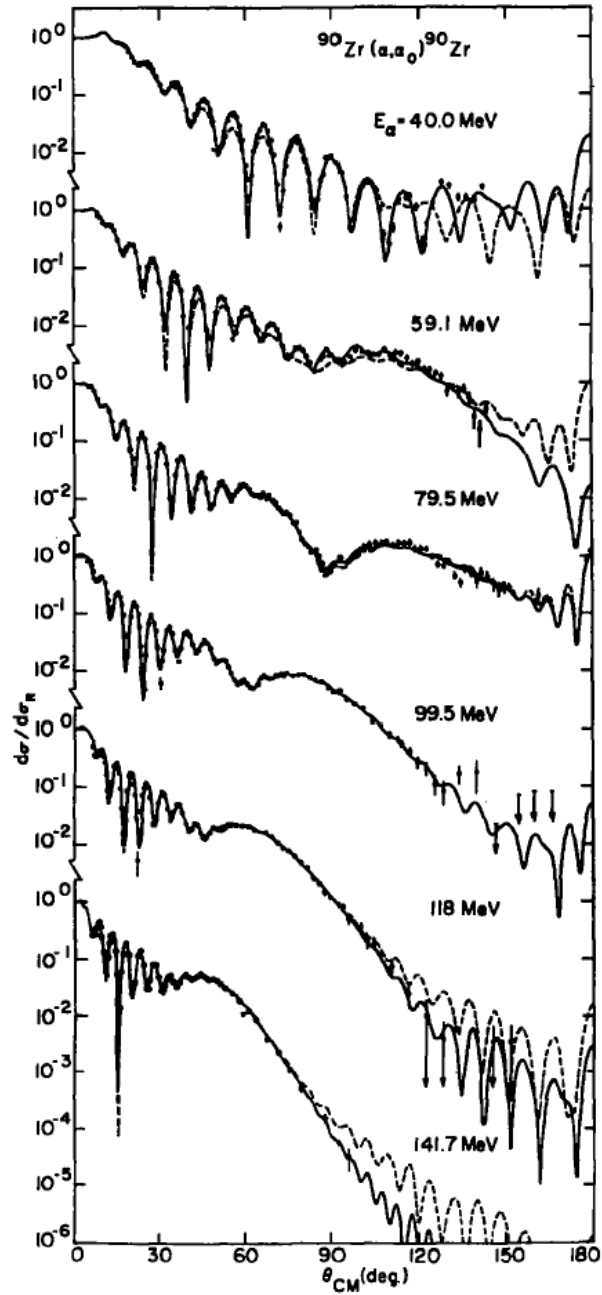}
 \vspace{-1.5cm}
  \caption{Elastic \aZr scattering data measured at $E_{\rm lab}=
  40 - 118$ MeV by Put and Paans \cite{Pu77} and at the higher energy of
  141.7 MeV  by Goldberg {\sl et al.} \cite{Go74} in comparison with the OM
  results given by the set of deep real optical potentials. Illustration taken
  from Ref.~\cite{Pu77}.}
\end{center}
\label{fig19}
\end{figure}

The elastic and inelastic \aA scattering on different targets has been
precisely measured at $E_{\rm lab}=104$ MeV by Karlsruhe group (see, e.g.,
Ref.~\cite{Ba89} for an overview). Given the broad bump of the primary rainbow
observed at large angles, the Karlsruhe data have been used successfully to
probe the nuclear matter distribution and determine the real \aA OP at 104 MeV
by a ``model independent" method (see Fig.~21 below). Among light HI systems,
the elastic \cc scattering has been studied extensively since the 70's of the
last century and the nuclear rainbow pattern has been established and
investigated based on the elastic \cc data measured at energies from about 6 up
to 200 MeV/nucleon (see a detailed systematics of the elastic \cc data in
Ref.~\cite{Bra97}). While the \cc and \oo systems are quite `transparent' for
refractive effects to appear, the Mott interference caused by the boson
symmetry between the two identical nuclei  leads to rapidly oscillating elastic
cross sections at angles around $\Theta_{\rm c.m.}=90^\circ$, which in turn
distort the original Airy structures. The \oc system does not have the boson
symmetry and was suggested as a good candidate for the study of the nuclear
rainbow \cite{Bra97}. However, up to the late 90's, the available elastic data
for this system covered only limited angular intervals and did not allow to
reveal any feature of refractive scattering. This has motivated several
high-precision experiments on the elastic \oc scattering by Ogloblin {\sl et
al.} \cite{Oglob98,Oglob00} where the elastic \oc cross sections have been
measured accurately, to cover a large angular region, at $E_{\rm lab}=132, 170,
200, 230$ and 260 MeV. These data show clearly the diffractive and refractive
scattering patterns at small and large scattering angles, respectively. The
nearside/farside decomposition of the elastic \oc scattering amplitudes at
these energies \cite{Oglob00} has lead to about the same pronounced Airy
interference pattern as those observed in the symmetric \cc and \oo systems
(see Fig.~20). Although not distorted by the Mott interference, the observed
backward rise of the elastic cross section towards
 $\Theta_{\rm c.m.}=180^\circ$ has been shown \cite{szilner02} to partially
originate from the elastic $\alpha$-particle transfer between the projectile and
target, and this gives again some additional interference structures.
\begin{figure}[ht]
 \begin{center}\vspace{0cm}
 \includegraphics[width=0.65\textwidth]{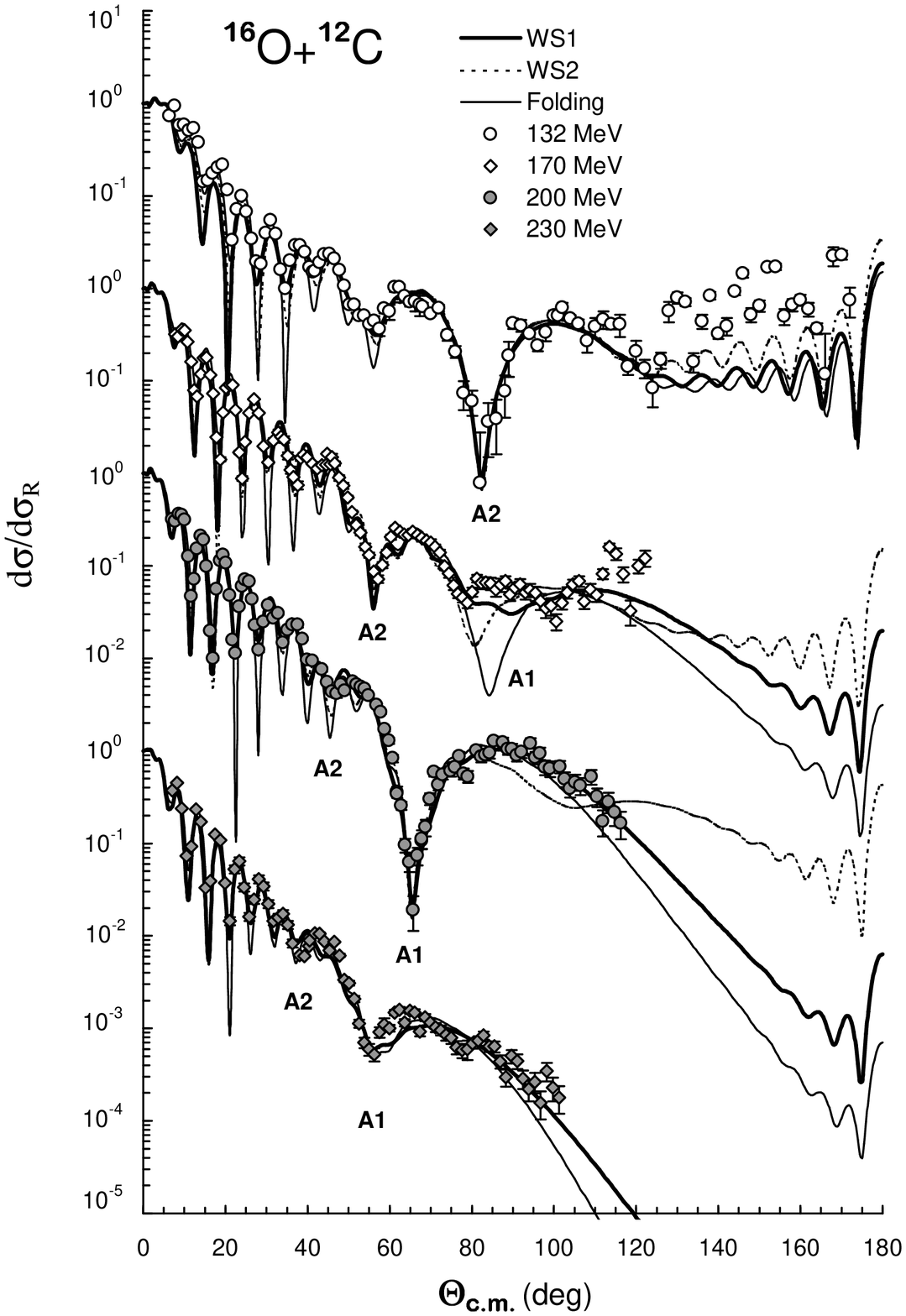}\vspace{-1cm}
  \caption{The elastic \oc scattering data at $E_{\rm lab}=132, 170, 200$
and 230 MeV in comparison with the OM fits given by the folding potential and
two different families of the Woods-Saxon potential. A1 and A2 are the first and
second Airy minima generated by the folding potential. Illustration taken from
Ref.~\cite{Oglob00}.}
\end{center}
\label{fig20}
\end{figure}
Elastic \oc scattering at lower energies were measured by the Strasbourg group
and the higher-order Airy oscillatory pattern has been established \cite{Szi01}
which is similar to that observed in the \oo and \cc systems. With a rather
weak absorption in the \oc system, the evolution of the Airy minima in the
elastic \oc scattering at medium \cite{Oglob00} and low \cite{Szi01} energies
can be consistently described by the energy dependent real OP given by the
folding model.

We finally note that the rainbow pattern has also been observed in the elastic
scattering of $^{6,7}$Li at energies up to around 50 MeV/nucleon
\cite{Mic85,Nad93,Nad95} and $^9$Be at 18 MeV/nucleon \cite{Sat83b}. For
example, the elastic $^6$Li scattering on the light targets has shown a broad
exponential falloff in the cross section at large angles which was identified
as the ``shoulder" of the primary rainbow maximum. The OM analysis
\cite{Sat83b} of the elastic $^9$Be scattering from $^{12}$C and $^{16}$O
targets has found a rather weak primary Airy minimum in the cross section at
large angles. Since this Airy structure was strongly damped and scarcely
visible, it was referred to as the rainbow ``ghost" \cite{Sat83b}. The less
pronounced rainbow structure observed in these cases compared to those observed
earlier in the elastic \aA scattering at about the same energies is due to a
stronger absorption caused, in particular, by the breakup of these
($\alpha$-clustered) projectiles \cite{Sak86,Kho95b}. See more discussions on
the refractive $^6$Li scattering in the Sec.~\ref{sec8} below.

\section{Theoretical basis of the \AA OP}
\label{sec4} As discussed above, the nuclear rainbow can only be properly
identified and studied based on a \emph{realistic} choice of the \AA optical
potential. In general, the optical potential is an effective interaction $U(R)$
between the two colliding nuclei (whose centres of mass are separated by the
distance $R$) which is used in the following (one-body) Schr\"{o}dinger
equation for elastic scattering
\begin{equation}
\left[-{\hbar^{2}\over 2\mu}\nabla^{2}+U(R)+V_C(R)-E\right]
 \chi({\bi R})=0.   \label{ep1}
\end{equation}
Here $E$ is the energy of relative motion in the centre-of-mass (c.m.) system,
$\mu$ is the reduced mass of the two colliding nuclei, and $V_C(R)$ is the
Coulomb potential. The spin and isospin dependence of $U(R)$ is neglected for
the simplicity of discussions. It is assumed in the OM calculation that the two
nuclei remain in their ground states during the elastic scattering and
higher-order effects due to the coupling to other non-elastic reaction channels
are  taken into account by the imaginary part $W$ of the OP which describes the
loss of incident flux (\emph{absorption}) into the open non-elastic channels.
The elastic scattering cross section is then obtained \cite{Sat83} using the
solution $\chi({\bi R})$ of Eq.~(\ref{ep1}), with appropriate boundary
conditions. One can see from Eq.~(\ref{eq2}) that the use of a complex OP in
Eq.~(\ref{ep1}) is analogous to the introduction of a complex index of
refraction (used in the optics to describe the propagation of light through an
absorbing medium), so that the ``rainbow" interpretation of the elastic \AA
scattering can well be justified based on the solution of Eq.~(\ref{ep1}).

The simplest procedure of an OM analysis is to adopt a phenomenological
functional form for $U(R)$ and adjust its parameters until the calculated
elastic cross section agrees with the measurement. The most widely used
functional form for the OP is that based on the Woods-Saxon form factor
\cite{Sat83} which is a default option in all the available OM codes for the
\AA and \nA scattering. For some weakly absorbing, refractive \aA and light HI
systems, it was possible to determine the phenomenological WS parameters of the
OP without discrete ambiguity. Actually, after decades of the use of WS shapes
for the \AA OP, it became certain that the squared WS shape (WS2) is the
physically more preferable one. This WS2 shape is also close to the shape of
the microscopic real OP given by the folding model. Therefore, if one uses the
folding model to calculate the \AA OP for the study of elastic
\emph{refractive} \AA scattering, valuable information on the effective
nucleon-nucleon (NN) interaction can be obtained if realistic nuclear wave
functions of the projectile and target nuclei are available.

\subsection{Feshbach's reaction theory for the \AA OP }
A microscopic theory for the nuclear OP, as usually discussed in the
literature, is just an approach to predict the \nA or \AA OP starting
essentially from the NN interaction between the nucleons in the system and
realistic nuclear wave functions for the projectile and target nuclei. In such
a formulation, the rigorous microscopic foundation can be established only for
the \nA OP based on the G-matrix studies of nuclear matter \cite{Ma91}. In
particular, the G-matrix method has been used to construct the effective
in-medium NN interaction for the microscopic folding calculation
\cite{Jeu77,Bri77} of the \nA OP at low and medium energies. The interaction
between two composite nuclei is a much more complicated many-body problem due
to the HI collision dynamics, and there is no truly microscopic theory for the
\AA OP like that for the \nA OP which is based on the G-matrix only. However,
an approximate approach to the microscopic understanding of the \AA OP can be
formulated \cite{Bra97} within the framework of the reaction theory by Feshbach
\cite{Fe92}.

Let us expand the total wave function for the two colliding nuclei in terms of
the complete set of internal wave functions of the projectile ($a$) and target
($A$) nuclei as
\begin{equation}
\Psi=\sum_{mn}\chi_{mn}({\bi R})\psi^{(a)}_m(\xi_a)\psi^{(A)}_n(\xi_A),
 \label{ep2}
 \end{equation}
where $\chi_{mn}({\bi R})$ describes the relative motion of the colliding
system when projectile and target are in states labelled by $m$ and $n$,
respectively. With the ground states labelled by $m=0$ and $n=0$, elastic
scattering is described by $\chi_{00}({\bi R})$. Within the OM frame, the \AA
OP should generate $\chi_{00}({\bi R})$ when used in Eq.~(\ref{ep1}). In
general, the expansion (\ref{ep2}) must be inserted into the many-body
Schr\"{o}dinger equation which gives an infinite set of coupled equations for
$\chi_{mn}(\bi R)$ after the integration over internal coordinates $\xi_a$ and
$\xi_A$. By using Feshbach's projection operator \cite{Sat83,Fe92}, we obtain
the equivalent effective interaction between the two nuclei $U$ which acts in
the elastic channel only and, hence, can be used in Eq.~(\ref{ep1}) to
determine $\chi_{00}(\bi R)$.
\begin{equation}
 U=V_{00}+\lim_{\epsilon \rightarrow 0}\sum_{\alpha\alpha'}
 {\kern -1pt ^\prime}
 V_{0\alpha}\Big({1\over E-H+i\epsilon}\Big)_{\alpha\alpha'} V_{\alpha'0}.
\label{ep3}
\end{equation}
Here $V_{\alpha\alpha'}$ is the first-order interaction between the two nuclei,
where $\alpha=mn$ stands for a pair of internal states of the projectile and
target. The primed sum runs over all the pair states excluding $\alpha=0\equiv
00$. The first term of (\ref{ep3}) is real and can be evaluated within the
double-folding approach \cite{Sat79,Kho00,KhoSat}
\begin{equation}
 V_{00}=V_{\rm F}\equiv(\psi^{(a)}_0\psi^{(A)}_0|V|\psi^{(a)}_0\psi^{(A)}_0),
 \label{ep4}\end{equation}
where the round brackets denote integration over the internal coordinates
$\xi_a$ and $\xi_A$ of the two nuclei being in their ground states
$\psi^{(a)}_0$ and $\psi^{(A)}_0$, respectively. We can rewrite Eq.~(\ref{ep3})
now as
\begin{equation}
U=V_{\rm F} + \Delta U.
 \label{ep5}
 \end{equation}
Here, $\Delta U$ is often referred to as the \emph{dynamic polarization
potential} (DPP) which arises from couplings to all the open non-elastic
channels. Depending on the energy and the binding structures of the two
colliding nuclei, the `polarizing' contribution by the energy-dependent and
complex $\Delta U$ to the total \AA optical potential $U$ can be quite
substantial \cite{Bra97,Sat83,Sak86}. Im$\Delta U$ is the main source of the
absorption (the imaginary part of the OP) due to transitions to the open
non-elastic channels. $\Delta U$ also contributes to the real part of the OP but
Re$\Delta U$, which originates from virtual excitations of the two nuclei, is
at least one order of magnitude smaller than $V_{F}$ \cite{Bra97,Sak86}.
Furthermore, $\Delta U$ is nonlocal because the system that is excited into a
non-elastic channel at position ${\bi R}$ returns, in general, to the elastic
channel at another position ${\bi R'}\neq{\bi R}$. Since the direct (one-step)
elastic scattering occurs via the first-order folded potential $V_{\rm F}$, the
weaker the absorption caused by the DPP the stronger the refractive scattering
which can lead to the appearance of the rainbow pattern at appropriate incident
energies.

\subsection{The double-folding model}
It is clear from the discussion above that $V_{00}$, the first term in
Eq.~(\ref{ep3}), is the key quantity for our understanding of the
nucleus-nucleus interaction when elastic scattering proceeds directly in one
step. Among various models for the \AA OP, the double-folding model (see
Refs.~\cite{Bra97,Sat79,KhoSat} and references therein) has been used most
widely as a simple microscopic method to calculate $V_{F}$ starting from an
appropriately chosen effective NN interaction between nucleons in the system.
Further input for the folding calculation are the realistic nuclear density
distributions of the projectile and target nuclei which are deduced either
directly from electron scattering data or from an appropriate nuclear structure
model (See Fig.~33 below for an illustrative overview of the folding model
analysis). The success of the double-folding model (DFM) in describing the
observed elastic scattering of many HI systems suggests that it indeed produces
the dominant part of the real \AA OP. Let us now briefly discuss the main
features of a recent version of the DFM that has been used in our OM analyses
of the refractive \aA and \AA scattering \cite{Kho94,Kho95,Kho97} and
generalized for a consistent folding description of the elastic and inelastic
\AA scattering \cite{KhoSat}.

In the folding approach, the real \AA interaction $V$ is based on a sum of
effective (two-body) NN interactions $v_{ij}$ between nucleon $i$ in the
projectile $a$ and nucleon $j$ in the target $A$
\begin{equation}
  V=\sum_{i\in a,j\in A}v_{ij}.
\label{ef1}
\end{equation}
Although the individual internal ground state (g.s.) wave functions
$\psi^{(a)}_0(\xi_a)$ and $\psi^{(A)}_0(\xi_A)$ in Eq.~(\ref{ep2}) are each
taken to be antisymmetrized, the Pauli principle still requires the total wave
function $\Psi$ also to be antisymmetric under interchange of nucleons between
the two nuclei. If we restrict this exchange of projectile and target nucleons
to the one-nucleon exchange process known as the single-nucleon knock-on
exchange (SNKE), then the effective NN interaction $v_{ij}$ in Eq.~(\ref{ef1})
should be replaced by
\begin{equation}
  v_{ij}(1-P_{ij}) = v_{\rm D}+v_{\rm EX}P_{ij}^{x},
\label{ef2}
\end{equation}
\begin{equation}
{\rm where}\ \ v_{\rm D}\equiv v^{({\rm D})}_{ij}=v_{ij}\ \ {\rm and}\ \
 v_{\rm EX}\equiv v^{(\rm EX)}_{ij}=-v_{ij}P_{ij}^{\sigma}P_{ij}^{\tau}.
 \label{ef3}
\end{equation}
Here $P_{ij}^{x}$, $P_{ij}^{\sigma}$ and $P_{ij}^{\tau}$ represent the operators
for the exchange of spatial, spin and isospin coordinates of the nucleon pair,
respectively. Due to the exchange of the spatial coordinates, the first term of
Eq.~(\ref{ep3}) now consists of two components: the local direct part and
\emph{nonlocal} exchange part
\begin{equation}
 V_{00}=V_{\rm F}\equiv\Big(\psi^{(a)}_0\psi^{(A)}_0\Big|
 \sum_{i\in a,j\in A}v_{ij}(1-P_{ij})\Big|\psi^{(a)}_0\psi^{(A)}_0\Big)=
 V_{\rm F}^{(\rm D)}+V_{\rm F}^{(\rm EX)}.
 \label{ef4}\end{equation}
As can be seen from Eq.~(\ref{ep3}) the imaginary part of the OP should be
constructed from an appropriate theory for the dynamical polarization potential
$\Delta U$. This is, however, a complicated task and lies well beyond the
framework of the DFM. In many cases of elastic HI scattering, it was found
sufficient to simply treat the effective NN interaction as having a complex
strength determined from the analysis of elastic scattering data and the real
and imaginary parts of the \AA OP have essentially the same radial shape.
However, it has been found from the analyses of refractive $\alpha$-nucleus
scattering \cite{Sat97} and of light HI scattering \cite{Bra97,Br97} that the
imaginary potential is definitely required to have a different shape, and the
ratio of the imaginary to the real potential tends to peak near the nuclear
surface but becomes relatively weak in the interior. Therefore, it is common to
resort to a hybrid approach by using the DFM to generate the real part of the
OP but to use a phenomenological (local) Woods-Saxon form factor $W(R)$ for the
imaginary part \cite{KhoSat,Kho94,Kho95,Kho97}. Sometimes, like in the case of
refractive \oo scattering, the WS imaginary potential needs to be composed of
two terms (volume + surface potentials) in order to match independently the
absorption at small $R$ and that at large $R$, in the surface region
\cite{Kho00,Kho95,Kho97}.

In general, the real folded potential ({\ref{ef4}) supplemented by a local WS
imaginary potential should be inserted into Eq.~(\ref{ep1}) and, given a
nonlocal exchange potential $V_{\rm F}^{(\rm EX)}$, one needs to solve an
integro-differential equation for the scattering wave function \cite{Love78}.
However, from a practical point of view a reliable local approximation for
$V_{\rm F}^{(\rm EX)}$ is highly desirable. In this case, not only the OM
calculation is much simpler, but also the comparison with the (local)
phenomenological OP is much more direct which is particularly essential in the
study of the nuclear rainbow scattering. The direct part of the folded potential
is local and obtained from a double-folding integral over $v_{\rm D}$ and the
g.s. densities of the two nuclei as
\begin{equation}
 V_{\rm F}^{(\rm D)}(R)=\int\rho_a({\bi r}_a)\rho_A({\bi r}_A)
 v_{\rm D}(s)d^3r_a d^3r_A, \ \ {\bi s}={\bi r}_A-{\bi r}_a+{\bi R}.
\label{ef5}
\end{equation}
We use a local WKB approximation \cite{Sat83} for the change in the
relative motion wave function induced by the exchange of spatial coordinates of
each interacting nucleon pair
\begin{equation}
 \chi_{00}({\bi R}+{\bi s})\approx\exp\left(\frac{i{\bi K}(R){\bi s}}{M}\right)
 \chi_{00}({\bi R}), \label{ef5a}
\end{equation}
where the recoil factor is $M=aA/(a+A)$, with $a$ and $A$ being the mass numbers
of the projectile and target, respectively, and ${\bi K}(R)$ is the local
momentum of relative motion at the internuclear distance $R$. Then, the
following local expression for the exchange folded potential can be obtained
\cite{Bri77,Sin75,Cha86,Kho88}
\begin{equation}
\fl\ \ V_{\rm F}^{(\rm EX)}(R)=\int\rho_a ({\bi r}_a,{\bi r}_a +{\bi s})
 \rho_A({\bi r}_A,{\bi r}_A -{\bi s})v_{\rm EX}(s)
 \exp\left(\frac{i{\bi K}(R){\bi s}}{M}\right)d^3r_ad^3r_A.
 \label{ef6}\end{equation}
The local momentum of relative motion must be determined self-consistently
through the total \emph{real} OP as
\begin{equation}
 K^2(R)={{2\mu}\over{\hbar}^2}[E-V_{\rm F}^{(\rm D)}(R)-
 V_{\rm F}^{(\rm EX)}(R)-V_C(R)]. \label{ef7}
\end{equation}
The calculation of $V_{\rm F}^{(\rm EX)}(R)$ still contains a self-consistency
problem and involves an explicit integration over the \emph{nonlocal} nuclear
density matrices of the projectile and target. In practice, the nuclear g.s.
densities are usually available in the local form. Therefore, the DFM
calculation of the exchange potential has been done
\cite{KhoSat,Kho94,Kho95,Kho97} using a realistic approximation for the nonlocal
density matrix \cite{Ca78,Kho01} that has been adopted earlier in the folding
calculations of the \nA OP \cite{Bri77,Sin75}. This local approximation for the
density matrix was shown \cite{Kho01} to be of around 1\% accuracy in the DFM
results for the \aA OP.

We note that a much simpler zero-range approximation for the SNKE has been used
in numerous double-folding calculations. In this approach \cite{Sat79,Lov75},
the knock-on exchange potential is included by adding a zero-range
pseudo-potential to the interaction $v_{ij}$ in Eq.~(\ref{ef1}). Namely,
\begin{equation}
  v_{ij}(1-P_{ij})\rightarrow v_{ij}(s) + \hat{J}(E)\delta({\bi s}),
 \label{ef8}\end{equation}
which immediately makes the exchange potential $V_{\rm F}^{(\rm EX)}$ local.
Here, the strength $\hat{J}(E)$ of the pseudo-potential has been obtained by
calibrating against ``exact" calculations of the exchange potential for the \nA
scattering \cite{Lov75}. Although this zero-range approximation has been used
with some success in DFM calculations of the HI optical potential at low
energies \cite{Sat79} where the data are sensitive only to the OP at the surface
(near the strong absorption radius), it has been shown to be inadequate
\cite{Kho88} in the case of rainbow scattering where the data are sensitive to
the real OP over a wider radial domain. A very recent study by Hagino {\sl et
al.} \cite{Hag06} has also shown that the finite-range treatment of the
localized exchange potential (\ref{ef6}) gives the OM results very close to
those obtained from the exact solution of integro-differential equation
\cite{Love78} using the explicit nonlocal $V_{\rm F}^{(\rm EX)}$, while the
zero-range prescription (\ref{ef8}) gives a large discrepancy with the exact
results.

We also note here a simpler folding approach (see, e.g., Ref~\cite{Trache00})
to evaluate the \AA OP by the direct folding integration (\ref{ef5}) only,
using the so-called JLM effective interaction. Since the complex JLM
interaction was deduced (in a local density approximation) from the strengths
of both the direct and exchange components of the G-matrix for infinite nuclear
matter obtained by Jeukenne, Lejeune and Mahaux \cite{Jeu77}, the exchange term
(\ref{ef6}) of the folding potential is not explicitly treated in this
approach.

Another approximate treatment of the exchange effects has been developed
recently by the Sao Paolo group \cite{Cham97,Cham02} where the exchange
non-locality is effectively taken into account by an exponential dependence of
the potential strength on the local relative-motion momentum (\ref{ef7}), as
suggested some 40 years ago by Perey and Buck \cite{Per62} for the \nA OP. In
this case, the \AA OP is also evaluated by the direct folding integration
(\ref{ef5}) only, using an effective NN interaction fitted empirically to a
global systematics of HI elastic data \cite{Cham02}.

\subsection{Other microscopic approaches}

Besides the double-folding model, there are at least two other important
microscopic approaches developed in the past to study the \AA OP, which are the
resonating group method (RGM) (see, e.g., Refs.~\cite{Bu77,Le79,Ho91}) and the
nuclear matter approach by the T\"ubingen group \cite{Tref85,Kho90}.

For the \AA elastic channel, the RGM freezes the two nuclei in their ground
states, like the DFM, but takes full account of the exchange of projectile and
target nucleons. A Schr\"odinger equation for the relative-motion wave function
$\chi_{00}({\bi R})$ is then obtained with a highly nonlocal RGM ``kernel".
Besides a local direct term which is just the double-folded potential
 $V_{\rm F}^{(\rm D)}$, the full antisymmetrization results in a hierarchy
of nonlocal exchange terms, according to the number of nucleons exchanged. In
practical calculations, an accurate localization procedure has been used to
yield local potentials equivalent to these exchange terms \cite{Ho91}. After
localization, the RGM approach predicts a deep real OP which becomes shallower
with the increasing energy as the attractive exchange term becomes weaker, in
the same manner as that given by the DFM calculation \cite{Kho00,Kho94}. These
RGM studies also show that the SNKE contribution is the largest of the exchange
terms and dominates for peripheral collisions due to the long range of
single-nucleon exchange. In this sense, the more rigorous RGM method provides a
solid theoretical justification for the DFM which only takes into account the
SNKE for the exchange effects. The full antisymmetrization makes the RGM quite
complicated and it can be used to estimate the real OP for very light systems
only. Moreover, the inclusion of a realistic density dependence into the
effective NN interaction (a very important ingredient in the folding
calculation as explained below) is also a technical difficulty in the RGM due
to its rigorous treatment of the exchange.

The basic physical picture of the nuclear matter (NM) approach
\cite{Tref85,Kho90} is that the \AA collision is locally represented by the
collision of two pieces of the NM whose densities are the local densities of
the target and projectile. The momentum distribution of the colliding system
with densities $\rho_a$ and $\rho_A$ is represented by two Fermi spheres with
radii
 $k_{F_a}=(1.5\pi^2\rho_a)^{1/3}$ and $k_{F_A}=(1.5\pi^2\rho_A)^{1/3}$ whose
centres are separated by the \emph{asymptotic} momentum of the (nucleon)
relative motion $\hbar k=\sqrt{2mE/M}$, where $m$ is the nucleon mass and $M$
is the recoil factor used in Eq.~(\ref{ef5a}). It is easy to see that
$k=K_\infty/M$, where $K_\infty$ is derived from Eq.~(\ref{ef7}) at
$R\to\infty$ where one can neglect the OP. The exchange effects imposed by the
Pauli principle, known as Pauli blocking in the G-matrix study of the NM, lead
to a modification of the shape of two Fermi spheres \cite{Tref85} when they
overlap, i.e., when $k<k_{F_a}+k_{F_A}$.  The OP between two nuclei separated
by a distance $R$ is then defined as the difference of the total energy of the
system at $R$ from that at infinity
\begin{equation}
 U(R,k)=E(R,k)-E(\infty,k),
 \label{ef9}
 \end{equation}
with the potential energy calculated from the G-matrix given by the
Bethe-Goldstone equation for two Fermi spheres colliding in momentum space
\cite{Tref85}. Although the nuclear matter approach can give some estimates for
both the real and imaginary OP based on the complex Bethe-Goldstone G-matrix,
the ansatz (\ref{ef9}) remains questionable \cite{Bra97,Ho91} because the total
energy $E(R,k)$ does not determine just the relative motion while the two
nuclei remain in their ground states and includes, in general, a wide range of
excited states so that $U(R,k)$ cannot be used in Eq.~(\ref{ep1}) to describe
elastic scattering. The OP obtained by this method has also properties that
conflict with the global systematics of \AA OP. Namely, its real part is quite
shallow at low incident energies and becomes deeper as the energy increases
\cite{Tref85,Kho90}, while empirically the real \AA OP has been found to be
deep at low energies and to become shallower as the energy increases
\cite{Bra97}. A possible reason for this inadequacy is given below.

\section{Nuclear rainbow and preference of a deep real OP}\label{sec5}

If the depth of the real \nA OP at low and medium incident energies is known to
be around 40-50 MeV for a wide range of target masses, based on both the
microscopic G-matrix calculations and the phenomenological OM analyses of \nA
elastic scattering, the depth of HI optical potential has been uncertain for
years. The question ``Is the HI optical potential deep or shallow?" has often
been one of the basic questions posed in the studies of HI scattering
\cite{Bra97}. As discussed above in Sec.~\ref{sec2}, the main reason that
hindered our knowledge about the shape of the OP is the strong absorption which
is typical for most HI systems \cite{Sat79}, especially, those involving medium
to heavy nuclei. However, the careful OM studies of rainbow scattering observed
in the \aA \cite{Go72,Go74,Pu77} and light HI systems \cite{Bra97,Br97} seem to
show unambiguously that the physically realistic real OP must be ``deep".

Typically, the OM analyses of the elastic (refractive) \aA  scattering data
using various forms of the OP such as the standard Woods-Saxon potential
\cite{Go73,Go74}, spline functions \cite{Pu77} or that deduced from a
model-independent analysis (MIA) of the elastic \aA data \cite{Gi87,At96}, have
always resulted in a weakly absorbing imaginary potential and a deep real
potential which is close to that predicted by the folding model. As an
illustration, Fig.~21 shows the real OP for the \aCa system at 104 MeV given by
the MIA using a series of Fourier-Bessel functions \cite{Gi87} in comparison
with that predicted by the DFM using different inputs for the effective,
density dependent NN interaction. It turned out that the MIA potential agrees
best with the double-folded potentials given by the most realistic choice of
the density dependent NN interaction (CDM3Y6 and BDM3Y1 in Table~1). This kind
of comparison is very helpful for the justification of the folding model as a
reliable tool to predict the real \AA OP. We note further that, in contrast to
many cases of the elastic HI scattering, the real \aA OP has no ``family"
problem (the existence of different potential families which give nearly the
same OM fit to the elastic data), especially, when the systematic behavior of
the volume integral of the OP is taken into account \cite{At96}.
\begin{figure}[ht]
 \begin{center}
 \vspace{-1cm}
  \includegraphics[scale=0.55]{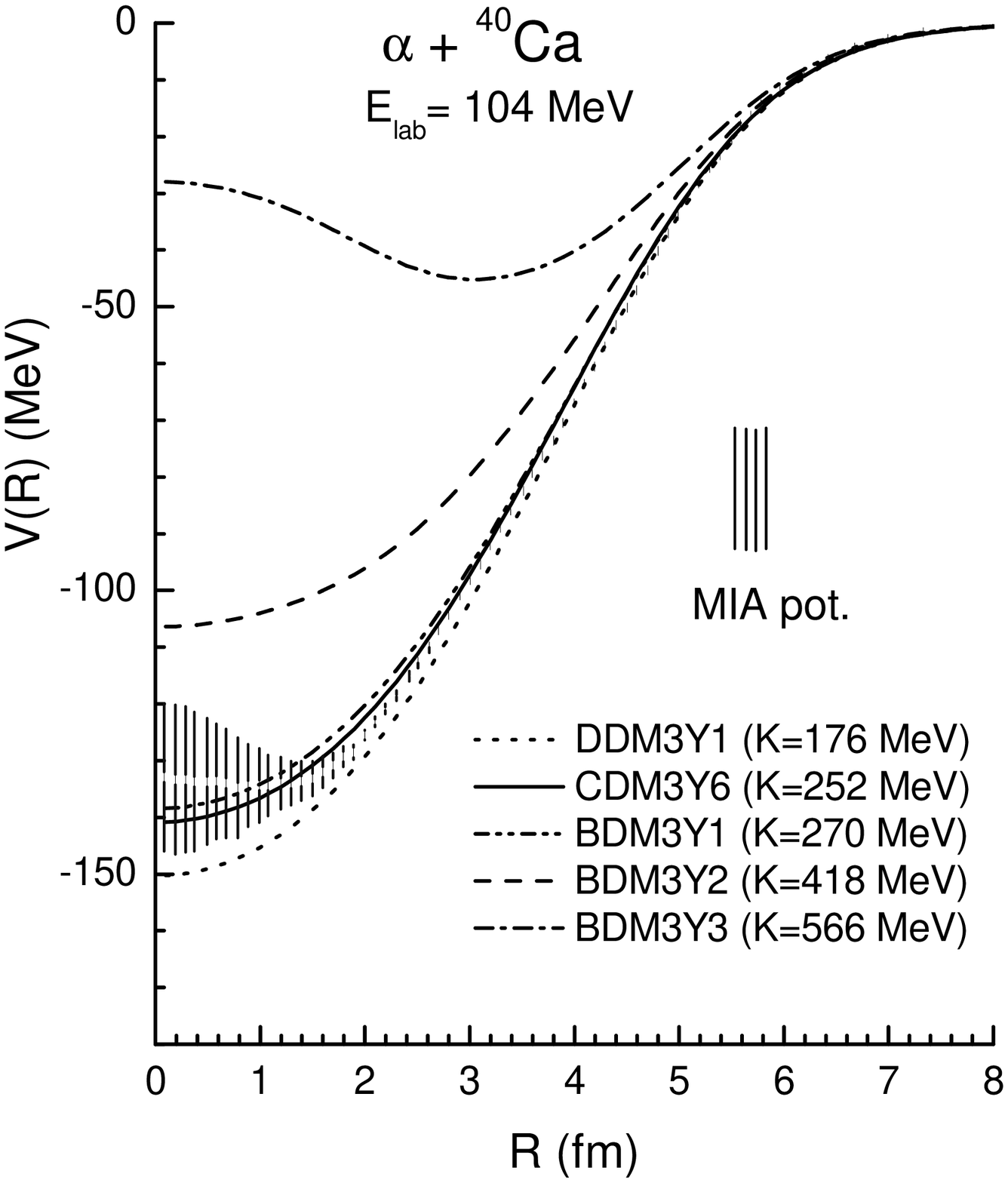}\vspace{-2cm}
  \caption{Radial shapes of the real OP for the \aCa system at 104 MeV given by
a model-independent analysis (hatched area) \cite{Gi87} and the double-folded
potentials obtained with different density dependent NN interactions (which give
the corresponding values of the nuclear incompressibility $K$ in the HF
calculation of nuclear matter \cite{Kho97}).}
\end{center}
\label{fig21}
\end{figure}

Although there still exist ambiguities in the depth of HI optical potential for
heavy systems due to the strong absorption, numerous OM studies of the nuclear
rainbow scattering patterns observed in light HI systems like \cc, \oc and \oo
have lead to a rather unique OP systematics for these systems
\cite{Bra96,Br97}. In particular, Brandan and McVoy \cite{Br97} have shown that
in the ``rainbow" energy range of 6 - 100 MeV/nucleon, the central depth of the
real OP for light HI systems is $V(R=0)\approx 100-300$ MeV, and the ratio of
the imaginary to the real parts of the OP was found to be  $W(R)/V(R)\ll 1$,
for both small and large distances $R$ (which reflects the internal and
far-tail transparency of the OP), and $W(R)/V(R)\approx 1$ in the surface
region. Such a ``deep" real OP agrees closely with that predicted by the
double-folding calculation \cite{Kho94,Kho95,Kho97}. The double-folded
potential has been shown \cite{Kho00,Nicoli} to give the correct order of the
Airy oscillation in the observed rainbow patterns of the elastic \oo cross
section. The use of a deep real OP was also found necessary to explain
consistently the shape of the low-energy resonances as well as the bound \cc
cluster states in $^{24}$Mg \cite{Kon98} and the \oo cluster states in $^{32}$S
\cite{Ohkubo02}. Here only the deep potential can generate the correct number
of nodes  for the total (antisymmetrized) wave function of the cluster state
that is \emph{not} Pauli-forbidden. Thus, a consistent description of the
low-energy resonances as well as the bound cluster states has been achieved
only with a deep real OP which is a continuation of the deep real OP found
necessary to explain the nuclear rainbow scattering at higher energies. As a
result, one can ascribe this deep potential to a \emph{mean-field} potential
\cite{Kon98} which is similar to the nucleon mean-field potential used in a
consistent study of low-energy \nA elastic scattering and single-nucleon bound
states \cite{Ma91}.
\begin{figure}[ht]
 \begin{center}
 \includegraphics[width=0.6\textwidth]{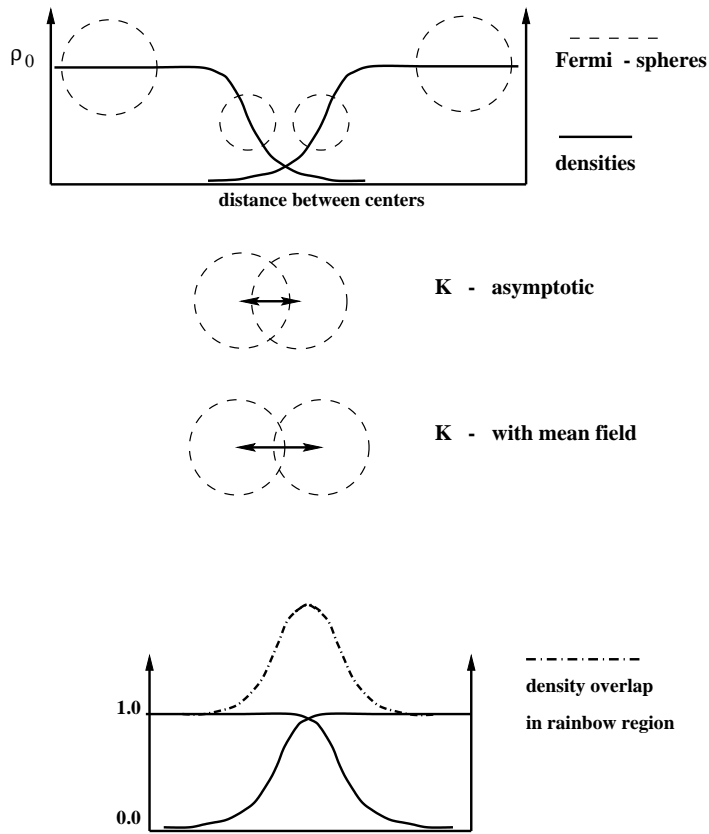}
 \vspace{5mm}
  \caption{Nuclear densities of the two colliding nuclei and two corresponding
Fermi spheres (dashed circles) separated by the relative-motion momentum $K$. If
$K$ is taken to be the local relative momentum (\ref{ef7}) determined
self-consistently by a mean-field attractive potential then it becomes larger as
the potential gets deeper at low energies and, thus, suppresses the Pauli
distortion of the two Fermi spheres in the momentum space \cite{Sub01}.}
   \end{center}
  \label{fig22}
\end{figure}

The preference of a deep mean-field potential for the real \AA OP, whose depth
and radial shape are consistent with the folding results \cite{Bra96,Br97}, has
been confirmed by a detailed study of the Pauli blocking effects in the
double-folding formalism by Soubbotin {\it et al.} \cite{Sub01}. The main
concept of this ``Pauli distorted double-folding model" (PDDFM) \cite{Sub01} is
illustrated in upper part of Fig.~22. Namely, the overlapping of the two
nuclear densities at short distances in the coordinate space is treated
self-consistently by a Pauli distortion of the two corresponding Fermi spheres
in the momentum space which are separated by the local relative-motion
momentum. Such a treatment of the Pauli distortion of the two Fermi spheres (to
prevent them from overlapping with each other) has been developed in the past
by Tuebingen group \cite{Tref85,Kho90} in their study of HI collision. The only
difference is that in the PDDFM the two Fermi spheres are separated by the
local nucleon relative momentum $k=K(R)/M$, determined self-consistently by the
local real OP at a given radial distance $R$ using Eq.~(\ref{ef7}). In the
nuclear matter approach \cite{Tref85,Kho90} this distance in momentum space is
equal the asymptotic relative momentum at infinity $k=K_\infty/M$. In the
spirit of the local density approximation (widely used in the NM studies of the
\nA OP) the use of the asymptotic relative momentum for the separation between
the two Fermi spheres is not appropriate and leads to a strong repulsion
between the two Fermi spheres at low energies. As a result, the real \AA OP
obtained in the NM approach is quite shallow at low energies and becomes deeper
as the energy increases \cite{Tref85,Kho90}, in a contradiction with the
established systematics \cite{Bra97}. The consistent use of the local nucleon
relative momentum $k=K(R)/M$ in the PDDFM calculation of the exchange potential
has confirmed \cite{Sub01} the earlier prediction of the DFM
\cite{Kho94,Kho95,Kho97} of the \emph{mean-field} type real OP which is deep at
low energies and becomes shallower with the increasing energy, as shown in
Fig.~23. An attractive real OP, which is deep at small impact parameters $R$,
generates correspondingly a large local momentum $K(R)$ that suppresses the
Pauli distortion of the two Fermi spheres in a \emph{boot-strap} manner (see
upper part of Fig.~22). Soubbotin {\sl et al.} \cite{Sub01} also found that the
maximal effect by the Pauli distortion appears at the sub-surface distances of
3-5 fm and it accounts, in part, for the renormalization factor of the real
folded potential required by the OM fit to the elastic \AA scattering data. In
general, the Pauli distortion \cite{Sub01} can lead also to the excitation of
the two colliding nuclei which induces a loss of flux from the elastic channel
to other channels. However, the transformation of the Pauli distortion in the
momentum space into the real excitation of the two nuclei depends strongly on
their internal structure, and such a Pauli excitation is expected to be less
significant for a system of two strongly bound (closed-shell) nuclei. This
explains again why the refractive elastic \AA scattering with pronounced
rainbow pattern has been observed only in the elastic scattering of the
``robust" $\alpha$-particle and of the light HI systems involving strongly
bound nuclei like $^{12}$C and $^{16}$O.
\begin{figure}[ht]
 \begin{center}
 \vspace{-1cm}
\includegraphics[width=0.6\textwidth]{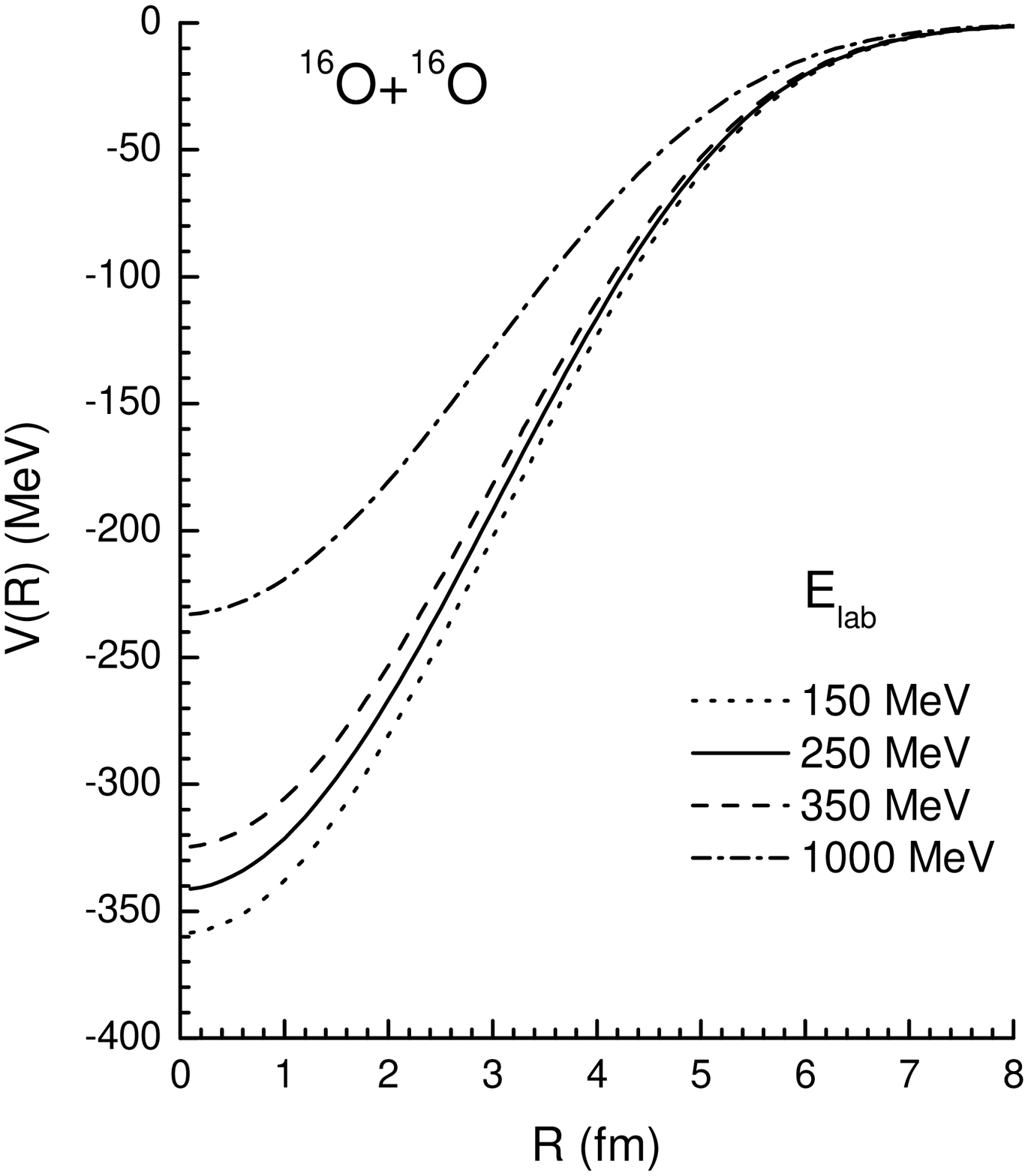}
\vspace{-2cm} \caption{Radial shapes of the real OP for the \oo system at
incident energies of 150, 250, 350 and 1000 MeV predicted by the DFM using the
CDM3Y6 version (see Table~1) of the density dependent M3Y-Paris interaction
\cite{Kho97}.}
   \end{center}
  \label{fig23}
\end{figure}

Thus, the agreement of the DFM results with the global systematics for the real
\AA OP has been shown to have a solid physical origin, which further justifies
the use of the DFM to probe the effective NN interaction as well as the wave
functions (or nuclear densities) of the two colliding nuclei in the folding
analysis of the refractive, nuclear rainbow scattering data.

\section{Rainbow scattering as a probe for the density dependence of
 in-medium NN interaction}
 \label{sec6}

Given correct nuclear densities as inputs for the folding calculation, it
remains necessary to have an appropriate in-medium NN interaction for a
reliable prediction of the (real) \AA OP. To evaluate an in-medium NN
interaction starting from the free NN interaction still remains a challenge for
the nuclear many-body theory. For example, the sophisticated
Brueckner-Hartree-Fock calculations which include the two- and three-nucleon
correlations cannot describe simultaneously the equilibrium density and the
binding energy of normal nuclear matter, unless the higher-order correlations
as well as relativistic effects are taken into account \cite{Muether}.
Therefore, most of the ``microscopic" nuclear reaction calculations so far
still use different kinds of the effective NN interaction. Such interactions
can be roughly divided into two groups. In the first group one parameterizes
the effective interaction directly as a whole, like the Skyrme forces, leaving
out any connection with the realistic free NN interaction. In the second group
one parameterizes the effective NN interaction in a functional form, amendable
to the folding calculation, based on the results of a nuclear many-body
calculation using the realistic free NN potential. Very popular choices in the
second group have been the so-called M3Y interactions which were designed by
the MSU group to reproduce the G-matrix elements of the Reid \cite{Be77} and
Paris \cite{An83} free NN potentials in an oscillator basis (further referred
to as M3Y-Reid and M3Y-Paris interaction, respectively). The original (spin-
and isospin independent) M3Y interaction is density independent and given in
terms of Yukawa functions as follows
\begin{eqnarray}
\fl\hbox{M3Y-Reid:}\hskip 1cm  v_{\rm D}(s)=7999.0\frac{\exp(-4s)}{4s}-
 2134.25\frac{\exp(-2.5s)}{2.5s},  \nonumber \\
\fl\hskip 1cm  v_{\rm EX}(s)= 4631.38\frac{\exp(-4s)}{4s}-
  1787.13\frac{\exp(-2.5s)}{2.5s} -7.8474\frac{\exp(-0.7072s)}{0.7072s};
\label{m3yR}\end{eqnarray}
\begin{eqnarray}
\fl\hbox{M3Y-Paris:}\hskip 1cm v_D(s)=11061.625\frac{\exp(-4s)}{4s}-
 2537.5\frac{\exp(-2.5s)}{2.5s},  \nonumber \\
\fl\hskip 1cm v_{\rm EX}(s)= -1524.25\frac{\exp(-4s)}{4s}-
 518.75\frac{\exp(-2.5s)}{2.5s}
 -7.8474\frac{\exp(-0.7072s)}{0.7072s}.
 \label{m3yP} \nonumber \\
\end{eqnarray}
The Yukawa strengths in Eqs.~(\ref{m3yR}) and (\ref{m3yP}) are given in MeV,
and $s$ is the distance between the two interacting nucleons. These
interactions, especially, the M3Y-Reid version have been used with some success
in the DFM calculation of the HI optical potential at low energies
\cite{Sat79}, with the elastic data usually limited to the forward scattering
angles and, thus, sensitive to the OP only at the surface. However, in cases of
refractive (rainbow) \AA scattering where the elastic data are sensitive to the
\AA OP over a much wider radial domain as discussed above in Sect.~\ref{sec3},
the density independent M3Y interactions failed to give a good description of
the data.
\begin{figure}[ht]
 \begin{center}
  \vspace{-4cm}
 \includegraphics[angle=0,scale=0.8]{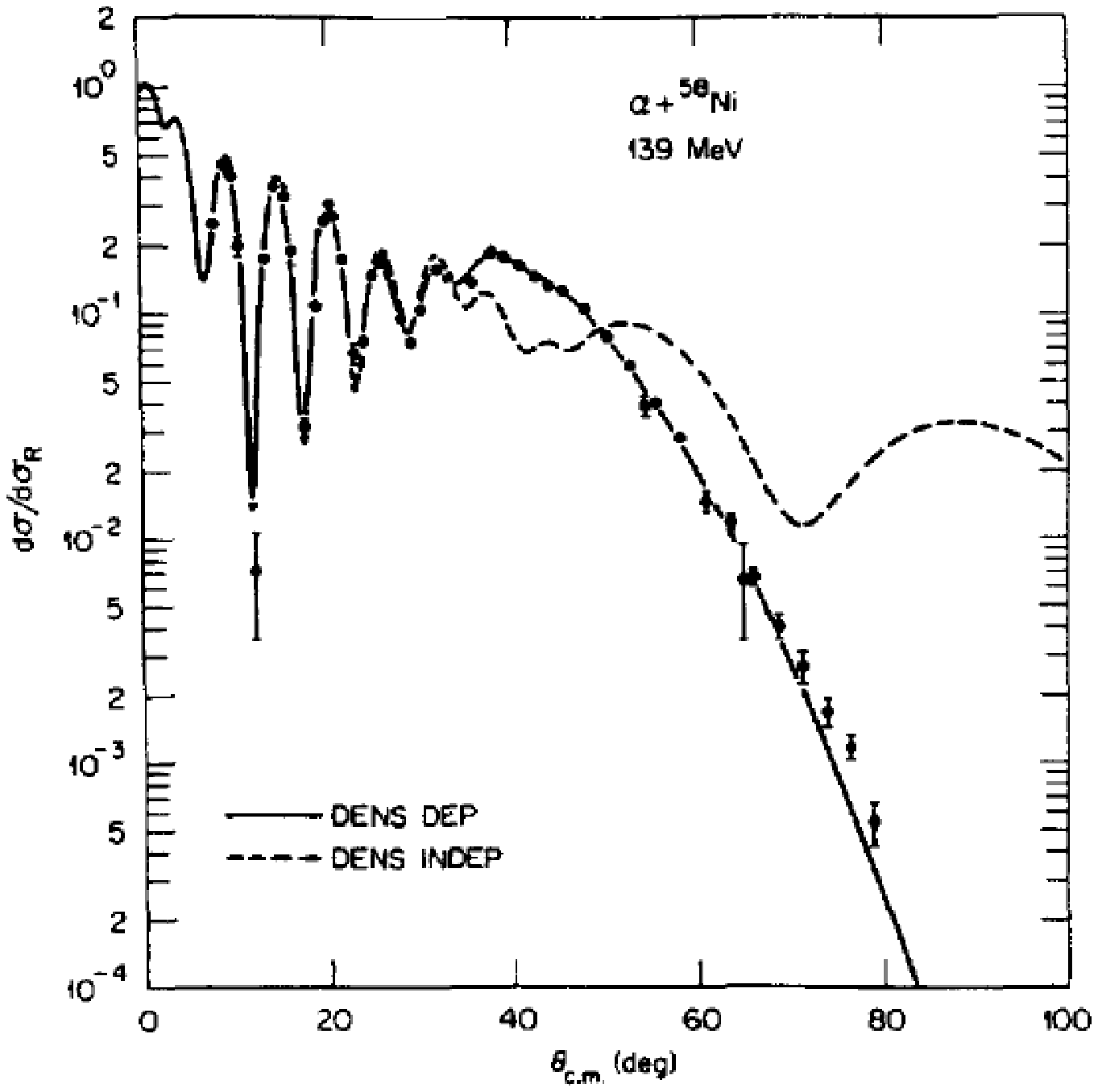}
  \vspace{-6cm}
\caption{OM fits to the elastic \aNi scattering data at 139 MeV \cite{Go73}
given by the folded potential obtained with (solid curve) and without (dashed
curve) density dependence of the effective NN interaction. The density
dependence was empirically introduced \cite{Sat83} to reduce the depth of the
folded potential at small distances while leaving the potential at the surface
nearly unchanged. Illustration taken from Ref.~\cite{Sat83}.}
   \end{center}
  \label{fig24}
\end{figure}
Namely, the folded potential is too deep at small distances $R$ to reproduce
the elastic cross sections at the large angles. A typical example is shown in
Fig.~24 where the inclusion of a density dependence into the effective NN
interaction was found essential to describe the rainbow ``shoulder" observed in
the elastic \aNi scattering at 139 MeV \cite{Go73}. In terms of medium effects,
the inclusion of an explicit density dependence was needed to account for a
reduction in the strength of the \AA interaction that occurs at small $R$ where
the overlap density of the nuclear collision increases. An early version of the
density dependence of the M3Y-Reid interaction was constructed by Kobos {\sl et
al.} \cite{Ko82} based upon the G-matrix results obtained by Jeukenne, Lejeune
and Mahaux \cite{Jeu77}. It was dubbed the DDM3Y interaction and has been used
to improve the folding model description of the elastic \aA \cite{Ko82,Ko84}
and light HI \cite{Br88} scattering.

\subsection{Density dependent M3Y interaction in the Hartree-Fock calculation
of nuclear matter} The physical origin of the density dependence of effective
NN interaction can be very well illustrated \cite{Kh93} in a Hartree-Fock (HF)
calculation of nuclear matter. Namely, given the direct $v_{\rm D}$ and
exchange $v_{\rm EX}$ parts of the effective in-medium NN interaction one can
easily calculate \cite{Walecka} the total NM binding energy
\begin{eqnarray}
E=E_{kin.}+{1\over 2}\sum_{k\sigma\tau}\sum_{k'\sigma'\tau'}
 [& <\bi{k\sigma\tau,k'\sigma'\tau'}|v_{\rm D}|\bi{k\sigma\tau,k'\sigma'\tau'}>
 \nonumber \\
 & +<\bi{k\sigma\tau,k'\sigma'\tau'}|v_{\rm EX}|\bi{k'\sigma\tau,k\sigma'\tau'}>]
 \label{efnn1}
\end{eqnarray}
using plane waves for $|\bi{ k\sigma\tau}>$. Our HF calculation \cite{Kh93} of
the NM energy (\ref{efnn1}) has shown that the original density independent M3Y
interaction (\ref{m3yR})-(\ref{m3yP}) failed to saturate NM, leading to a
collapse. The HF method is the first order of nuclear many-body calculation and
the introduction of a density dependence into the original M3Y interaction
accounts, therefore, for higher-order NN correlations which lead to the NM
saturation. The earlier DDM3Y version of the density dependent M3Y-Reid
interaction \cite{Ko82} resulted in a correct NM binding energy of around 16
MeV but at the wrong density ($\rho_0\simeq$ 0.07 fm$^{-3}$ compared to the
empirical saturation density of about 0.17 fm$^{-3}$ as shown in Fig.~1 of
Ref.~\cite{Kh93}). We have further introduced several versions of the density
dependence of the M3Y-Reid and M3Y-Paris interactions \cite{Kho95,Kh93,Kh95} by
scaling them with an explicit density dependent function $F(\rho)$
\begin{equation}
 v_{\rm D(EX)}(\rho,s)=F(\rho)v_{\rm D(EX)}(s),
\label{efnn2}
\end{equation}
where $v_{\rm D(EX)}$ are the direct and exchange components of the M3Y
interactions defined in Eqs. (\ref{m3yR})-(\ref{m3yP}) and $\rho$ is the NM
density. $F(\rho)$ was taken to be either the exponential dependence \cite{Ko82}
or the power-law density dependence \cite{Be71}, and the parameters were
adjusted to reproduce the observed NM saturation properties in the HF
calculation (\ref{efnn1}). Although different versions of the density dependence
give the same NM saturation properties, they do result in different curvatures
of the NM binding energy curve near the saturation point (see Fig.~25), i.e.,
they are associated with different values of the NM incompressibility $K$ which
is determined as
\begin{equation}
 K=9{\rho^2}\frac{d^2[E/A]}{d\rho^2}\ \Bigr\vert_{\displaystyle\rho=\rho_0}.
 \label{efnn3}
\end{equation}
\begin{table}
\caption{Parameters of different density dependences $F(\rho)$,
Eq.~(\ref{efnn4}), associated with the M3Y-Reid (\ref{m3yR}) and M3Y-Paris
(\ref{m3yP}) interactions \cite{Be77,An83}. The values of the nuclear
incompressibility $K$ were obtained from the NM binding energy using
Eq.~(\ref{efnn3}).}
 \vskip 0.5cm
\begin{tabular}{|c|c|c|l|l|l|c|c|c|} \hline\small
Interaction & $v_{\rm D(EX)}(s)$ & $C$ & $\alpha$ & $\beta$ (fm$^3$) &
 $\gamma$ (fm$^{3n}$) & $n$ & $K$ (MeV) & Ref. \\ \hline
DDM3Y1 & Eq.~(\ref{m3yR}) & 0.2845 & 3.6391 & 2.9605 & 0.0 & 0 & 171
 & \cite{Kh93} \\
DDM3Y1 & Eq.~(\ref{m3yP}) & 0.2963 & 3.7231 & 3.7384 & 0.0 & 0 & 176
 & \cite{Kh95} \\
CDM3Y1 & Eq.~(\ref{m3yP}) & 0.3429 & 3.0232 & 3.5512 & 0.5 & 1 & 188
 & \cite{Kho97} \\
CDM3Y2 & Eq.~(\ref{m3yP}) & 0.3346 & 3.0357 & 3.0685 & 1.0 & 1 & 204
 & \cite{Kho97} \\
CDM3Y3 & Eq.~(\ref{m3yP}) & 0.2985 & 3.4528 & 2.6388 & 1.5 & 1 & 217
 & \cite{Kho97} \\
CDM3Y4 & Eq.~(\ref{m3yP}) & 0.3052 & 3.2998 & 2.3180 & 2.0 & 1 & 228
 & \cite{Kho97} \\
BDM3Y1 & Eq.~(\ref{m3yR}) & 1.2253 & 0.0 & 0.0 & 1.5124 & 1 & 232
 & \cite{Kh93} \\
CDM3Y5 & Eq.~(\ref{m3yP}) & 0.2728 & 3.7367 & 1.8294 & 3.0 & 1 & 241
 & \cite{Kho97} \\
CDM3Y6 & Eq.~(\ref{m3yP}) & 0.2658 & 3.8033 & 1.4099 & 4.0 & 1 & 252
 & \cite{Kho97} \\
BDM3Y1 & Eq.~(\ref{m3yP}) & 1.2521 & 0.0 & 0.0 & 1.7452 & 1 & 270
 & \cite{Kh95} \\
BDM3Y2 & Eq.~(\ref{m3yR}) & 1.0678 & 0.0 & 0.0 & 5.1069 & 2 & 354
 & \cite{Kh93} \\
BDM3Y2 & Eq.~(\ref{m3yP}) & 1.0664 & 0.0 & 0.0 & 6.0296 & 2 & 418
 & \cite{Kh95} \\
BDM3Y3 & Eq.~(\ref{m3yR}) & 1.0153 & 0.0 & 0.0 & 21.073 & 3 & 475
 & \cite{Kh93} \\
BDM3Y3 & Eq.~(\ref{m3yP}) & 1.0045 & 0.0 & 0.0 & 25.115 & 3 & 566
 & \cite{Kh95} \\
\hline
\end{tabular}\end{table}
To have different $K$ values in finer steps, we have introduced a hybrid of the
exponential and power-law forms for $F(\rho)$ \cite{Kho97} and the different
density dependences of the M3Y-Reid and M3Y-Paris \cite{Kho97,Kh93,Kh95} can
all be written in the following form
\begin{equation}
 F(\rho)=C[1+\alpha\exp(-\beta\rho)-\gamma\rho^n]. \label{efnn4}
\end{equation}
The parameters $C,\alpha,\beta,\gamma$ and $n$ were chosen in each case to
reproduce the NM saturation properties in the HF calculation (\ref{efnn1}) and
they are given in Table 1. The HF results for the total NM energy obtained with
several versions of the density dependent M3Y-Paris interaction are shown in
Fig.~25, and one can see that different density dependences (resulting in
different $K$ values) lead to different slopes of the NM binding energy curve
at high NM density, i.e., different nuclear equations of state (EOS). We note
that $F(\rho)$, when used in the folding calculation of the real OP, needs to
be scaled by an energy dependent factor, $g(E)\approx 1-0.003\varepsilon$ and
$1-0.002\varepsilon$ for the M3Y-Paris and M3Y-Reid interaction, respectively,
where $\varepsilon$ is the bombarding energy per nucleon (in MeV). The
inclusion of $g(E)$ was found \cite{Kh93} necessary to account for the
empirical energy dependence of the nucleon-nucleus OP.
\begin{figure}[ht]
 \begin{center}
 \vspace{-1.0cm}
\includegraphics[width=0.65\textwidth]{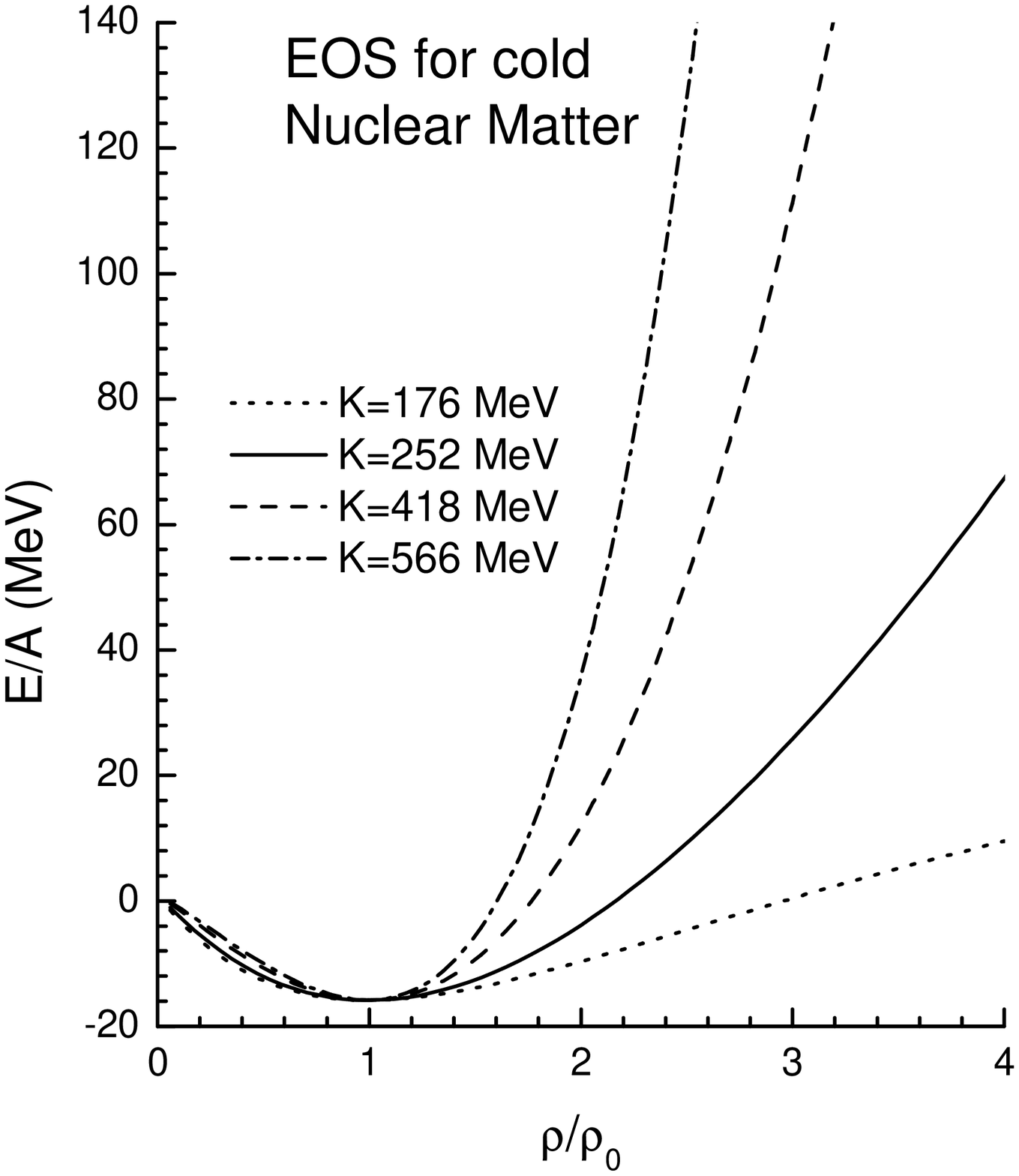}
\vspace{-2cm} \caption{Nuclear matter binding energy as function of the NM
density given by the HF calculation (\ref{efnn1}) using 4 choices of the
density dependent M3Y-Paris interaction (named as DDM3Y1, CDM3Y6, BDM3Y2 and
BDM3Y3 in Table 1). $\rho_0\approx 0.17$ fm$^{-3}$ is the NM saturation
density. These density dependent interactions are associated with the nuclear
incompressibility $K$ ranging from 176 to 566 MeV \cite{Kho95,Kho97}.}
   \end{center}
  \label{fig25}
\end{figure}

\subsection{Probing the nuclear EOS in the folding model analysis of refractive
\AA scattering}
One of the main goals of the study of HI collisions remains the determination of
the nuclear EOS, which is important in both nuclear physics and astrophysics.
Different types of the EOS are usually distinguished by different values of the
nuclear incompressibility $K$. Many attempts in this direction have been made in
the study of high-energy central HI collisions where one hopes to deduce from
the measured transverse flows and particle spectra (nuclear fragments) some
information on the incompressibility $K$ of high density matter formed in the
compression stage of such a reaction. Various transport models have been
successfully used in reproducing such data, but in many cases the results still
remain inconclusive concerning the EOS.
\begin{figure}[ht]
 \begin{center}
 \vspace{-3cm}
\includegraphics[angle=0,scale=0.6]{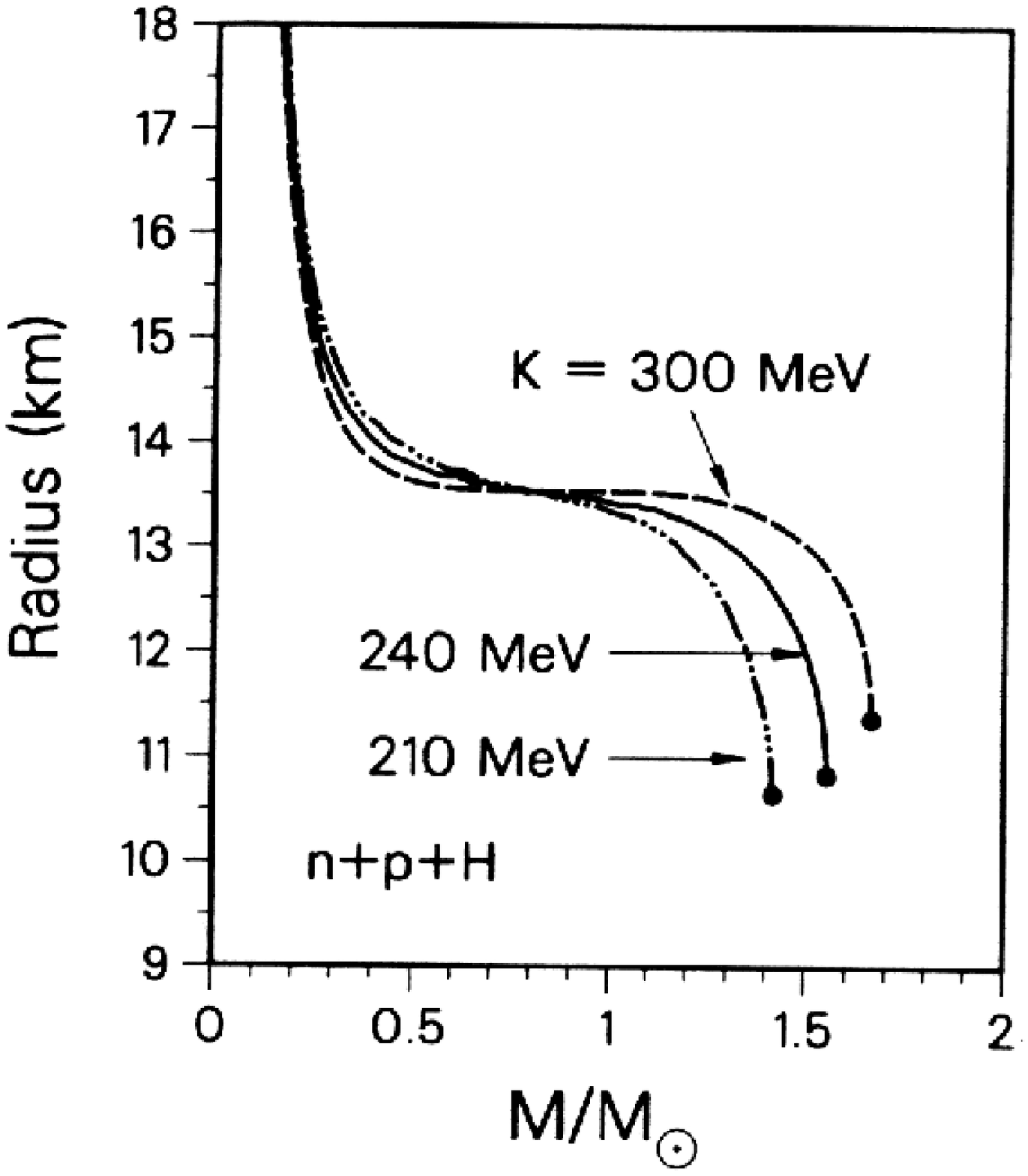}
\vspace{-3cm} \caption{Relation between the radius and mass of the neutron star
given by the EOS's corresponding to different nuclear incompressibilities $K$.
In all cases the neutron star matter is in equilibrium between neutrons,
protons, hyperons and leptons. The end point of each curve marks the final
value for the neutron star radius. Illustration taken from
Ref.~\cite{Glend00}.}
   \end{center}
  \label{fig26}
\end{figure}

In general, we need a well-defined (and sensitive to $K$) quantity which can be
measured with high precision. We need further an effective NN interaction which
reproduces, on one hand, the basic NM properties, and on the other hand, can be
used as a basic input in the description of the considered experimental
quantity. With this interaction one should be able to generate different $K$
values by varying parameters of its density dependence, so that one can
directly test the sensitivity of the considered quantity to the nuclear
incompressibility $K$. It turned out that the folding model analysis of
high-precision nuclear rainbow scattering data can be used as an independent
method to determine the nuclear incompressibility $K$. This is the reason why
we have parameterized several density dependences of the M3Y interaction as
summarized in Table 1. In the early 80's, a very soft EOS (with $K$ around 160
- 180 MeV) was thought to be sufficient to allow a prompt explosion in
supernovae \cite{Ba85}, but more recent numerical hydrodynamical studies
indicate that this is not the case and the constraint by the observed neutron
star mass requires \cite{Sw94,Glend00} higher $K$ values  around 240 MeV.
Fig.~26 shows the mass-radius relation given by three model calculations of the
neutron star which are distinguished by different values of the nuclear
incompressibility $K$ of the symmetric NM \cite{Glend00}. Since the neutron
star mass is well constrained to around 1.5 solar mass by the observed radio
pulsar masses, the realistic $K$ values should lie within the range of 210-300
MeV as shown in Fig.~26. Some studies of high-energy central HI collisions
suggest quite high $K$ values, e.g., the determination of $K$ based upon the
production of hard photons in HI collisions has led to an estimate of $K\approx
290\pm 50$ MeV \cite{Sc96}. All this has motivated us to study in more detail
the sensitivity of refractive \AA scattering data to the $K$ value and, thus,
to determine it with more precision by using different density dependent M3Y
interactions (given in Table~1).

\begin{figure}[ht]
 \begin{center}\vspace{-1cm}
\includegraphics[width=0.7\textwidth]{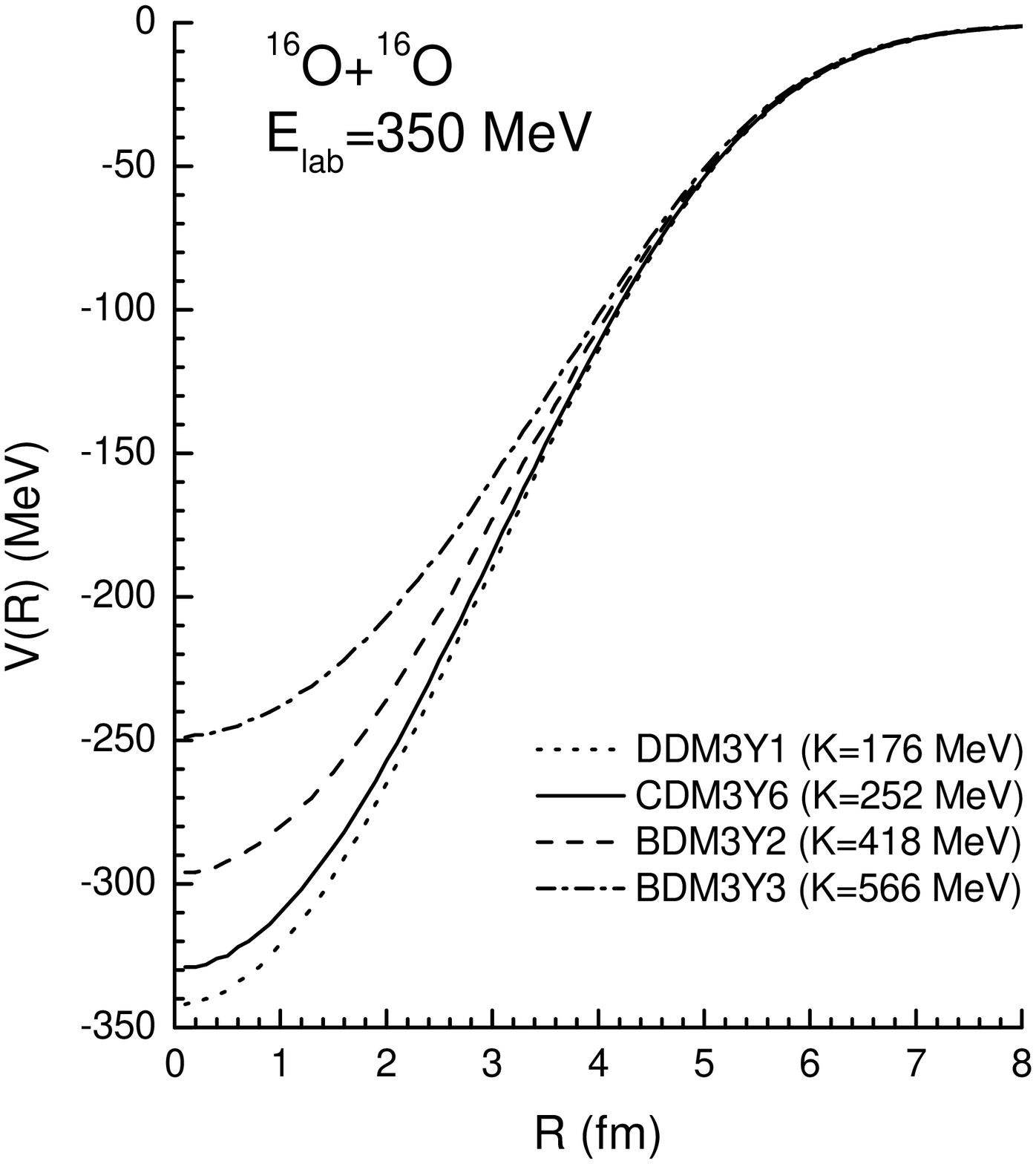} \vspace{-2cm}
\caption{Radial shapes of the real OP for the \oo system at $E_{\rm lab}=350$
MeV predicted by the DFM using the same versions of the density dependent
M3Y-Paris interaction as those used in the HF calculation of NM shown in
Fig.~25.}
   \end{center}
  \label{fig27}
\end{figure}

\begin{figure}[ht]
 \begin{center}
  \vspace{-1.5cm}
\includegraphics[width=0.7\textwidth]{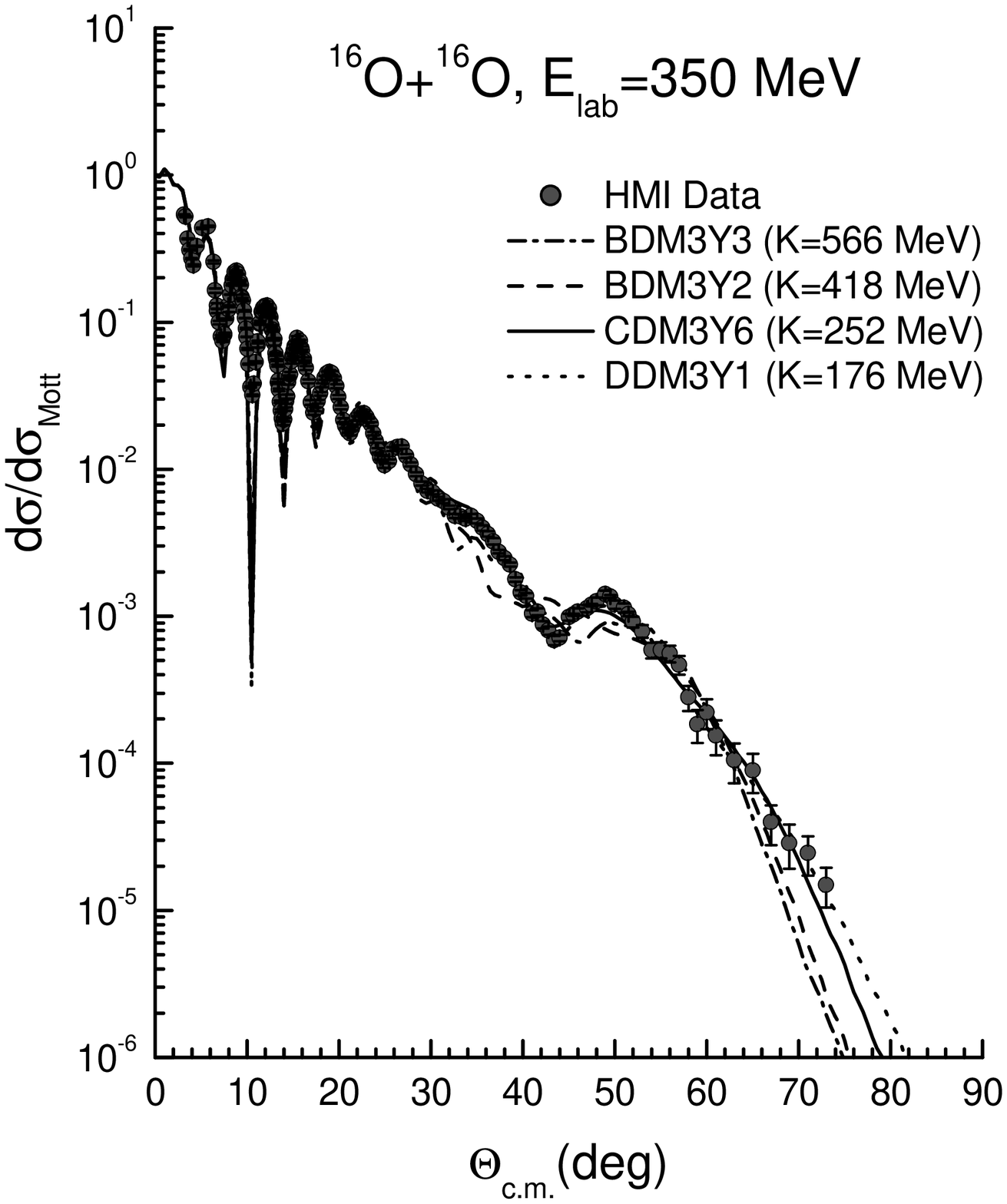}
 \vspace{-1.5cm}
\caption{OM description of the elastic \oo scattering data at $E_{\rm lab}=350$
MeV \cite{Sti89} given by the real folded potentials shown in Fig.~27 and an
absorptive WS imaginary potential taken from Ref.~\cite{Kho00}. The best-fit
version is CDM3Y6 which gives $K\approx 252$ MeV in the HF calculation of NM.}
   \end{center}
  \label{fig28}
\end{figure}

A typical example is presented in Fig.~27, where different density dependent
M3Y-Paris interactions (which give different EOS's shown in Fig.~25) are used
in the DFM to calculate the real \oo optical potential at 350 MeV. One can see
that the difference in the double-folded potentials is strongest at small
distances where the overlap density of the \oo system is large. Since the
primary rainbow pattern of the elastic \oo data at 350 MeV is quite sensitive
to small partial waves, as shown in Fig.~13, these data can be used to probe
the $K$ value by using a realistic imaginary WS potential obtained from the OM
systematics \cite{Kho00} and the real double-folded potentials given by
different density dependent M3Y interactions. The corresponding OM results are
shown in Fig.~28 where the CDM3Y6 interaction \cite{Kho97} has been found as
the most favorable interaction. This version of the density dependent M3Y-Paris
interaction gives $K\approx 252$ MeV in the HF calculation of symmetric NM.

\begin{figure}[hbt]
 \begin{center}\vspace{-3cm}\hspace{-4cm}
\includegraphics[angle=270,scale=0.6]{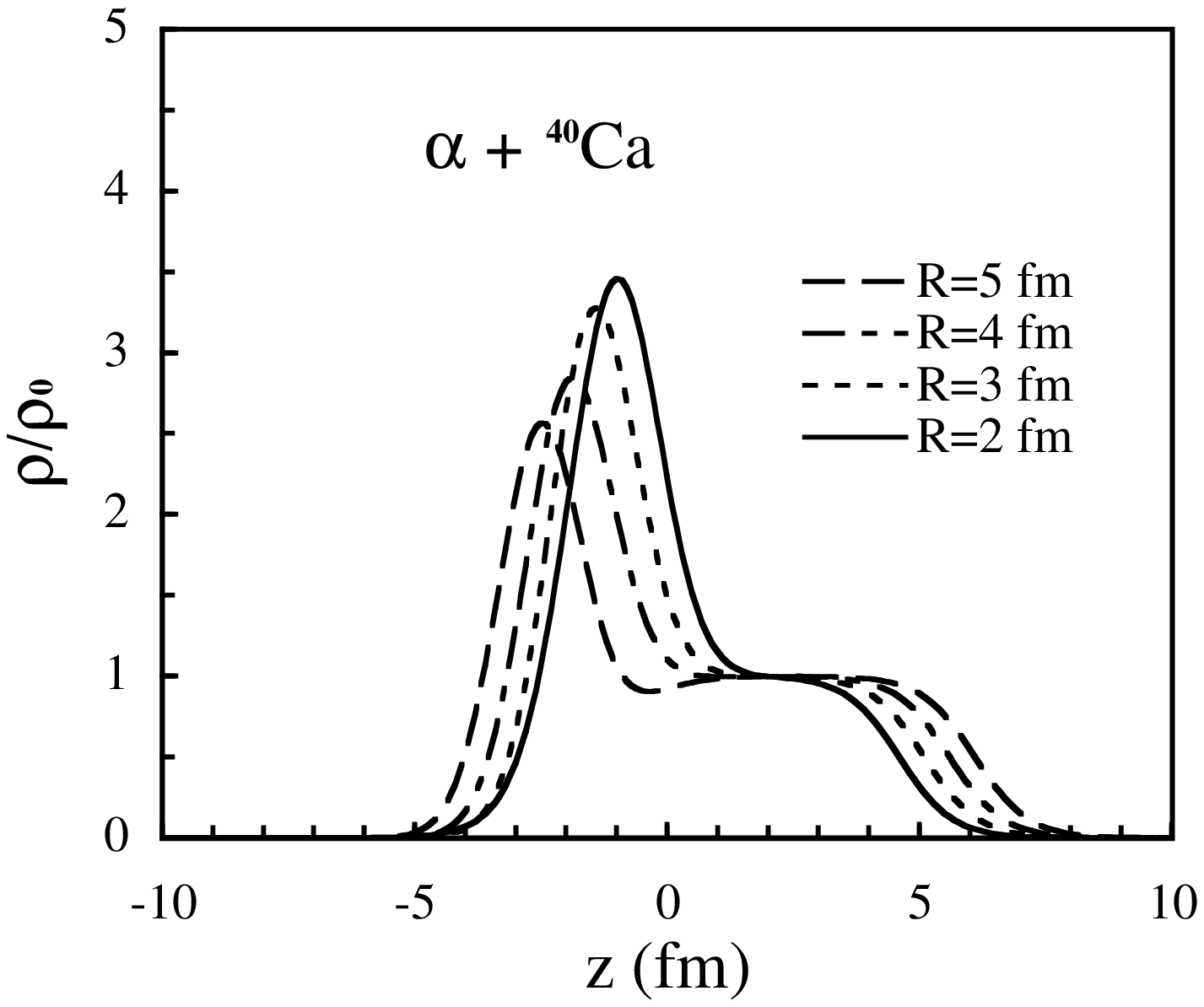}\vspace{-6cm}
\caption{The overlap density of the \aCa system at different internuclear
distances $R$. The z axis is directed along the line connecting the centres of
the two nuclei. Illustration taken from Ref.~\cite{Kho97}.}
   \end{center}
  \label{fig29}
\end{figure}
A similar (and quite unambiguous) conclusion about the realistic value of the
nuclear incompressibility $K$ has been reached in our folding analysis
\cite{Kho97,Kh95} of the refractive elastic \aA scattering data. The weak
absorption observed in the refractive \aA scattering at medium energies, with
the appearance of the nuclear rainbow pattern, offers a unique opportunity to
probe the density dependence of the effective NN interaction. A crucial point
in this connection is the very high and compact density profile of the $\alpha$
particle. Given a density as high as $\rho\simeq 2\rho_0$ in the centre of the
$^4$He nucleus \cite{Si76}, the total density for the $\alpha$ particle
overlapping a target nucleus may reach as much as 3$\rho_0$ (as shown in Fig.
29).
\begin{figure}[ht]
 \begin{center}
  \vspace{-1cm}
\includegraphics[scale=0.55]{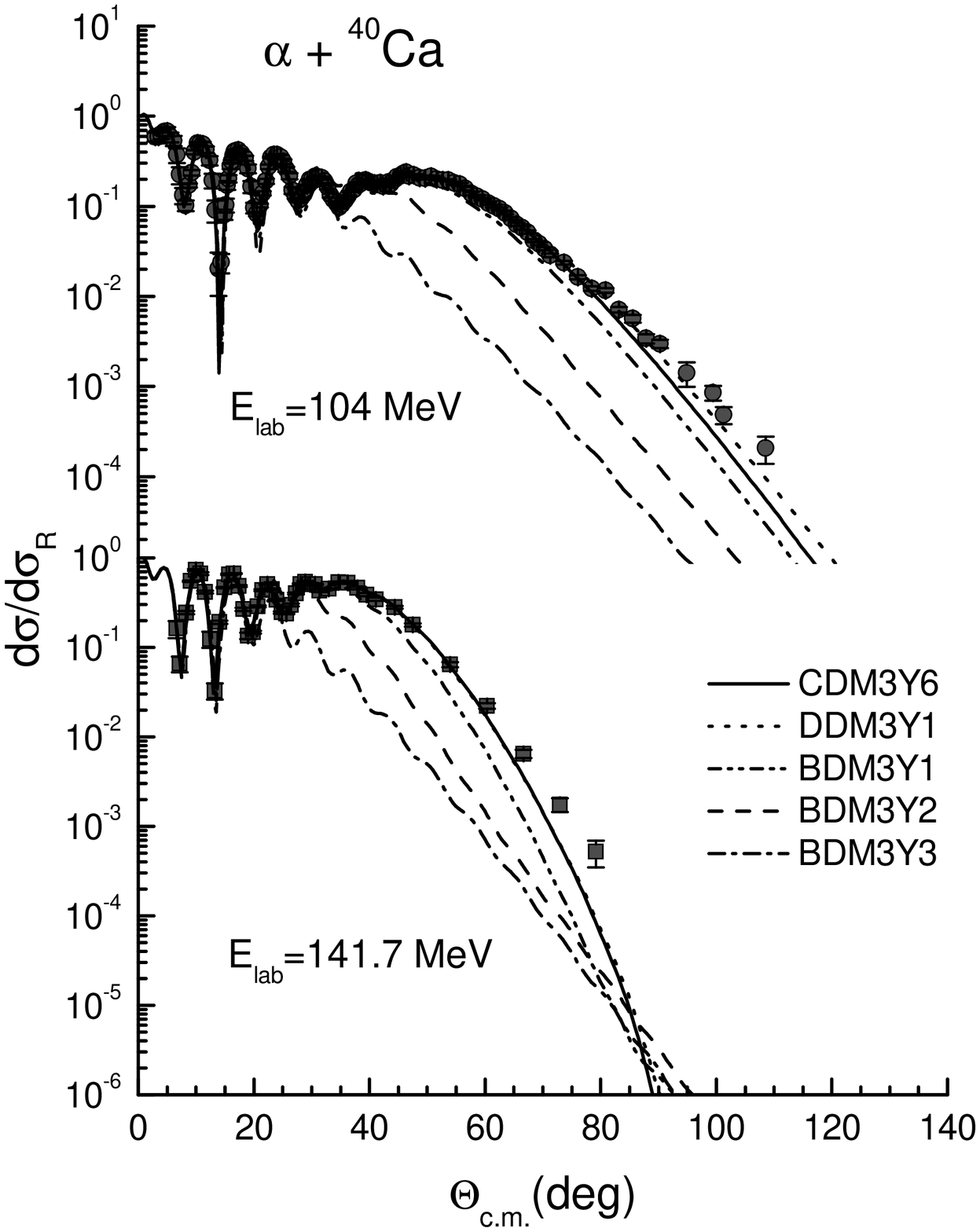}
 \vspace{-1.5cm}
\caption{OM description of the elastic \aCa scattering data at $E_{\rm lab}=104$
\cite{Gil80} and 147 MeV \cite{Go74} given by the different real folded
potentials shown in Fig.~21 and an imaginary WS potential determined from the OM
systematics \cite{Ba89,Gi87}.}
   \end{center}
  \label{fig30}
\end{figure}
From Fig.~21 one can see that such a high \aCa overlap density results in a
significant difference in the double-folded potential already at a separation
distance of $R=4$ fm. The real \aA OP can be very well determined at such a
radius if the bombarding energy is sufficiently high for the appearance of the
primary rainbow maximum in the elastic cross section. The results of our folding
analysis of the elastic \aCa scattering data at $E_{\rm lab}=104$ \cite{Gil80}
and 147 MeV \cite{Go74} are shown in Fig.~30 where one can easily deduce the
most appropriate density dependence of the M3Y-Paris interaction. The double
folded \aCa potentials are compared with the real OP deduced from a
model-independent Fourier-Bessel analysis in Fig.~21. One can see that the shape
of the real OP is determined rather well for $R$ down to about 2 fm. At this
distance the difference between different folded potentials is so obvious that
one can exclude immediately the BDM3Y2 and BDM3Y3 interactions (which give
rather high $K$ values in the HF calculation) as unrealistic ones.

To validate the results presented above, it is important to discuss the
approximation for the overlap density in the DFM calculation. We recall that the
folding model generates the first-order term of the Feshbach optical potential
(\ref{ep4}) which is further used in the OM equation (\ref{ep1}) to obtain the
relative-motion wave function of the two nuclei remaining in their ground
states. Given the antisymmetrization of the dinuclear system accurately taken
into account, a reasonable approximation for the total density $\rho$ of the two
overlapping nuclei is the sum of the two g.s. densities. For example, in the
calculation of the direct folded potential (\ref{ef5}) the overlap density
$\rho$ in $F(\rho)$ is taken as the sum of the two g.s. densities at the
position of each nucleon
\begin{equation}
  \rho_{a+A}=\rho_a ({\bi r}_a) + \rho_A({\bi r}_A).
 \label{ef0}
\end{equation}
Such an assumption, dubbed as Frozen Density Approximation (FDA), gives
naturally the overlap density $\rho_{a+A}$ reaching up to twice the NM
saturation density $\rho_0$ at small internuclear distances. The FDA has been
widely used in the folding calculations with density dependent NN interaction
\cite{Sat79,KhoSat,Ko82,Ko84,Br88} for the elastic \AA scattering when the
energy is not too low. Any density rearrangement that might happen during the
collision would lead to the nuclear states different from the ground states,
and thus contribute to the higher-order dynamic polarization potential $\Delta U$
of Eq.~(\ref{ep5}). In general, the FDA reproduces very well the observed
reduction of the attractive strength of the real OP at small distances, like
example shown in Fig.~24. The use of FDA is, therefore, crucial for the probe
of the density dependence of the effective NN interaction in the folding model
analysis of refractive \AA scattering.
\begin{figure}[ht]
 \begin{center}
  \vspace{-1cm}
\includegraphics[width=0.7\textwidth]{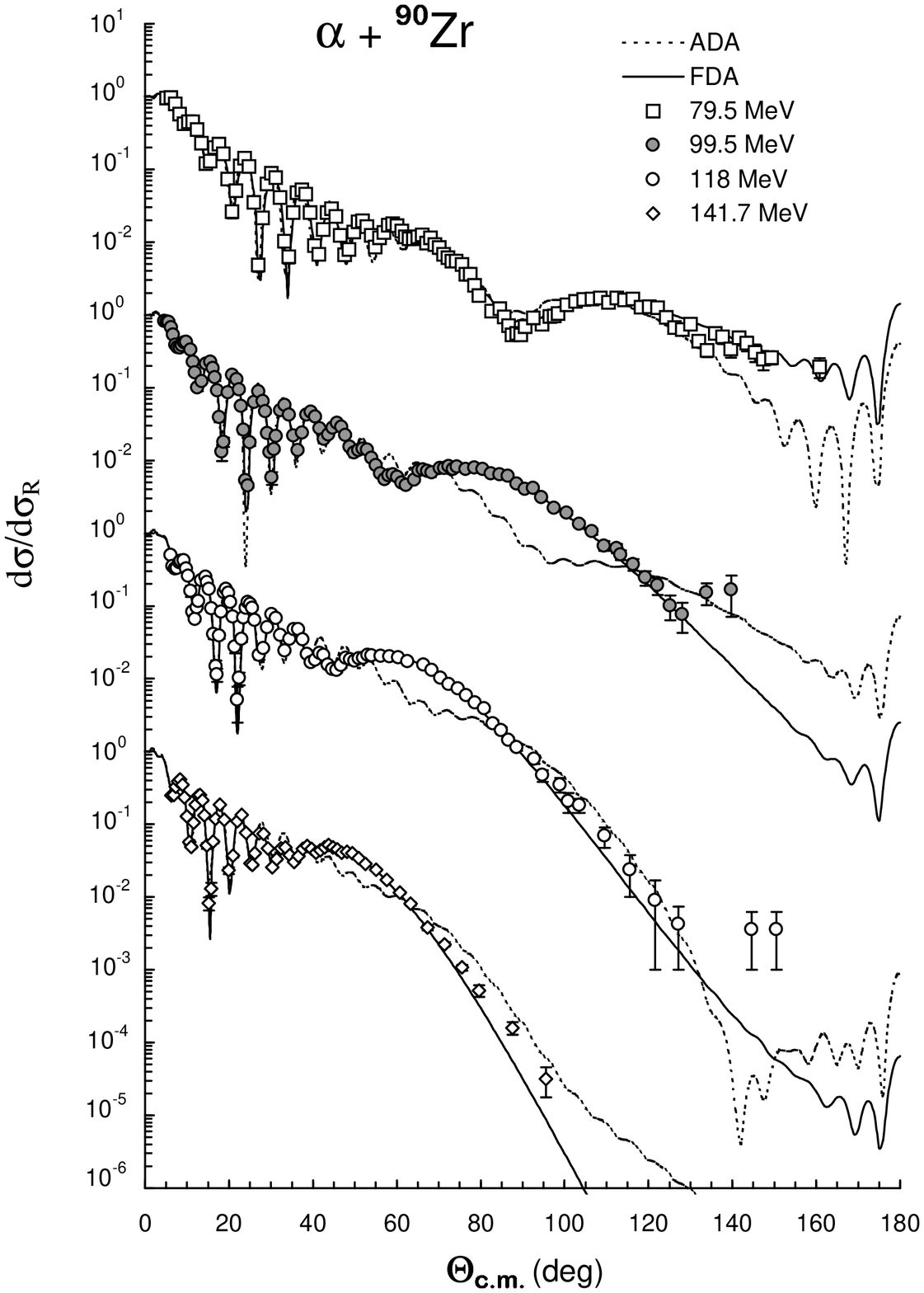}
 \vspace{-1.5cm}
\caption{Elastic \aZr scattering data at $E_{\rm lab}=79.5, 99.5, 118$ MeV
\cite{Pu77} and 141.7 MeV \cite{Go74} in comparison with the OM fits given by
the CDM3Y6 double-folded potentials obtained with two different approximations
for the overlap density: Average Density Approximation (dotted curves) and
Frozen Density Approximation (solid curves). Illustration taken from
Ref.~\cite{Kho01}.}
   \end{center}
  \label{fig31}
\end{figure}
We note that other approximations for $\rho_{a+A}$, based on the geometric or
arithmetic averages of the two local densities, have also been used in the
folding calculation. For example, in the JLM double-folding calculation
\cite{Trache00} one has adopted the arithmetic average of the two densities for
$\rho_{a+A}$ to prevent it from becoming larger than $\rho_0$, because the JLM
parameters \cite{Jeu77} were determined for $\rho\leq\rho_0$ only. Such an
Average Density Approximation (ADA) has been carefully compared with the FDA in
Ref.~\cite{Kho01} using the density-dependent CDM3Y6 interaction, and the ADA
was found not appropriate to account for the observed reduction of the real \aA
OP at small distances. Since the ADA gives a much smaller overlap density
compare to the FDA, the folding potential calculated using the ADA is more
attractive and significantly deeper than that given by FDA (see Fig.~13 in
Ref.~\cite{Kho01}). One can see in Fig.~31 that the excessive depth of the
double-folded potential given by the ADA results in the failure of this
potential to describe the observed rainbow pattern in the elastic \aZr data at
large angles. The use of the ADA in the JLM double-folding calculation is also
the most likely reason why the JLM folding potential fails to correctly
describe the rainbow ``shoulder" seen in the $^{6,7}$Li+$^{12,13}$C elastic
data at large angles (see Fig.~6 of Ref.~\cite{Trache00}). Thus, at the
``rainbow" energies, the FDA should be more appropriate for the overlap density
in the double-folding calculation (\ref{ef5})-(\ref{ef6}), where the exchange
term is explicitly treated. A very recent double-folding analysis of the
elastic \aA scattering using the JLM interaction by Furumoto and Sakuragi
\cite{Fur06} has shown consistently that the renormalization factor of the real
folded potential is around 0.7, a value normally observed only for a loosely
bound projectile, such as $^6$Li or $^{11}$Li, but \emph{not} for the robust
$\alpha$ particle. The geometric average used by these authors for the local
density gives $\rho_{a+A}=\sqrt{\rho_a\rho_A}$ so that the overlap density does
not exceed $\rho_0$. As a result, the double-folded \aA potential is also too
deep in such a ``factorizing" treatment of the di-nuclear density and, hence,
needs a renormalization factor significantly smaller than unity. The authors of
Ref.~\cite{Fur06} have concluded that an improvement of the parametrization for
the density dependence of the JLM interaction is necessary to resolve the
substantial renormalization problem in the folding analysis of the \aA
scattering.

\begin{figure}[ht]
 \begin{center}
 \vspace{-2cm}\hspace{-3cm}
\includegraphics[angle=270,scale=0.6]{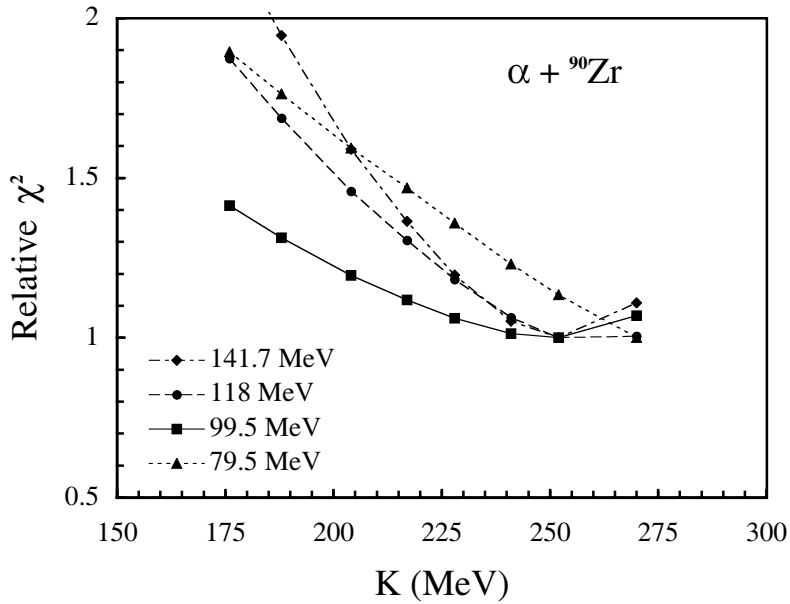}
\vspace{-6cm} \caption{Relative $\chi^2$ values (in ratio to the lowest
$\chi^2$ value obtained in each case) of the OM fits to the elastic \aZr
scattering data \cite{Go74,Pu77}, versus the corresponding $K$ values
associated with the different density dependent M3Y-Paris interactions (see
Table~1) used in the DFM calculation. The lines are only to guide the eye.
Illustration taken from Ref.~\cite{Kho97}.}
   \end{center}
  \label{fig32}
\end{figure}
From a detailed folding analysis \cite{Kho97} of the strongly refractive \aZr
rainbow scattering data \cite{Go74,Pu77} shown in Fig.~19, we have established
a systematic behavior of the $\chi^2$ value (of the OM fit) which reaches a
clear minimum with the CDM3Y5 or CDM3Y6 versions of the M3Y-Paris interaction
(see Fig.~32). These density dependent M3Y-Paris interactions give values of
the nuclear incompressibility with $K\approx 241$ and $252$ MeV, respectively.
This result clearly indicates that a very soft EOS (with $K$ around 180 MeV) is
less realistic than a slightly stiffer EOS (with $K\simeq 250$ MeV). To
conclude this discussion, an overview of the method used to probe the nuclear
incompressibility $K$ in the folding model analysis of the refractive elastic
\AA scattering is schematically illustrated in Fig.~33.
\begin{figure}[htbp]
 \begin{center}
 \vspace*{0cm}
 \includegraphics[width=0.65\textwidth]{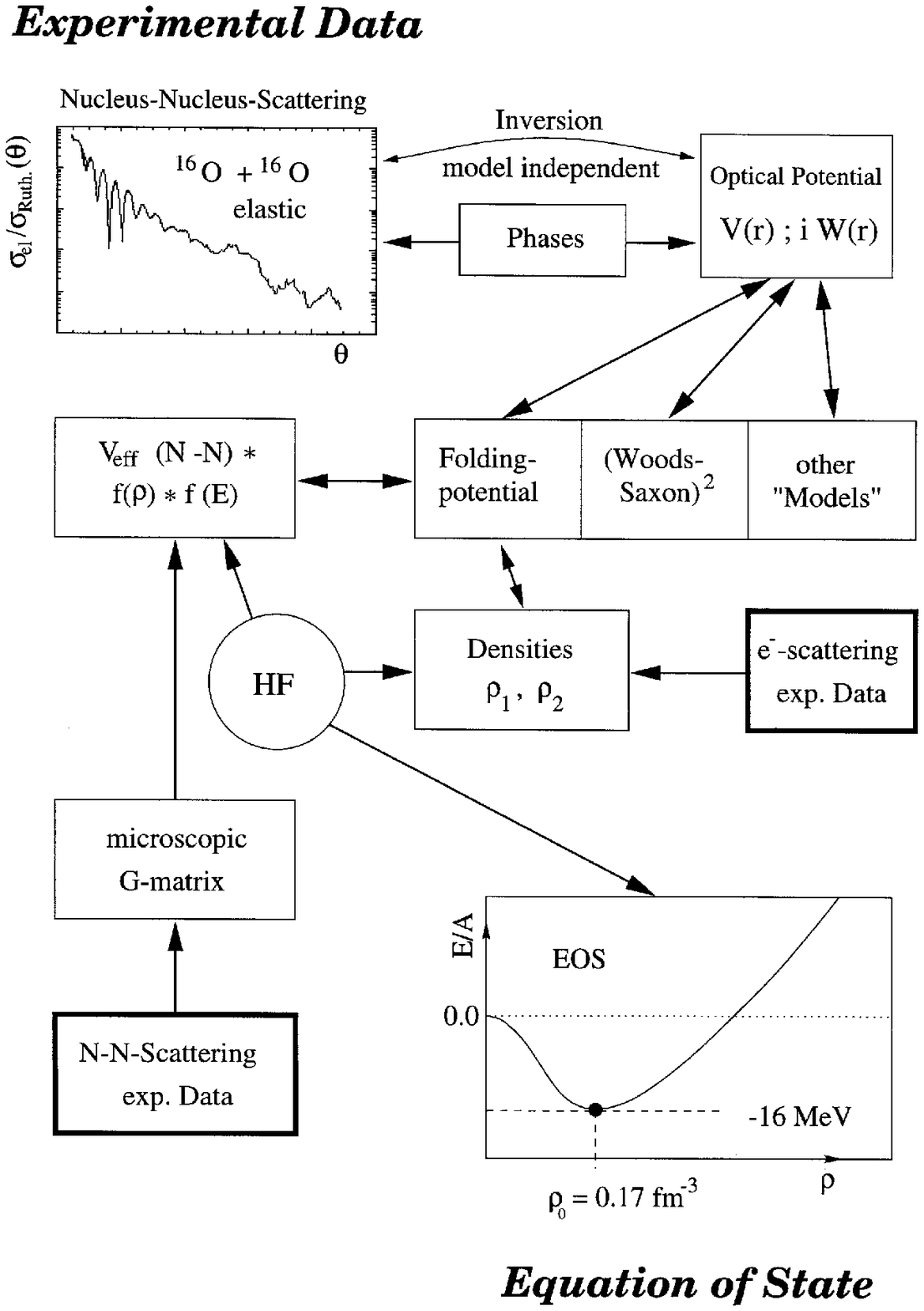}
  \vspace{0cm}
  \caption{The schematic link between the experimental data of elastic \AA
  scattering, nucleus-nucleus OP, nuclear densities and the effective
  NN interaction used in the folding calculation. Given parameters of the
  density dependence chosen to reproduce the NM saturation properties in the HF
  calculation (\ref{efnn1}), the nuclear incompressibility $K$
 (a key to specify the nuclear EOS) can be probed in the folding analysis
  of the \emph{refractive} \AA scattering.}
   \end{center}
  \label{fig33}
\end{figure}

Finally, it is complementary to make a brief comment on the method to deduce
the $K$ value from the study of isoscalar giant monopole resonance (ISGMR) in
medium-mass nuclei. The most recent experimental development has made it
possible to measure the ISGMR energies $E_0$ with high precision ($\Delta E_0
\sim 0.1-0.3$ MeV) \cite{You04}. However, in order to deduce a realistic $K$
value, one needs to compare the measured $E_0$ with the ISGMR energy and
strength predicted by a nuclear structure model. Due to the approximations
usually made in structure calculations, one has an additional uncertainty in
the calculated $K$ value, which must be added quadratically to the experimental
error \cite{Col04a}. Therefore, even if one uses the accurately measured
$E_0\approx 13.96\pm 0.20$ MeV for the ISGMR in $^{208}$Pb \cite{You04}, the
uncertainty in the extracted $K$ value remains significant ($K\approx 200-300$
MeV). Although a clear correlation between the ISGMR energy and $K$ value has
been well established since the original work by Blaizot {\it et al.}
\cite{Bla80}, until a few years ago, the extraction of $K$ value was plagued by
a critical dependence on nuclear models. Namely, the correlations between $E_0$
and $K$ were different for different families of functionals used in the
structure calculations, like RPA based on the Skyrme or Gogny forces or the
Relativistic Mean Field (RMF) calculations. Recently, it has been shown by
Col\`o {\it et al.} \cite{Col04a} that the discrepancy between the RPA results
obtained with Skyrme forces and those obtained with Gogny forces disappears if
the selfconsistency violation (the neglect of Coulomb and spin-orbit residual
terms) is properly corrected. Then, the fully self-consistent RPA calculations
based on Skyrme forces do not give $K\approx 210$ MeV as quoted in
Ref.~\cite{Bla80}, but predict $K\approx 235$ MeV which agrees well with the
RPA result obtained with Gogny forces. The $K$ value given by the RPA
calculations is slightly lower than that given by the relativistic RMF
calculations \cite{Vret03}, which predict $K\approx 250-270$ MeV. Such a
difference is now also understood as caused by different behaviors of the
symmetry energy within these models \cite{Pie02,Agr03,Col04b}. Guided by
realistic physics inputs, one can deduce from these structure calculation, that
$K\approx 240 \pm 20$ MeV, which is very close to that deduced from the folding
model studies of the \aA and \AA rainbow scattering (see Fig.~32).

\section{Rainbow features in other quasi-elastic scattering channels}
 \label{sec7}
Although the nuclear rainbow pattern has been observed mainly in the elastic,
refractive \AA scattering, it was natural to expect that rainbow features
appear also in other quasi-elastic reactions like the inelastic scattering,
nucleon transfer and charge exchange channels.
\begin{figure}[htbp]
 \begin{center}
\includegraphics[angle=0,scale=0.6]{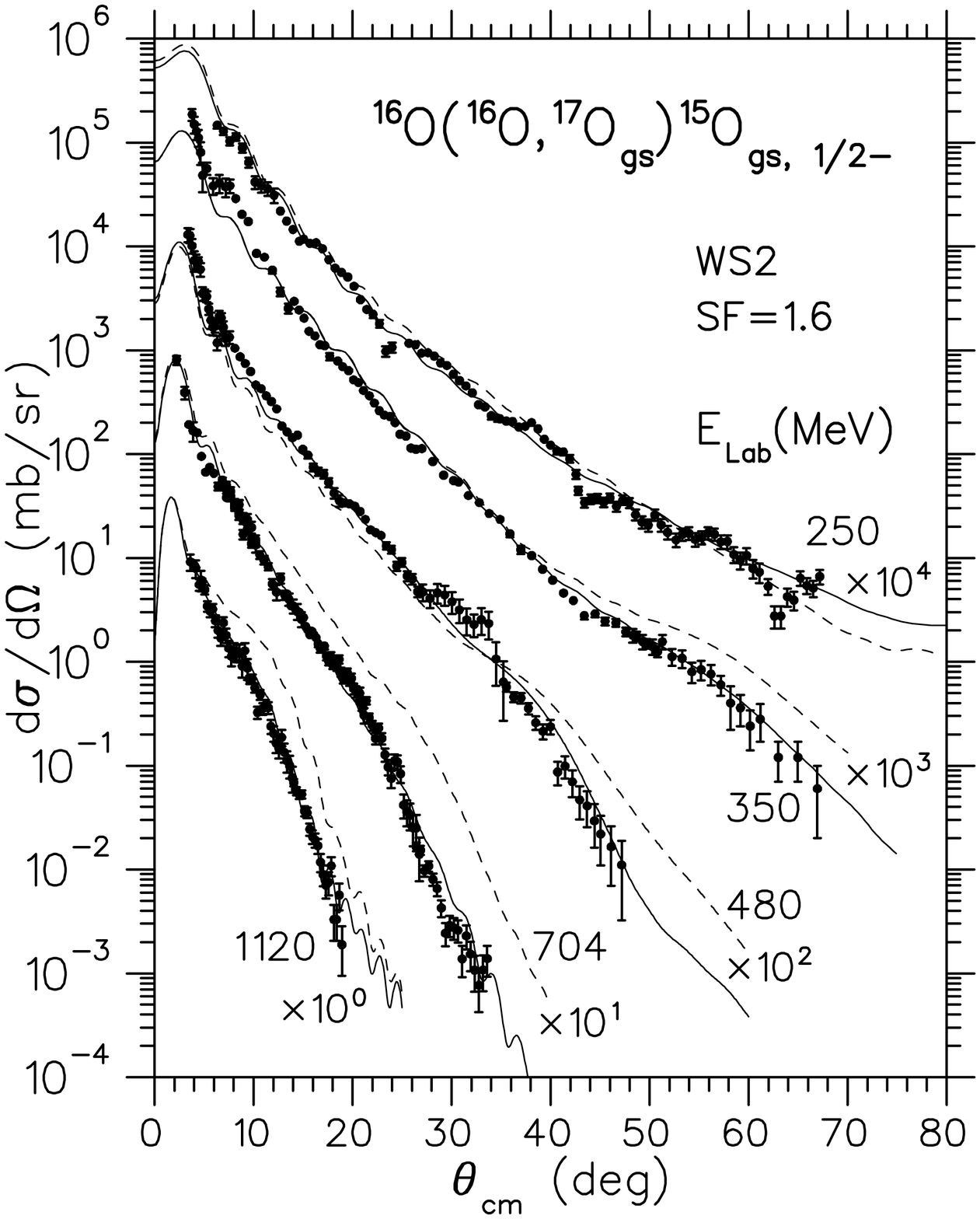}
\caption{Results of the DWBA calculation for the \o17o15 one-neutron transfer
reaction to the $^{15}$O$_{{1/2}^-}$ ground state at $E_{\rm lab}=250\to 1120$
MeV in comparison with the data \cite{Boh02}. The dashed curves were obtained
with the same complex OP for the \7o5o exit channel as that used for the
entrance \oo channel. The solid curves were obtained with a more absorptive OP
in the \7o5o exit channel. Illustration taken from Ref.~\cite{Boh02}}
  \end{center}
 \label{fig34}
\end{figure}
In general, due to a stronger absorption, these non-elastic rainbow patterns
(which have no counterparts in the optical rainbow) should be less pronounced
and harder to observe experimentally. The rainbow effects have been
investigated, e.g., in the inelastic scattering and one-neutron transfer
reactions measured with $^{12,13}$C+$^{12}$C systems at the energy of 20
MeV/nucleon \cite{Boh85}. While the refractive effects were found much weaker
in the inelastic $^{12,13}$C+$^{12}$C scattering, some remnant of the nuclear
rainbow has been identified in the one-neutron transfer
\begin{figure}[htb] \begin{center}
 \vspace{-0.5cm}
\includegraphics[angle=0,scale=0.60]{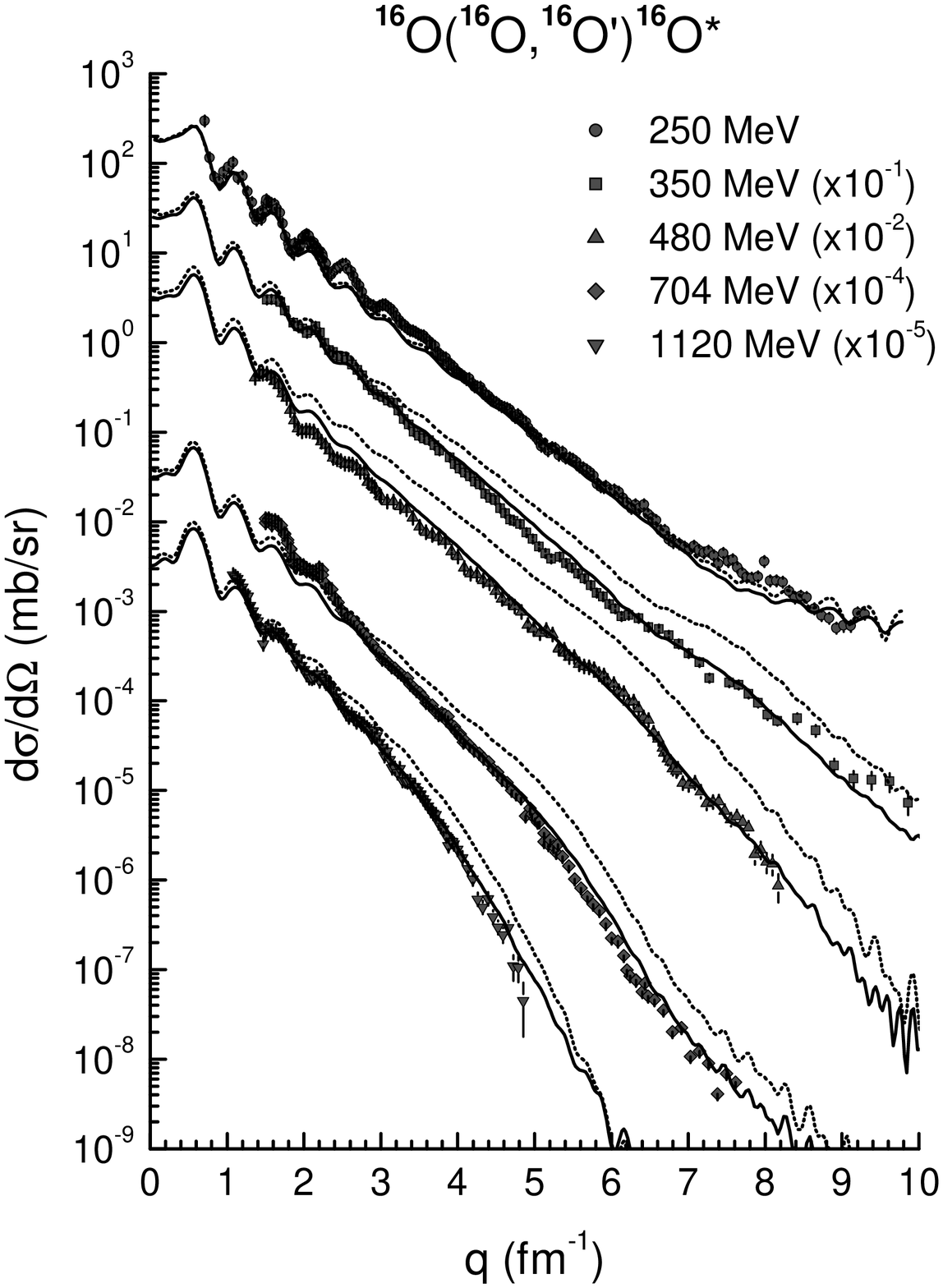}
 \vspace{-1cm}
\caption{The DWBA description of the total (2$^++3^-$) inelastic \InOO
scattering data at $E_{\rm lab}=250-1120$ MeV given by the real folded form
factors \cite{Kho05}. The dotted curves were obtained with the same complex OP
for the entrance and exit channels, and the solid curves were obtained with a
more absorptive OP in the exit channels. The cross sections are plotted versus
the momentum transfer $q=2k\sin(\Theta_{\rm c.m.}/2)$, where $k$ is the wave
number of the projectile. Illustration taken from Ref.~\cite{Kho05}}
\end{center}\label{fig35}
\end{figure}
$^{12}$C($^{12}$C,$^{13}$C)$^{11}$C channel \cite{Boh85}. These refractive
one-neutron transfer data were shown by Satchler \cite{Sat89} to be rather
sensitive to the shape of OP used in the transfer calculation and could be
used, therefore, to reduce the OP ambiguities for the \cc system.
\begin{figure}[htb]
\begin{center} \vspace{-1cm}
\includegraphics[angle=0,scale=0.6]{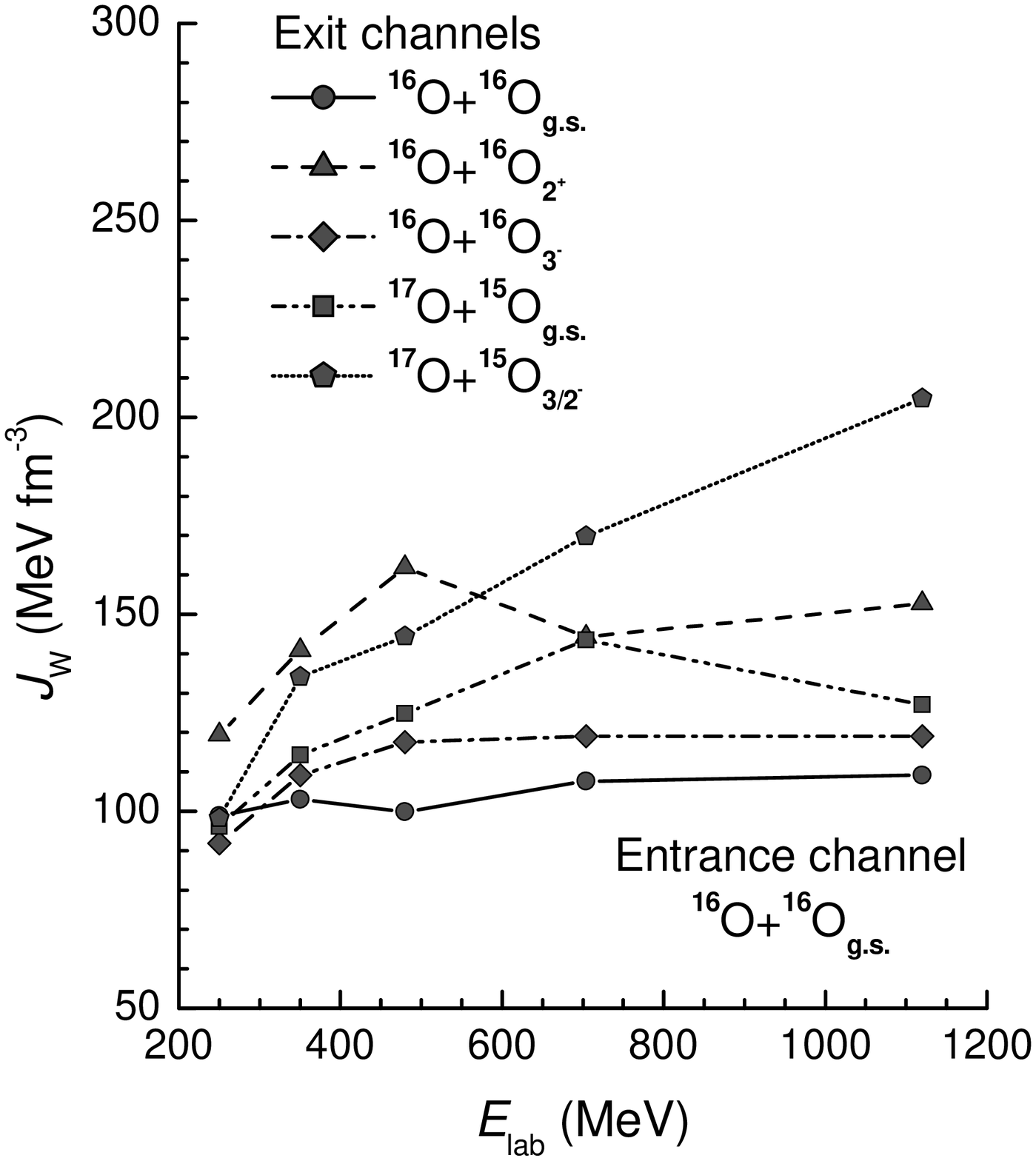}\vspace{-2.5cm}
\caption{Energy dependence of the volume integral $J_{\rm W}$ per interacting
nucleon pair (see Eq.~(17) in Ref.~\cite{Kho00}) of the best-fit WS imaginary
OP for the entrance channel \oo and exit channels of inelastic scattering \2x
and \3x and of one-neutron transfer to the ground state $^{15}$O$_{\rm g.s.}$
and excited state $^{15}$O$_{{3/2}^-}$. The lines are only to guide the eye.
Illustration taken from Ref.~\cite{Kho05}} \label{fig36}\end{center}
\end{figure}

With the strong nuclear rainbow pattern observed in the elastic \oo scattering
\cite{Kho00}, the one-neutron pickup \o17o15 reaction has been measured, in
parallel with the elastic \oo scattering, at the energies $E_{\rm lab}=250 \to
1120$ MeV \cite{Sti89,Boh93,Bar96,Nuo98}. These data were obtained for the
transitions to $^{15}$O in the 1p$_{1/2}$ ground state and in the 1p$_{3/2}$
excited state at 6.176 MeV, with $^{17}$O remaining in the 1d$_{5/2}$ ground
state in both cases. A detailed analysis \cite{Boh02} of the \o17o15 data based
on the full finite-range distorted wave Born approximation (DWBA) has found a
clear remnant of the nuclear rainbow in the \o17o15 transfer channel to the
g.s. state $^{15}$O$_{{1/2}^-}$, especially at the  rainbow energy of 350 MeV
(see Fig.~34). This rainbow remnant is, however, suppressed in the transfer
channel to the excited state $^{15}$O$^*_{{3/2}^-}$ due to a much stronger
absorption in the \x7o5o exit channel. Beside the elastic scattering and
transfer channels, the inelastic \InOO scattering has also been measured at
$E_{\rm lab}=250 \to 1120$ MeV  and the data were analyzed \cite{Kho05} in the
DWBA using the OP and inelastic form factor given by the folding model
\cite{KhoSat}. Although the refractive pattern of the inelastic \InOO
scattering was found to be much weaker compared to that observed in the elastic
scattering channel, the remnant of the rainbow could still be traced in the
inelastic scattering cross section up to $E_{\rm lab}=704$ MeV (see Fig.~35).
\begin{figure}[bht]
 \begin{center}\vspace{-0.5cm}
\includegraphics[angle=0,scale=0.30]{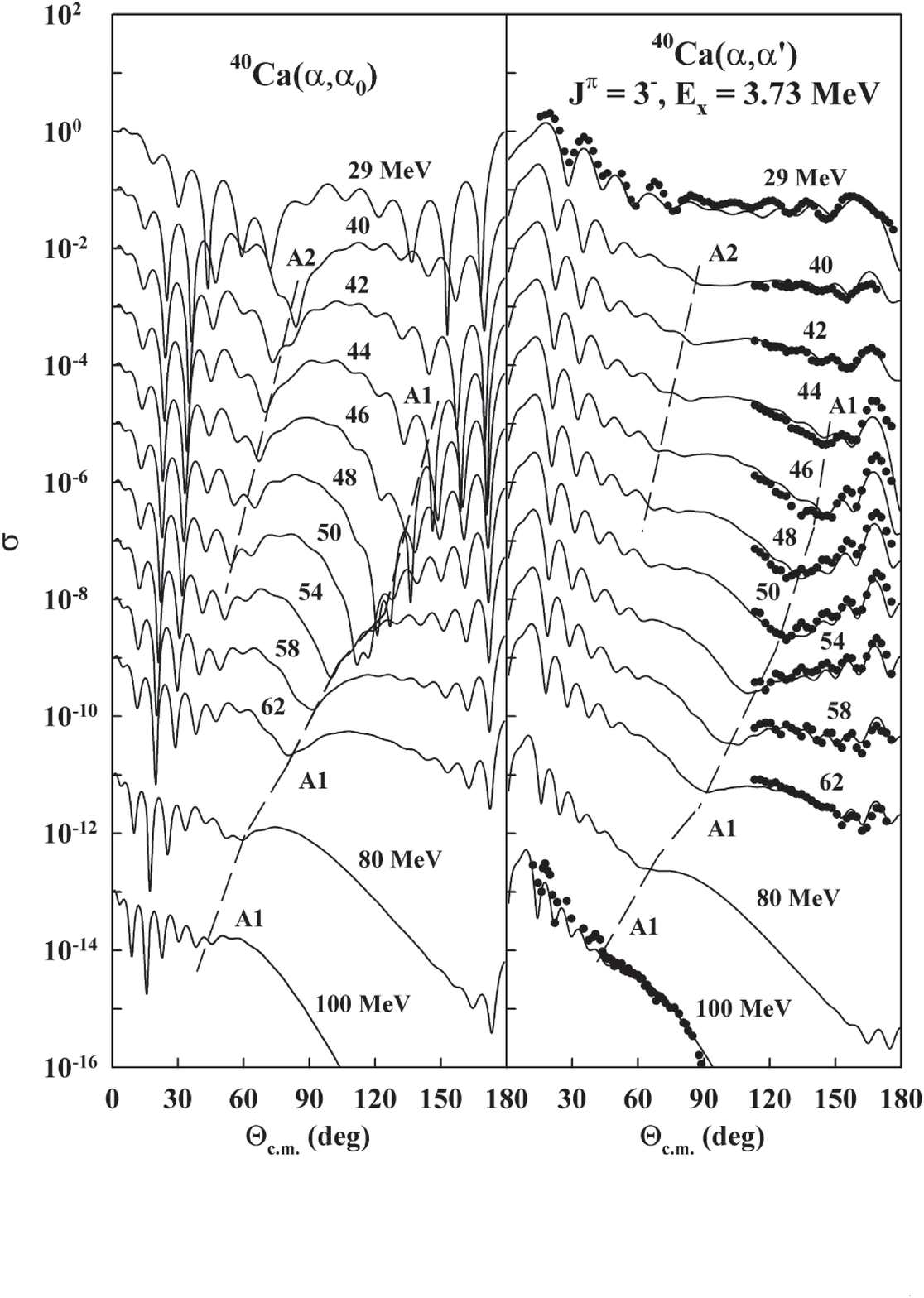}\vspace{-2cm}
\caption{The OM and DWBA results for the elastic (left) and inelastic ($J^\pi =
3^-, E_x = 3.73$~MeV, right) \aCa scattering, respectively,
 at the incident energies between 28 MeV and 100 MeV. The dashed lines show
the evolution of first (A1) and second (A2)
Airy minima. Illustration taken from Ref.~\cite{Michel04A}.}
   \end{center}
  \label{fig37}
\end{figure}

The weaker rainbow pattern found in the inelastic \oo scattering and \o17o15
transfer reaction is due to the enhanced absorptions in the exit channels
\cite{Kho05}. Namely, the DWBA description of the inelastic data covering a
large angular range and about 6 orders of the cross-section magnitude required
consistently an increased absorption in the exit channels of the inelastic
scattering and transfer reaction (see Fig.~36). This important result indicates
the need to have a realistic choice for the OP not only in the entrance but
also in the exit channel. The use of the same complex OP in both the entrance
and exit channels might lead to a large uncertainty in the deduced transition
strength if one follows the standard method of scaling the inelastic FF to
match the DWBA results to the measured angular distributions. This effect
should be essential in the study of the quasi-elastic \AA scattering induced by
unstable nuclei, where the partitions in the entrance and exit channels are
very differently bound.

Given the broad rainbow ``shoulder" clearly observed in the elastic \aA
scattering at refractive energies as discussed in Sec.~\ref{sec3}, its remnant
was seen also in the inelastic \aA scattering at similar energies \cite{Kho87}.
Recently, it has been shown by Michel and Ohkubo \cite{Michel04A,Michel04B}
that the Airy structure of the nuclear rainbow in the inelastic \aA scattering
is caused by the same interference mechanism as that established for the
elastic \aA scattering. For example, from the OM and DWBA results for the
elastic and inelastic \aCa scattering at different energies shown in Fig.~37,
one can see the evolution of the first Airy minimum (A1) in the measured
inelastic scattering cross sections at 44 to 50 MeV. As in the elastic case,
the inelastic scattering amplitude can also be decomposed into the
internal-wave and the barrier-wave components which lead essentially to the
same Airy interference pattern as that found in the elastic scattering. Note
that at each energy, the inelastic Airy minimum shows up at a slightly larger
angle than its elastic counterpart, due to the energy loss to the excitation of
the target state (compare locations of the Airy minima in the left and right
panels of Fig.~37).

Inelastic \aA rainbow scattering can also be used to study the
$\alpha$-particle condensation of nuclei. For example, the  0$^+$ (7.65 MeV)
state of $^{12}$C, the famous Hoyle state in the nucleosynthesis of Carbon, has
a dilute three-$\alpha$ cluster structure which has been discussed to be an
$\alpha$-condensate state \cite{Tohsaki01,Kokalova06}. This dilute structure
has been shown to lead to a more pronounced Airy structure (with a higher-order
Airy minimum) in the cross section of inelastic \aC scattering at $E_{\rm
lab}=140$ to 240 MeV \cite{Ohkubo04} compared to that in the elastic \aC
scattering at the same energies. Such effect is a strong indication that the
refraction caused by a big volume of the dilute 0$^+$ state is stronger than
that caused by the compact ground state of $^{12}$C, as illustrated in Fig.~38.
\begin{figure}[hbt]
 \begin{center}\vspace{-3cm}
\includegraphics[angle=0,scale=0.6]{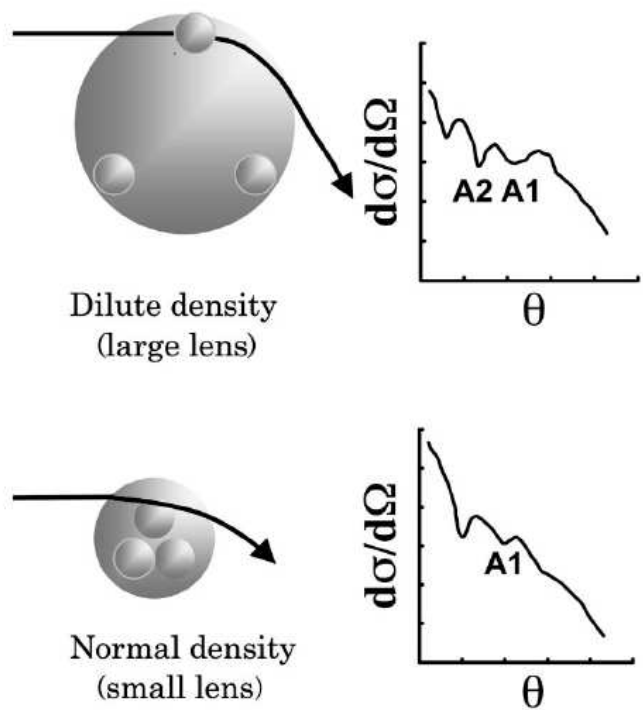}\vspace{-3cm}
\caption{Refraction in the \aC rainbow scattering from the ground state (small
lens) and the 0$^+$ (7.65 MeV) state (large lens) and the resulting Airy
structures in the angular distributions. Illustration taken from
Ref.~\cite{Ohkubo04}.}
   \end{center}
  \label{fig38}
\end{figure}

Finally, we note that the rainbow-like structure has also been observed in the
charge exchange reactions, like ($^{6}$Li,$^{6}$He) at $E_{\rm lab}=93$ MeV
\cite{Dem87}, ($^{3}$He,$t$) at $E_{\rm lab}=38$ and 60 MeV \cite{Bur94,Bur03}.
However, there has been no systematic study of the evolution of the rainbow
pattern with the incident energies as was done for the elastic, inelastic
scattering and one-neutron transfer reaction measured with the \oo system as
discussed above. It is, therefore, highly desirable to have more measurements
of the charge exchange reactions at refractive energies in order to have a
complete understanding of the nuclear rainbow structure in the quasi-elastic
nuclear scattering.

\section{Nuclear rainbow scattering studies with the loosely-bound and/or
 unstable nuclei}
\label{sec8}

\begin{figure}[hbt]
 \begin{center}\vspace{-1cm}
\includegraphics[angle=0,scale=0.60]{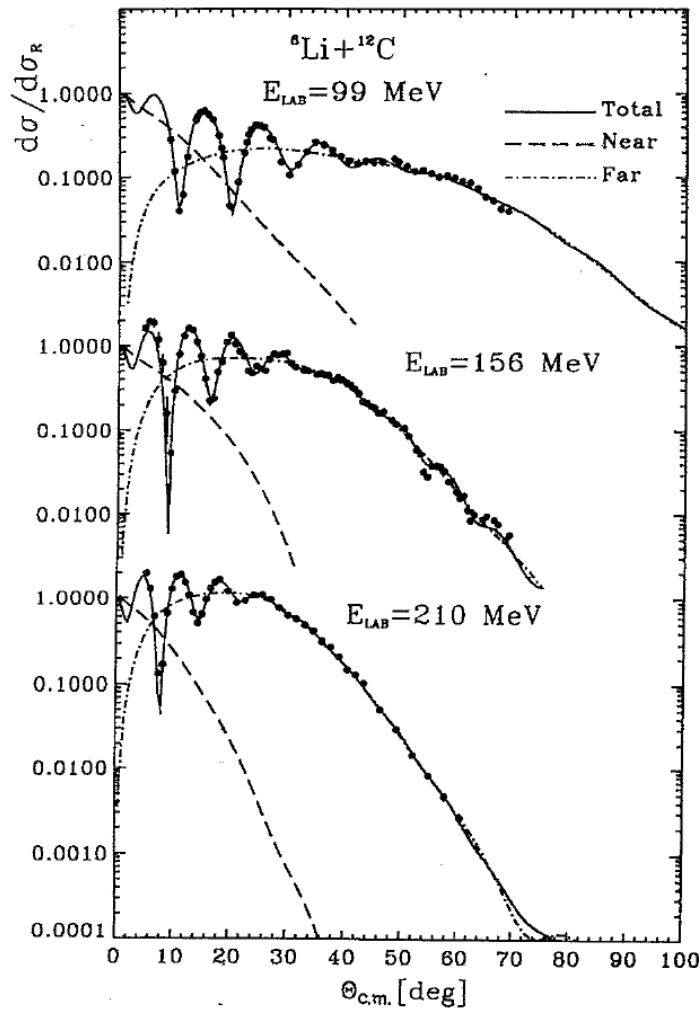}\vspace{-1.5cm}
\protect\caption{Decomposition of the elastic \Li6C scattering cross section
(solid curves) at $E_{\rm lab}=99, 156$ and 210 MeV \cite{Kho95b} into the
nearside (dashed curves) and farside (dash-dotted curves) components using
Fuller's method \cite{Ful75}. The dynamic polarization contribution from the
breakup of $^6$Li projectile to the OP was explicitly taken into account by a
spline potential. Illustration taken from Ref.~\cite{Kho95b}.}
   \end{center}
  \label{fig39}
\end{figure}
In general, the absorption becomes stronger for the loosely bound and/or
unstable nuclei and the rainbow-like structure is suppressed and harder to
observe. There are, however, interesting exceptions observed for the loosely
bound $^{6,7}$Li and $^9$Be nuclei for which the rainbow pattern (or its
remnant) has been observed in the elastic scattering as discussed in
Sec.~\ref{sec3}. It has been shown for the \Li6C system \cite{Kho95b} that even
after the contribution of the dynamic polarization potential caused by the
breakup of $^6$Li \cite{Sak86} is taken into account (which significantly
reduced the attractive strength of the real OP at the surface) the total \Li6C
optical potential still remains strongly refractive and can give rise to the
nuclear rainbow pattern in the elastic scattering. One can see from the
nearside/farside decomposition of the elastic \Li6C scattering amplitude
\cite{Kho95b} shown in Fig.~39 that the \Li6C system is indeed strongly
refractive, with a dominant farside (rainbow) scattering contribution at the
large angles.
\begin{figure}[hbt]
 \begin{center}\vspace{-0.5cm}
\includegraphics[angle=0,scale=0.35]{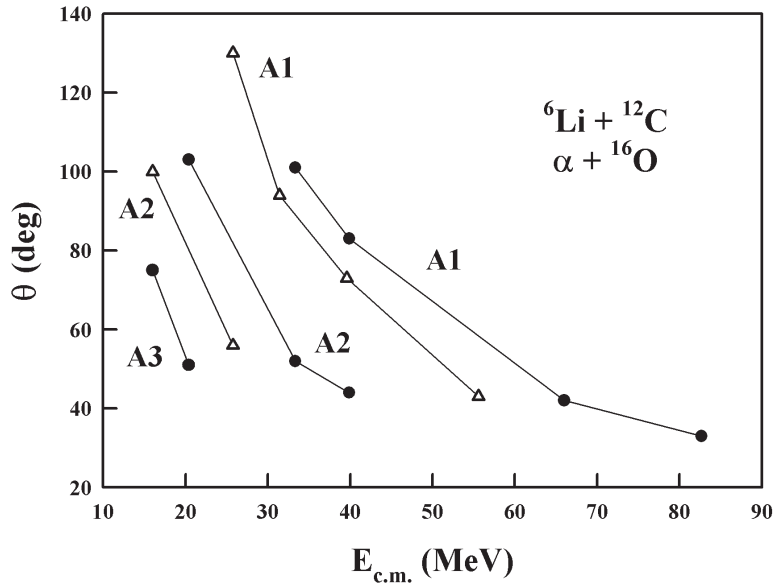}\vspace{0cm}
\caption{Energy dependence of the Airy minima location found in the \Li6C (dots)
and $\alpha+^{16}$O (triangles) elastic angular distributions. A1, A2 and A3 are
the first, second and third Airy minima, respectively. Illustration taken from
Ref.~\cite{Michel05}.}
   \end{center}
  \label{fig40}
\end{figure}

Recently, the elastic $^{6,7}$Li scattering has been shown to have the Airy
structure very similar to that observed in the \aCa and \oo elastic scattering
\cite{Michel05}. In Fig.~40 the angular locations of the Airy minima in the
elastic \Li6C scattering at different energies is compared with those found in
the elastic $\alpha$+$^{16}$O scattering. One can see that these two systems
have rather similar behaviors of the Airy minima, although for the former the
DPP caused by the breakup has reduced significantly the attractive strength of
the real OP at the surface and sub-surface distances \cite{Kho95b}. A strong
refractive behavior has also been established by Carstoiu {\it et al.}
\cite{Car04} in the large-angle elastic $^{6,7}$Li scattering from $^{9}$Be and
$^{12,13}$C targets. All this suggests that the refractive effect seen in
elastic $^{6,7}$Li scattering is not accidental and the most likely explanation
is that these projectiles have a well established $\alpha$-cluster structure
($\alpha+d$ and $\alpha+t$ for $^6$Li and $^7$Li, respectively). The observed
refractive pattern is probably due to a strong contribution by the
$\alpha$-core during the scattering process, and an explicit three-body
solution for the elastic $^{6,7}$Li scattering should give a more definitive
conclusion on the role of the $\alpha$-core.
\begin{figure}[hbt]
 \begin{center}\vspace{-3cm}
\includegraphics[angle=0,scale=0.6]{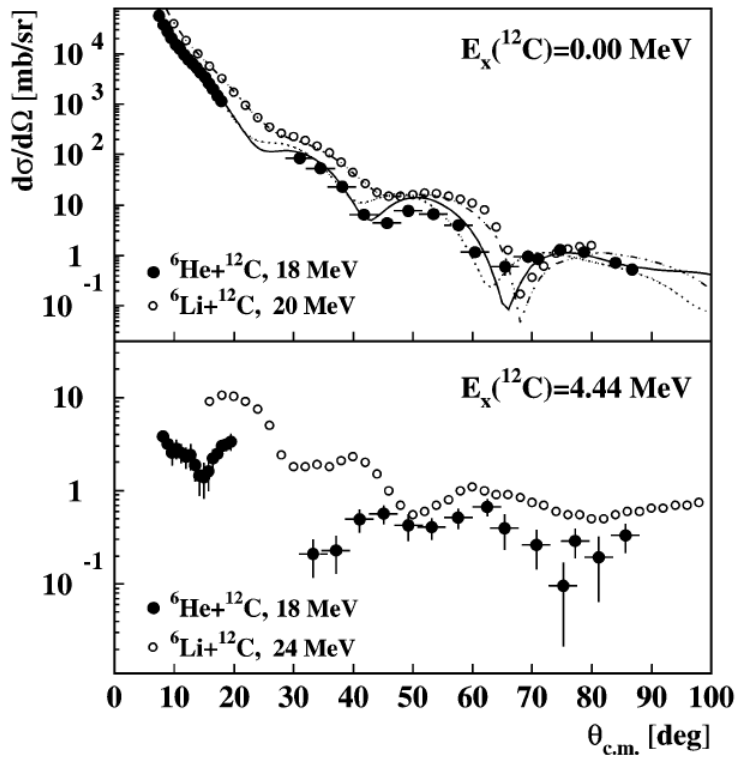}\vspace{-3cm}
\caption{The \He6C elastic and inelastic angular distributions measured at
$E_{\rm lab}=18$ MeV \cite{Milin04} compared to the \Li6C elastic data at 20
MeV and inelastic data at 24 MeV. The dash-dotted curve shows the OM
description of the elastic \Li6C scattering using the same nuclear OP as that
used for the $^6$He+$^{12}$C system (full curve). Illustration taken from
Ref.~\cite{Milin04}.}
\end{center}\label{fig41}
\end{figure}

$^{6}$He is one of the most studied unstable nuclei and it also has a well
established $\alpha$-cluster structure ($\alpha$-core + $2n$-halo). Therefore,
the similarity between the recently measured elastic \He6C scattering data at
$E_{\rm lab}=18$ MeV by Milin {\it et al.} \cite{Milin04} and elastic \Li6C
scattering data at about the same energy is very interesting (see Fig.~41). The
fact that the same nuclear OP gives reasonable description to both the \He6C
and \Li6C elastic scattering indicates that the contribution by the $\alpha$
core is nearly the same in both cases with a small shift in the elastic cross
section (obviously caused by the difference between the $2n$-halo and deuteron
state, and that between the two Coulomb potentials). The elastic \He6C
scattering at 38.3 MeV/nucleon has also been measured at GANIL by Lapoux {\it
et al.} \cite{Lapoux02}, but the data cover mainly the forward angles.
Nevertheless, a similarity between the \He6C and \Li6C elastic scattering at
this medium energy was found and the same realistic \aC potential has been used
successfully in both cases to estimate the DPP caused by the breakup of
$^{6}$He and $^6$Li \cite{Lapoux02}. All this suggests that the \He6C system
should be also strongly refractive with a significant farside (or
internal-wave) contribution at large scattering angles. The future measurement
to find the rainbow structure in the elastic \He6C and/or $^6$He+$^{16}$O
scattering would be very helpful, not only to give more information on the
nuclear rainbow but also provide valuable scattering data to further probe the
$2n$-halo wave function of this Borromean nucleus.
\begin{figure}[hbt]
 \begin{center}
\includegraphics[angle=0,scale=0.5]{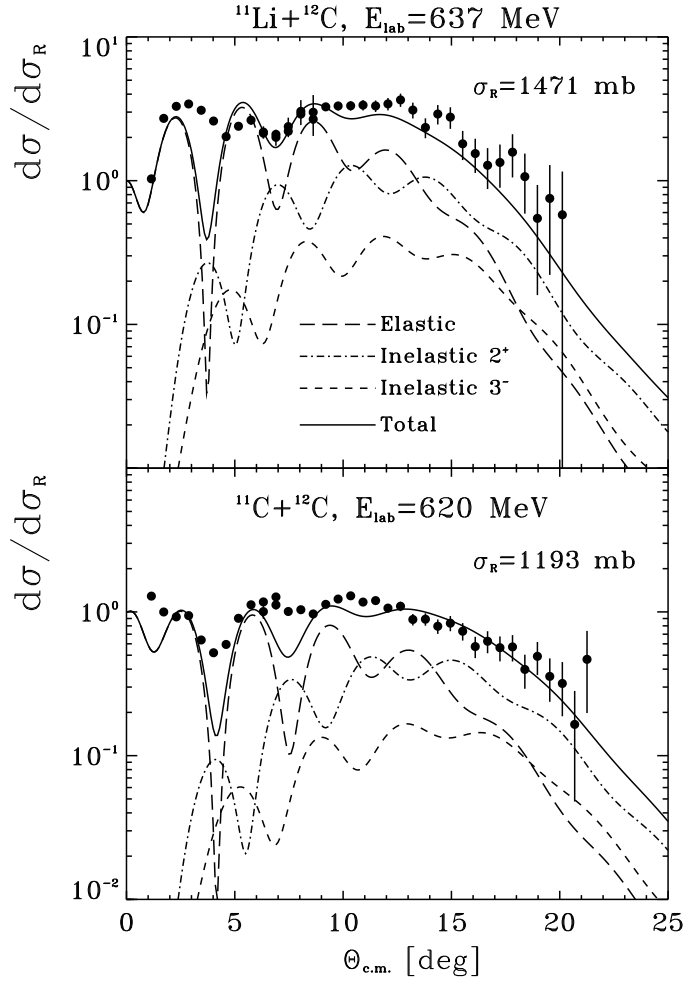}\vspace{0.5cm}
\caption{Total, elastic and inelastic scattering cross sections given by the
semi-microscopic DWBA calculation \cite{Khoa95} compared with the quasi-elastic
data \cite{Kol92} for the $^{11}$Li+$^{12}$C system at $E_{\rm lab}= 637$ MeV
and $^{11}$C+$^{12}$C system at 620 MeV. The \LiC11 optical potential was added
by a (complex) dynamic polarization potential to account for the breakup of
$^{11}$Li projectile. Illustration taken from Ref.~\cite{Khoa95}.}
\end{center}\label{fig42}
\end{figure}

The unstable $^{11}$Li nucleus has also been studied extensively during the
last two decades. Given the fact that light HI systems like \cc or \oo show
strong refractive effects as discussed above, the \Li6C system was considered
as the most likely case where the refractive rainbow pattern might show up in
the elastic scattering. Such refractive scattering data, if measured
accurately, could allow one to probe the halo structure ($^9$Li-core +
$2n$-halo) of $^{11}$Li in details. Short after some exploratory, theoretical
studies were made for the \LiC11 elastic scattering \cite{Sat91,Yab92}, the
measurement of quasi-elastic scattering of $^{11}$Li and $^{11}$C from the
$^{12}$C target at $E_{\rm lab}=637$ and 620 MeV, respectively, has been
performed by Kolata {\sl et al.} \cite{Kol92}. Although the enhanced refraction
predicted by Satchler {\sl et al.} \cite{Sat91} for the \LiC11 system has been
confirmed, no evidence was found for an Airy minimum in the farside scattering
\cite{Kol92}. These interesting data have inspired number of OM analyses
\cite{Mer93,Khoa95,Coo95,Car96,Kho96} as well as the few-body calculations
\cite{Tho93,Kha95} which treated explicitly the projectile breakup $^{11}$Li
$\to\ ^9$Li$ + 2n$.
\begin{figure}[hbt]
 \begin{center}\vspace{-2cm}
\includegraphics[angle=270,scale=0.6]{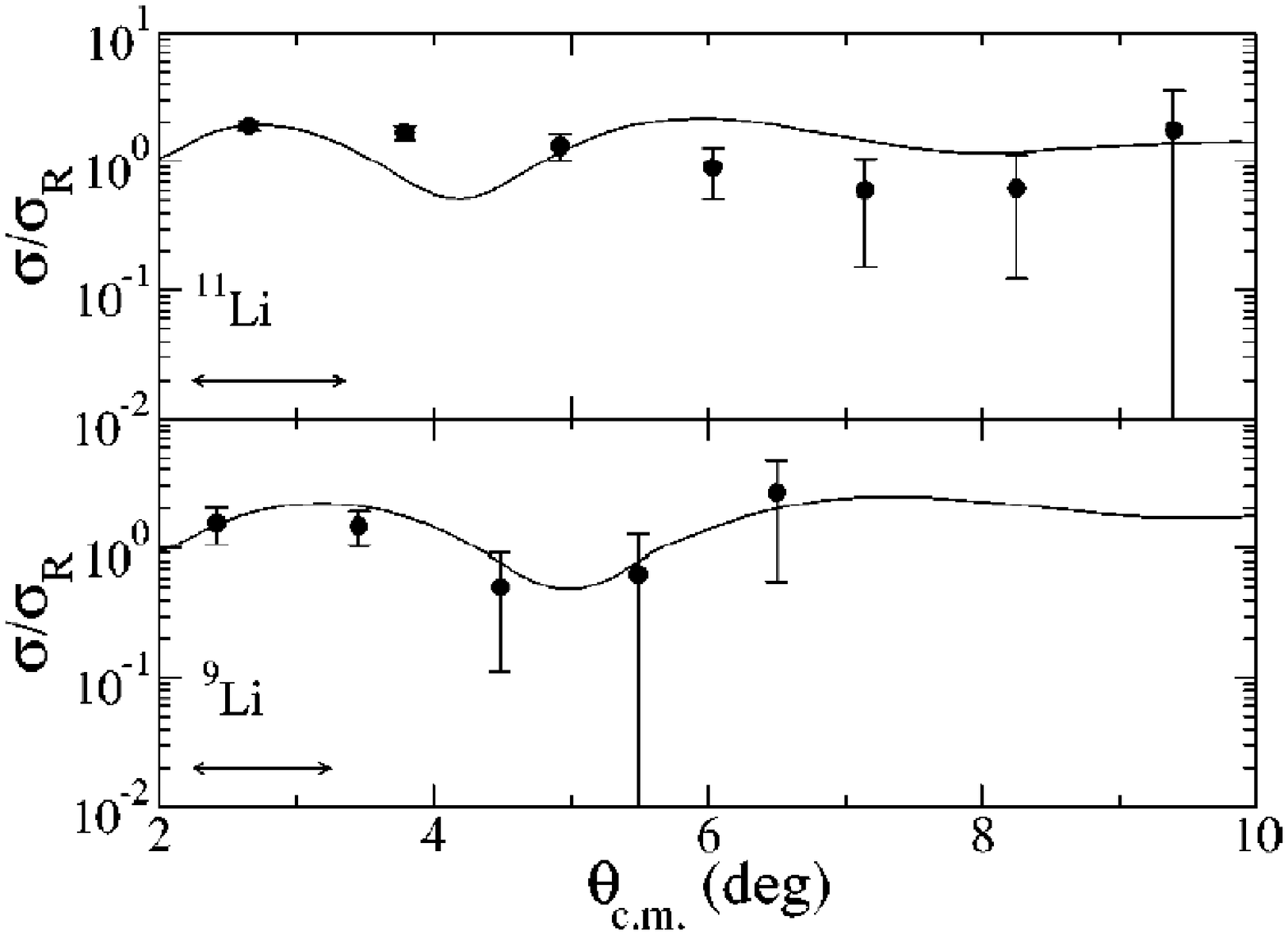}\vspace{-1cm}
\caption{Elastic \LiC11 angular distribution measured at 50 MeV/nucleon
\cite{Pet03} compared to the Glauber-model calculation folded with the
experimental resolution. Illustration taken from Ref.~\cite{Pet03}.}
\end{center}\label{fig43}
\end{figure}
Since the experimental energy resolution \cite{Kol92} did not allow for the
separation of the true elastic scattering from inelastic scattering to the
low-lying excited states, one usually adds the cross sections of inelastic
scattering to the lowest 2$^+$ and 3$^-$ states of $^{12}$C to the elastic
scattering cross section for a direct comparison with the data. For example,
our DWBA analysis \cite{Khoa95} of the elastic and inelastic scattering of
$^{11}$Li and $^{11}$C from $^{12}$C, using the semi-microscopic OP added by a
realistic DPP to account for the breakup effect, has given a good description
of both the quasi-elastic scattering data and total reaction cross section (see
Fig.~42). However, most of the OM analyses of the elastic \LiC11 scattering
$E_{\rm lab}=637$ MeV predict a deep nearside/farside interference minimum at
forward angles (see upper part of Fig.~42) which is absent in the measured
angular distribution \cite{Kol92}. Since this interference minimum has been
observed in experiment \cite{Kol92} for the $^{11}$C+$^{12}$C system (lower
part of Fig.~42), a question arises whether the absence of this minimum in the
\LiC11 case is an effect due to the weakly bound nature of $^{11}$Li. To solve
this puzzle, a new experiment on the elastic scattering of $^{9,11}$Li from the
$^{12}$C target at energy around 50 MeV/nucleon has been performed recently by
Peterson {\sl et al.} \cite{Pet03} using the S800 spectrograph at MSU. The
improved energy resolution in this experiment was sufficient to exclude the
inelastic contributions to the scattering at the forward angles, and the
measured elastic \LiC11 cross section shows again no evidence of a
nearside/farside interference minimum (see upper part of Fig.~43). Therefore,
the interference minimum is either missing or greatly attenuated in this case.
Since the deep interference minimum  predicted for the \LiC11 system (see
dashed curve in upper part of Fig.~42) is associated with the Fraunhofer
crossover $\bar\Theta$ (see discussion in Sec.~\ref{sec2}), the spacing between
this minimum and the nearest maximum should be $\Delta\Theta_{\rm c.m.}\simeq
2^\circ$.  However, after averaging over the angular bin, one could obtain only
two data points within such $\Delta\Theta_{\rm c.m.}$ spacing (see Fig.~43) and
it is likely that the interference minimum is so narrow and highly attenuated
that could not be determined from the present experiment \cite{Pet03}. It is
interesting to note that this interference minimum is still present in the
results of the Glauber-model calculation even after the calculated cross
section was folded with the experimental angular resolution (see upper part of
Fig.~43). Given a strong refractive effect predicted \cite{Kho96} for the
elastic $^{11}$Li scattering from Carbon target at energy around 30 MeV/nucleon
(see, e.g, Fig.~13 in Ref.~\cite{Kho96}), it is highly desirable to have
further high-precision measurement for the elastic \LiC11 scattering at 30 - 50
MeV/nucleon, which would not only give a final answer to this intriguing
question but also provide valuable scattering data to further probe the
$2n$-halo wave function of $^{11}$Li.

We finally note that the refractive pattern predicted by us \cite{Kho96} for
the elastic $^8$He scattering around 30 MeV/nucleon has been confirmed in the
elastic $^8$He+$^4$He scattering data measured at 26 MeV/nucleon by Wolski {\it
et al.} \cite{Wolski02}, where one observed a broad shoulder-like maximum of
the elastic cross section at large angles which is dominated by the farside
scattering. These data are, however, not complete and lack the small-angle
diffractive part which makes an accurate OM or folding model analysis
difficult. It is obvious that more experiments for the elastic scattering of
light unstable nuclei, such as the $\alpha$-clustered projectiles or unstable
p-shell nuclei, are strongly needed to reveal further features of the nuclear
rainbow which can then be used to probe their exotic structures in the OM
analysis.

\section{Summary}

We have presented an overview of the nuclear rainbow, a fascinating phenomenon
observed in the elastic $\alpha$- and light HI scattering at medium energies
which can well be understood based on the basic concepts of the optical model
description of the elastic scattering. Despite the striking similarity in the
interference structure between the nuclear rainbow and atmospheric rainbow, the
former has proven to be much harder to observe experimentally. It is important
to stress that the occurrence of the rainbow pattern in the \aA and light HI
elastic scattering is due to a \emph{strong} mean field caused by the two
nuclei overlapping each other. The attractive strength of the nucleon mean
field (or nucleon OP) is much weaker and the rainbow pattern is, therefore,
never observed in the \nA scattering. Moreover, the following three main
physical conditions must be met for a clear nuclear rainbow to be observed.

First, the real optical potential must be strongly attractive to cause the
refractive scattering. Such a deep \AA OP has been shown to be due to a strong
mean-field attraction at small dinuclear distances or at high overlap
densities. Second, the absorption in the nucleus-nucleus system must be weak
(which is the case for the tightly bound $\alpha$-particle or for some light
p-shell nuclei), so that the farside trajectories (passing through small
distances or high overlap densities) can survive in the elastic scattering
channel. Third, the incident energy should be high enough for these farside
trajectories to appear in the elastic scattering cross section at medium and
large scattering angles. Note that if the energy is too low, the scattering is
dominated by the diffractive Fraunhofer pattern in the whole observable angular
range, with the primary rainbow shifted into the unphysical angular region
beyond $\Theta_{\rm c.m.}=180^{\circ}$. However, if the energy is too high, the
refractive part moves to the forward angles and mixes with the diffractive part
of the elastic cross section and the rainbow features become difficult to
extract.

The semiclassical decomposition of the observed angular distributions into the
nearside and farside (or barrier-wave and internal-wave) contributions has
proven to be an important and powerful tool to understand the interference
structure of the nuclear rainbow. The same technique also allows us to probe
the \AA OP at different interaction distances based on the observed scattering
pattern. In particular, the nearside/farside interference at forward angles
give information about the OP at the surface. At larger angles, only the
farside scattering survives which gives rise to the Airy oscillation pattern of
the nuclear rainbow. Since the Airy structure is caused by the \AA interaction
at small distances, the rainbow scattering data of high accuracy can be used to
determine the \AA OP (and the associated potential family) with much less
ambiguity.

Given a weak absorption associated with the rainbow pattern in the elastic \aA
and \AA scattering, the rainbow scattering data can be used to probe the
density dependence of the in-medium NN interaction based on the folding model
analysis of the elastic scattering. Most of the elastic rainbow scattering data
were found to be best described by a \emph{deep} real OP given by the
double-folding calculation using a density dependent M3Y interaction which
gives a nuclear incompressibility $K \approx 230-260$ MeV in the HF calculation
of nuclear matter. This result confirms a rather soft EOS for symmetric nuclear
matter and the deduced $K$-value agrees very well with that of the latest
nuclear structure studies of the isoscalar giant monopole resonances in
medium-mass nuclei.

The refractive rainbow-like structures were also observed in other quasi-elastic
scattering reactions, as well as in the elastic scattering measured with the
loosely bound or unstable nuclei. However, more experiments on the
quasi-elastic scattering induced by the light unstable nuclei are needed to
learn more on the fascinating features of the nuclear rainbow, which can be
used to probe the wave functions of these unstable isotopes.

\section*{Acknowledgments}
The authors are indebted to many colleagues for their helpful discussions and
correspondences on this interesting topic during the last two decades. In
particular, D.T.K. is grateful to Ray Satchler for his numerous discussions
during the years of collaboration which have shed light on many issues discussed
in this topical review. A clarifying comment by Gianluca Col\`o on the nuclear
structure studies of the giant monopole resonance is also appreciated. The
present research has been supported, in part, by the Alexander-von-Humboldt
Stiftung of Germany, Hahn-Meitner-Institut Berlin, the German Ministry of
Research and Technology BMFT under the contract 06OB472D/4, Vietnam Natural
Science Council and Vietnam Atomic Energy Commission (VAEC). S.O. has been
supported by the Japan Society for Promotion of Science (JSPS) and the Yukawa
Institute for Theoretical Physics.

\end{document}